\documentclass[english,draftcls,onecolumn]{IEEEtran}

\usepackage{enumitem}

\usepackage{hyperref}
\usepackage[T1]{fontenc}
\usepackage[utf8]{inputenc}
\usepackage{verbatim}
\usepackage{float}
\usepackage[cmex10]{amsmath}
\usepackage{amssymb}
\usepackage{graphicx}
\usepackage{epstopdf}
\usepackage{stmaryrd}
\usepackage{cite}
\makeatletter
\floatstyle{ruled}
\newfloat{algorithm}{tbp}{loa}
\providecommand{\algorithmname}{Algorithm}
\floatname{algorithm}{\protect\algorithmname}

\usepackage{algorithm,algpseudocode}

\date{}

\@ifundefined{showcaptionsetup}{}{%
 \PassOptionsToPackage{caption=false}{subfig}}
\usepackage{subfig}
\makeatother

\usepackage{babel}
\newtheorem{thm}{Theorem}
\newtheorem{prop}{Proposition}
\newtheorem{lem}{Lemma}
\newtheorem{cor}{Corollary}
\newtheorem{rem}{Remark}

\newtheorem{conjecture}{Conjecture}
\newtheorem{defn}{Definition}
\newtheorem{example}{Example}

\newtheorem{fact}{Fact}

\begin{document}
\global\long\def\vct#1{\boldsymbol{#1}}%
\global\long\def\mat#1{\boldsymbol{#1}}%
\global\long\def\opvec{\text{vec}}%
\global\long\def\optr{\mbox{tr}}%
\global\long\def\opmat{\mbox{mat}}%
\global\long\def\opdiag{\mbox{diag}}%

\global\long\def\t#1{\widetilde{#1}}%
\global\long\def\h#1{\widehat{#1}}%
\global\long\def\abs#1{\left\lvert #1\right\rvert }%
\global\long\def\norm#1{\lVert#1\rVert}%

\global\long\def\inprod#1{\langle#1\rangle}%
\global\long\def\set#1{\left\{  #1\right\}  }%
\global\long\def\bydef{\overset{\text{def}}{=}}%

\global\long\def\teq#1{\overset{#1}{=}}%
\global\long\def\tleq#1{\overset{#1}{\leq}}%
\global\long\def\tgeq#1{\overset{#1}{\geq}}%
\global\long\def\tapprox#1{\overset{#1}{\approx}}%
\global\long\def\tle#1{\overset{#1}{<}}%

\global\long\def\EE{\mathbb{E}\,}%
\global\long\def\EEk{\mathbb{E}_{k}\,}%

\global\long\def\R{\mathbb{R}}%
\global\long\def\E{\mathbb{E}}%
\global\long\def\P{\mathbb{P}}%

\global\long\def\va{\boldsymbol{a}}%
\global\long\def\vb{\boldsymbol{b}}%
\global\long\def\vc{\boldsymbol{c}}%
\global\long\def\vd{\boldsymbol{d}}%
\global\long\def\ve{\boldsymbol{e}}%
\global\long\def\vf{\boldsymbol{f}}%
\global\long\def\vg{\boldsymbol{g}}%

\global\long\def\vh{\boldsymbol{h}}%
\global\long\def\vi{\boldsymbol{i}}%
\global\long\def\vj{\boldsymbol{j}}%
\global\long\def\vk{\boldsymbol{k}}%
\global\long\def\vl{\boldsymbol{l}}%

\global\long\def\vm{\boldsymbol{m}}%
\global\long\def\vn{\boldsymbol{n}}%
\global\long\def\vo{\boldsymbol{o}}%
\global\long\def\vp{\boldsymbol{p}}%
\global\long\def\vq{\boldsymbol{q}}%
\global\long\def\vr{\boldsymbol{r}}%

\global\long\def\vs{\boldsymbol{s}}%
\global\long\def\vt{\boldsymbol{t}}%
\global\long\def\vu{\boldsymbol{u}}%
\global\long\def\vv{\boldsymbol{v}}%
\global\long\def\vw{\boldsymbol{w}}%
\global\long\def\vx{\boldsymbol{x}}%
\global\long\def\vy{\boldsymbol{y}}%
\global\long\def\vz{\boldsymbol{z}}%

\global\long\def\vtheta{\boldsymbol{\theta}}%
\global\long\def\vxi{\boldsymbol{\xi}}%
\global\long\def\vdelta{\boldsymbol{\delta}}%
\global\long\def\veta{\vct{\eta}}%
\global\long\def\vlambda{\boldsymbol{\lambda}}%
\global\long\def\vbeta{\boldsymbol{\beta}}%

\global\long\def\mA{\boldsymbol{A}}%
\global\long\def\mB{\boldsymbol{B}}%
\global\long\def\mC{\boldsymbol{C}}%
\global\long\def\mD{\boldsymbol{D}}%
\global\long\def\mE{\boldsymbol{E}}%
\global\long\def\mF{\boldsymbol{F}}%
\global\long\def\mG{\boldsymbol{G}}%

\global\long\def\mH{\boldsymbol{H}}%
\global\long\def\mI{\boldsymbol{I}}%
\global\long\def\mJ{\boldsymbol{J}}%
\global\long\def\mK{\boldsymbol{K}}%
\global\long\def\mL{\boldsymbol{L}}%
\global\long\def\mM{\boldsymbol{M}}%
\global\long\def\mN{\boldsymbol{N}}%

\global\long\def\mO{\boldsymbol{O}}%
\global\long\def\mP{\boldsymbol{P}}%
\global\long\def\mQ{\boldsymbol{Q}}%
\global\long\def\mR{\boldsymbol{R}}%
\global\long\def\mS{\boldsymbol{S}}%
\global\long\def\mT{\boldsymbol{T}}%
\global\long\def\mU{\boldsymbol{U}}%

\global\long\def\mV{\boldsymbol{V}}%
\global\long\def\mW{\boldsymbol{W}}%
\global\long\def\mX{\boldsymbol{X}}%
\global\long\def\mY{\boldsymbol{Y}}%
\global\long\def\mZ{\boldsymbol{Z}}%

\global\long\def\mLa{\boldsymbol{\Lambda}}%
\global\long\def\mOm{\boldsymbol{\Omega}}%

\global\long\def\calS{\mathcal{S}}%
\global\long\def\calN{\mathcal{N}}%
\global\long\def\calL{\mathcal{L}}%
\global\long\def\calD{\mathcal{D}}%
\global\long\def\calV{\mathcal{V}}%
\global\long\def\calW{\mathcal{W}}%

\global\long\def\a{\alpha}%
\global\long\def\b{\beta}%
\global\long\def\m{\mu}%
\global\long\def\n{\nu}%
\global\long\def\g{\gamma}%
\global\long\def\s{\sigma}%
\global\long\def\e{\epsilon}%
\global\long\def\w{\omega}%
\global\long\def\veps{\varepsilon}%

\global\long\def\T{\top}%
\global\long\def\d{\text{d}}%

\global\long\def\nt{\left\lfloor nt\right\rfloor }%
\global\long\def\ns{\left\lfloor ns\right\rfloor }%
\global\long\def\textif{\text{if }}%
\global\long\def\otherwise{\text{otherwise}}%
\global\long\def\st{\text{s.t. }}%

\global\long\def\lmax{\lambda_{\max}}%
\global\long\def\lmin{\lambda_{\min}}%

\global\long\def\tvy{\t{\boldsymbol{y}}}%
\global\long\def\tc{\t c}%
\global\long\def\ttau{\t{\tau}}%
\global\long\def\tf{\t f}%
\global\long\def\th{\t h}%
\global\long\def\tq{\t q}%
\global\long\def\tz{\t z}%

\global\long\def\tva{\t{\boldsymbol{a}}}%
\global\long\def\tvw{\t{\boldsymbol{w}}}%
\global\long\def\tw{\t w}%
\global\long\def\aver#1#2{\mathtt{aver}(#1,#2)}%
\global\long\def\diffy#1{G(#1)}%

\global\long\def\diff{g}%
\global\long\def\diffi{g}%
\global\long\def\bdp{b}%

\global\long\def\opty#1{\eta(#1)}%

\global\long\def\meanc#1#2#3{\mathtt{mean}_{#3}(#1,\,#2)}%

\global\long\def\conde{\mathcal{T}}%

\global\long\def\asconv{\overset{a.s.}{\rightarrow}}%
\global\long\def\pconv{\overset{\mathbb{P}}{\rightarrow}}%
\global\long\def\wconv{\overset{\calL}{\rightarrow}}%

\global\long\def\iid{\overset{i.i.d.}{\sim}}%

\global\long\def\dmtx{\mA}%
\global\long\def\dmtxi{A}%

\global\long\def\sgl{\boldsymbol{\beta}}%
\global\long\def\noise{\boldsymbol{w}}%
\global\long\def\sgli{\beta}%
\global\long\def\est{\vb}%
\global\long\def\esti{b}%
\global\long\def\res{\vr}%
\global\long\def\err{\vv}%

\global\long\def\noisei{w}%
\global\long\def\sol{\widehat{\sgl}}%
\global\long\def\soli{\hat{\beta}}%
\global\long\def\equinoise{\mZ}%
\global\long\def\equinoisei{Z}%

\global\long\def\argmin#1{\underset{#1}{\text{argmin }}}%
\global\long\def\argmax#1{\underset{#1}{\text{argmax }}}%

\global\long\def\targmin#1{\text{argmin}_{#1}}%
\global\long\def\targmax#1{\text{argmax}_{#1}}%

\global\long\def\regu{J_{\vlambda}}%
\global\long\def\sregu{J}%
\global\long\def\comset{\mathcal{S}}%
\global\long\def\regulam#1{J_{#1}}%

\global\long\def\prox{\eta}%
\global\long\def\proxfree{f}%
\global\long\def\vprox{\veta}%
\global\long\def\tprox{\text{Prox}}%
\global\long\def\sgn{\text{sign}}%
\global\long\def\tproxl{\text{Prox}_{\vlambda}}%
\global\long\def\diag{\text{diag}}%

\global\long\def\Moreau{\mathcal{F}}%

\global\long\def\inte{\text{Int}}%
\global\long\def\bnd{\text{Bd}}%
\global\long\def\ndset{\mathcal{D}}%
\global\long\def\ndlset{\mathcal{B}}%

\global\long\def\cdfy{F_{Y}}%
\global\long\def\cdfay{F_{|Y|}}%
\global\long\def\icdfy{F_{Y}^{-1}}%
\global\long\def\icdfay{F_{|Y|}^{-1}}%

\global\long\def\yip{y_{i}^{(p)}}%
\global\long\def\lambdaip{\lambda_{i}^{(p)}}%
\global\long\def\uip{u_{i}^{(p)}}%
\global\long\def\lipset{\text{Lip}_{1}[0,\infty)}%

\global\long\def\ryip{y_{r,i}^{(p)}}%
\global\long\def\rlambdaip{\lambda_{r,i}^{(p)}}%
\global\long\def\ujp#1{u_{#1}^{(p)}}%

\global\long\def\rcsvy{\vy_{r}^{(p)}}%
\global\long\def\rcsvl{\vlambda_{r}^{(p)}}%
\global\long\def\rcsvg{\vg_{r}^{(p)}}%

\global\long\def\rcsyj#1{y_{r,#1}^{(p)}}%
\global\long\def\rcslj#1{\lambda_{r,#1}^{(p)}}%
\global\long\def\rcsgj#1{g_{r,#1}^{(p)}}%

\global\long\def\ltwospace{L_{2}(\R)}%
\global\long\def\unifdist{\text{Unif}}%

\global\long\def\indicatorfn{\mathbb{I}}%

\global\long\def\approxsol{\mathbb{\widetilde{\sgl}}}%
\global\long\def\approxdif{\mathring{\vv}}%

\global\long\def\approxsoli{\mathbb{\widetilde{\sgli}}}%
\global\long\def\approxdifi{\mathring{v}}%

\global\long\def\vapproxsol{\tilde{\vv}}%
\global\long\def\vapproxsoli{\tilde{v}}%

\global\long\def\bdball{\mathcal{B}}%
\global\long\def\outbdball{\mathcal{B}^{o}}%

\global\long\def\Lipndset{\mathcal{I}}%
\global\long\def\Lyspace{\mathcal{H}}%
\global\long\def\Realizableset{\mathcal{M}}%

\global\long\def\minsigmaI{I_{B}}%

\global\long\def\sg{\vs}%
\global\long\def\Sgobj{\mathcal{S}}%
\global\long\def\SgAOobj{S}%

\global\long\def\approxsg{\mathring{\vs}}%
\global\long\def\vapproxsgi{\mathring{s}}%

\global\long\def\dualball{C_{\vlambda}}%

\global\long\def\sigOpt{\sigma_{*}}%
\global\long\def\tauOpt{\tau_{*}}%
\global\long\def\theOpt{\theta_{*}}%
\global\long\def\YOpt{Y_{*}}%
\global\long\def\PsiOpt{\Psi_{*}}%

\global\long\def\optprox{\prox^{\text{opt}}}%
\global\long\def\optproxfree{\proxfree^{\text{opt}}}%
\global\long\def\optsig{\sigma_{\text{opt}}}%
\global\long\def\opttau{\tau_{\text{opt}}}%
\global\long\def\optlam{\Lambda_{\text{opt}}}%
\global\long\def\optY{Y_{0}}%

\global\long\def\stmajor{\prec_{S}}%
\global\long\def\majorized{\prec}%

\global\long\def\mealim{\mu}%
\global\long\def\meajoint#1#2{\mu_{#1,#2}}%
\global\long\def\meamarginal#1{\mu_{#1}}%

\global\long\def\Wspace{\mathcal{P}_{2}}
\global\long\def\Lamspace{\mathcal{P}_{\Lambda}}
\global\long\def\tLamspace{\widetilde{\mathcal{P}}_{\Lambda}}

\global\long\def\empiquant#1{F^{-1}_{#1}}%
\global\long\def\probthresh{q_{0}^{*}} %
\global\long\def\probthreshzero{q_{0}} %

\global\long\def\minsigsol{\sigma_{0}}%
\global\long\def\mintausol{\tau_{0}}%
\global\long\def\ythresh{y_{\text{th}}^{*}} %
\global\long\def\ythreshzero{y_{0}} %

\global\long\def\limpower{\mathcal{P}} %
\global\long\def\softfunset{\mathcal{F}_{\alpha,\sigma}} %
\global\long\def\toptsig{\tilde{\sigma}_{\text{opt}}}%

\global\long\def\toptsig{\sigma_{\text{opt},\alpha}}%
\global\long\def\toptY{Y_{0,\alpha}}%
\global\long\def\tminsigsol{\sigma_{0,\alpha}}%
\global\long\def\toptlam{\Lambda_{\text{opt},\alpha}}%
\global\long\def\tmintausol{\tau_{0,\alpha}}%
\global\long\def\tythreshzero{y_{0,\alpha}}%

\title{SLOPE for Sparse Linear Regression: Asymptotics and Optimal Regularization}

 \author{%
   \IEEEauthorblockN{Hong Hu and Yue M. Lu}
    \thanks{H. Hu and Y. M. Lu are with the John A. Paulson School of Engineering and Applied Sciences, Harvard University, Cambridge, MA 02138, USA. (e-mails: honghu@g.harvard.edu and yuelu@seas.harvard.edu). This work was supported by the Harvard FAS Dean's Fund for Promising Scholarship, and by the US National Science Foundation under grants CCF-1718698 and CCF-1910410. The preliminary version of this work has been presented at the 2019 Signal Processing with Adaptive Sparse Structured Representations (SPARS) workshop and the 2019 IEEE International Symposium on Information Theory (ISIT) \cite{Hu2019AsymptoticsAO}.}
 }



\maketitle


\begin{abstract}
In sparse linear regression, the SLOPE estimator generalizes LASSO by penalizing different coordinates of the estimate according to their magnitudes.
In this paper, we present a precise 
performance characterization of SLOPE in the asymptotic regime where the number of unknown
parameters grows in proportion to the number of observations. Our asymptotic characterization enables us to derive
the fundamental limits of SLOPE in both estimation and variable selection settings. We also provide a computational feasible way to optimally design the regularizing sequences such that the fundamental limits are reached. In both settings, we show that the optimal design problem can be formulated as certain infinite-dimensional convex optimization problems, which have efficient and accurate finite-dimensional approximations.
Numerical simulations verify all our asymptotic predictions. They demonstrate the superiority of our optimal regularizing sequences over other designs used in the existing literature.
\end{abstract} 

\section{Introduction}
\subsection{Motivation and Problem Setup}
In sparse linear regression, we seek to estimate a sparse vector $\sgl \in\mathbb{R}^{p}$ from
\begin{equation}\label{eq:model}
\vy=\dmtx\sgl+\noise,
\end{equation}
where $\dmtx\in\mathbb{R}^{n\times p}$ is the design matrix and $\noise$ denotes the observation noise.
In this paper, we study the \emph{sorted $\ell_{1}$ penalization estimator} (SLOPE) \cite{bogdan2015slope} (see also \cite{zhong2012efficient, zeng2014decreasing}), a new paradigm for sparse linear regression. Given a non-decreasing regularization sequence $\boldsymbol{\lambda} = [\lambda_1, \lambda_2, \ldots, \lambda_p]^\top$ with  $0\leq\lambda_{1}\leq\lambda_{2}\leq\cdots\leq\lambda_{p}$, SLOPE estimates $\sgl$ by solving the following optimization
problem
\begin{equation}
\widehat{\sgl} \in \argmin{\est} \frac{1}{2}\|\vy-\dmtx\est\|_{2}^{2}+\sum_{i=1}^{p}\lambda_{i}|\esti|_{(i)},\label{eq:slope_opt}
\end{equation}
where $|\esti|_{(1)}\leq|\esti|_{(2)}\leq\cdots\leq|\esti|_{(p)}$ is a reordering of the absolute values $\abs{\esti_1}, \abs{\esti_2}, \ldots, \abs{\esti_p}$ in increasing order. In \cite{bogdan2015slope}, the regularization term $\regu(\est) \bydef \sum_{i=1}^{p}\lambda_{i}|\esti|_{(i)}$ is referred to as the ``sorted $\ell_{1}$ norm'' of $\est$. The same regularizer was independently developed in a different line of work \cite{bondell2008simultaneous,zhong2012efficient, zeng2014decreasing, figueiredo2016ordered}, where the motivation is to promote group selection in the presence of correlated covariates.

The classical LASSO estimator is a special case of SLOPE. It corresponds to using a constant regularization sequence, \emph{i.e.}, $\lambda_1 = \lambda_2 = \cdots = \lambda_p = \lambda$. However, with more general $\vlambda$-sequences, SLOPE has the flexibility to penalize different coordinates of the estimate according to their magnitudes. This adaptivity endows SLOPE with some nice \emph{statistical} properties that are not possessed by LASSO. For example, it is shown in \cite{su2016slope,bellec2018slope}
that SLOPE achieves the minimax $\ell_{2}$ estimation rate
with high probability. When applied in variable selection problem, SLOPE is shown to control the false discovery rate (FDR) for orthogonal design matrices \cite{bogdan2015slope}, which is not the case for LASSO. In addition, the new regularizer $\regu(\est)$ is still a norm \cite{bogdan2015slope,zeng2014decreasing}. Thus, the optimization problem associated with SLOPE remains convex, and it can be efficiently solved by using \emph{e.g.}, proximal gradient descent \cite{zeng2014decreasing, bogdan2015slope}.

Although the flexible regularization of SLOPE creates the hope of potential performance enhancement, to fully realize SLOPE's potential, we have to carefully design the regularizing sequence $\vlambda$. Note that this is equivalent to specifying the empirical distribution of $\vlambda$. Popular choices in the previous works include delta distribution (\emph{i.e.}, LASSO), uniform distribution \cite{bondell2008simultaneous}, chi-distribution \cite{bogdan2013statistical}, etc. These regularization schemes are mostly devised based on statistical insights gained from simpler models and they are indeed superior than LASSO in several applications. However, the success of these regularizing sequences provide no quantitative answer to the following two questions:
\begin{enumerate}
  \item
  What is the fundamental limit of SLOPE?
  \item
  How to optimally design $\boldsymbol{\lambda}$ to reach the fundamental limit?
\end{enumerate}
The aforementioned studies on analyzing SLOPE provide very limited information for us to address the above two questions, since in these works, the SLOPE's performance is characterized in an order-wise manner, which contains loose constants.
What we need is an \emph{exact} performance characterization of SLOPE estimator, which is still absent in the existing literature. On the other hand, however, exact asymptotic analysis has been carried out
for LASSO \cite{bayati2012lasso,su2017false} and several other regularized regression techniques \cite{oymak2013squared,el2018impact, thrampoulidis2016ber, zheng2017does}, under certain statistical assumptions on the sensing matrix $\dmtx$. One key feature of all these results is that the performance in the originally high-dimensional model can be well-captured by some low dimensional problems, which are much easier to handle.
The technical hurdle that has precluded a similar treatment for SLOPE is that unlike all the regularizer considered in these works, the SLOPE norm $\regu(\vx)$ is \emph{non-separable}: it cannot be written as a sum of component-wise functions, \emph{i.e.}, $\regu(\vx) \neq \sum_{i=1}^p J_i(x_i)$. This makes a similar low-dimensional reduction more challenging.

\subsection{Main Contributions}
In this paper, we answer the questions raised above. Our main contributions are listed as follows:
\subsubsection{Asymptotic separability}
\begin{figure}[t]
\begin{centering}
\subfloat[\label{subfig:hist_stair}]{\begin{centering}
\includegraphics[clip,scale=0.42]{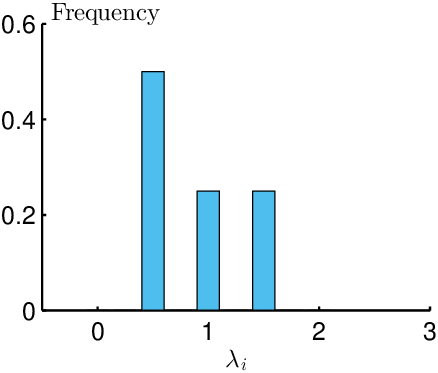}
\par\end{centering}
}
\hspace{0.0cm}
\subfloat[\label{subfig:prox_stair}]{\begin{centering}
\includegraphics[clip,scale=0.39]{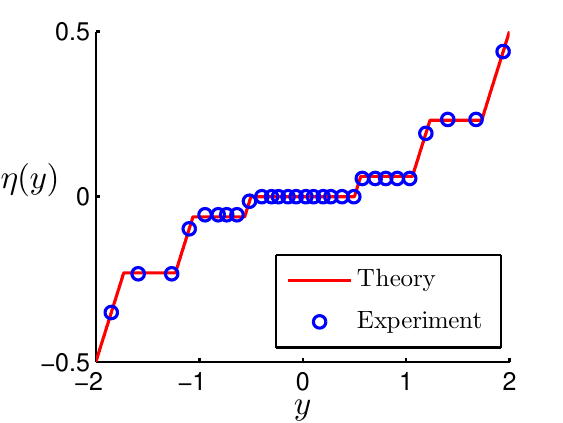}
\par\end{centering}
}
\hspace{0.0cm}
\subfloat[\label{subfig:hist_BHq}]{\begin{centering}
\includegraphics[clip,scale=0.42]{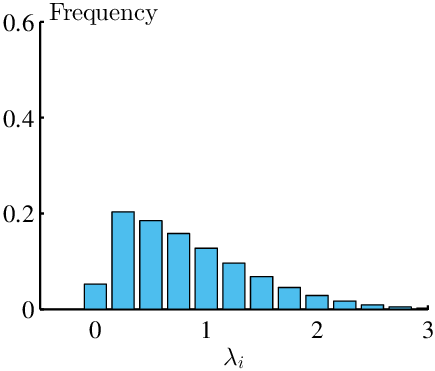}
\par\end{centering}
}
\hspace{0.0cm}
\subfloat[\label{subfig:prox_BHq}]{\begin{centering}
\includegraphics[clip,scale=0.39]{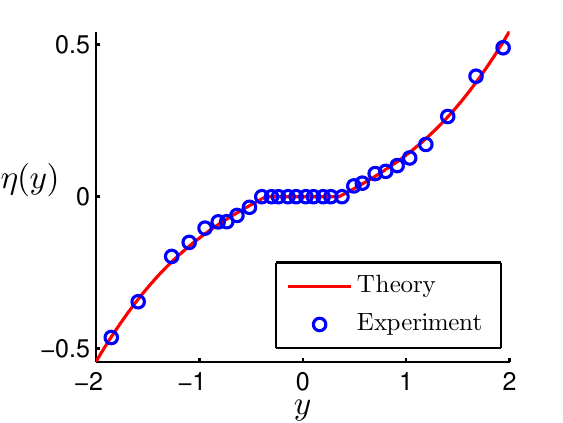}
\par\end{centering}
}
\par\end{centering}
\caption{(a) and (c): The histograms of two different $\vlambda$-sequences. (b) and (d): Sample points of $\left(y_{i},\,\left[\text{Prox}_{\boldsymbol{\lambda}}(\vy)\right]_{i}\right)$ (the blue dots) compared against the limiting scalar functions $\prox(y)$ (the red curves). In this experiment, $p = 1024$ and $y_i\iid\mathcal{N}(0,1)$. For better visualization, we randomly sample
3\% of all $\left(y_{i},\,\left[\text{Prox}_{\boldsymbol{\lambda}}(\vy)\right]_{i}\right)$.
\label{fig:prox_dist}}
\end{figure}

As mentioned above, the main obstacle in analyzing SLOPE asymptotics is the non-separability of SLOPE regularizer $\regu(\est) = \sum_{i=1}^{p}\lambda_{i}|\esti|_{(i)}$. We overcome this challenge by showing that the proximal operator of $\regu(\est)$ is \emph{asymptotically separable}. To be more concrete, we first give a technically light overview of this result. The \emph{proximal operator} of $\regu(\est)$ is defined as:
\begin{equation}
\label{eq:slope_prox0}
\tprox_{\vlambda}(\vy) \bydef \argmin{\vx} \frac{1}{2}\|\vy-\vx\|_{2}^{2}+\regu(\vx)
\end{equation}
In the case of LASSO, where we choose $\lambda_1 = \lambda_2 = \cdots = \lambda_p = \lambda$, characterizing $\text{Prox}_{\boldsymbol{\lambda}}(\vy)$ is easy, since the optimization in \eqref{eq:slope_prox0} is equivalent to $p$ scalar problems:
$\sum_{i=1}^{p}\min_{x_i}\frac{1}{2}(y_i-x_i)^{2}+\lambda |x_i|.$
Correspondingly, the proximal operator is \emph{separable}:
$[\text{Prox}_{\boldsymbol{\lambda}}(\vy)]_i = \sgn(y_i)\max(|y_{i}|-\lambda,0).$
In other words, the $i$th element of $\text{Prox}_{\boldsymbol{\lambda}}(\vy)$ is solely determined by $y_i$. However, this separability property does not hold for a general regularizing sequence. When $p$ is finite, $[\text{Prox}_{\boldsymbol{\lambda}}(\vy)]_i$ depends not only on $y_{i}$ but also on other elements of $\vy$. As one of the core results in this paper, we show that if the empirical distributions of $\vy$ and $\vlambda$ converge as $p\to\infty$, then
\begin{equation*}
\frac{1}{p}\,\|\text{Prox}_{{\vlambda}}(\vy)-\prox(\vy)\|^{2} \to 0,
\end{equation*}
where $\prox$ is a limiting scalar function that is uniquely determined by the limiting empirical measures of $\vy$ and $\vlambda$ (for the exact form, see Proposition \ref{prop:prox}). This result is illustrated in Fig.~\ref{fig:prox_dist}, where we compare the actual proximal operator $\text{Prox}_{\boldsymbol{\lambda}}(\vy)$ and the limiting scalar function $\prox(y)$, for two different $\vlambda$-sequences shown in Fig.~\ref{subfig:hist_stair} and Fig.~\ref{subfig:hist_BHq}. It can be seen that under a moderate dimension, the proximal operator $\text{Prox}_{{\tau\vlambda}}(\vy)$ can already be very accurately approximated by $\prox(\vy)$.

\subsubsection{Exact characterization}
The asymptotic separability allows us to obtain the exact characterization of SLOPE's performance in the linear asymptotic regime:
$n,\,p\to\infty$ and $n/p\to\delta$, under the assumption that sensing matrix $\dmtx$ is generated from i.i.d. Gaussian. On a high level, our main results show that the joint empirical distribution of $\{(\soli_{i},\sgli_{i})\}_{i=1}^{p}$ converges to a well-defined limiting measure (the precise description can be found in Theorem \ref{thm:asymp_char}). Note that the performance metrics of interests such as mean square error (MSE), type-I error, power are all functional of the empirical measure $\{(\soli_{i},\sgli_{i})\}_{i=1}^{p}$. Therefore, this makes it possible us to compute the high-dimensional limits of all these quantities.
Compared with the probabilistic bounds derived in previous work, our results are asymptotically \emph{exact}.
\subsubsection{Fundamental limits and optimal regularizagion}
The exact asymptotic characterization finally enables us to derive the fundamental limits of SLOPE in both estimation and variable selection tasks: (1) the minimum MSE that can be achieved by SLOPE; and (2) the highest possible power achievable under any given level of Type-I error.
Moreover, we show that in both cases, the optimal $\vlambda$ sequence can be obtained by solving certain infinite-dimensional convex optimization problems, which have efficient and accurate finite-dimensional approximations.
It is worth mentioning that a caveat of our optimal design is that it requires knowing the limiting empirical measure of $\sgl$ (\emph{e.g.}, the sparsity level and the distribution of its nonzero coefficients). For this reason, our results are \emph{oracle optimal}. However, it provides the first step towards optimal sequence designs under more realistic setting, where no or only limited information about $\sgl$ is available.

\begin{figure}[t]
\begin{centering}
\subfloat[MSE v.s. $\delta$\label{subfig:MSE_delta}]{\begin{centering}
\includegraphics[clip,scale=0.5]{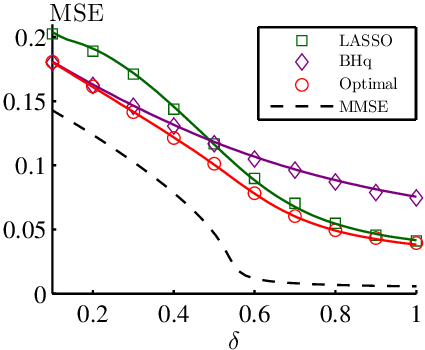}
\par\end{centering}
}
\hspace{0.0cm}
\subfloat[$\delta=0.1$\label{subfig:hist_opt_deltadot1}]{\begin{centering}
\includegraphics[clip,scale=0.5]{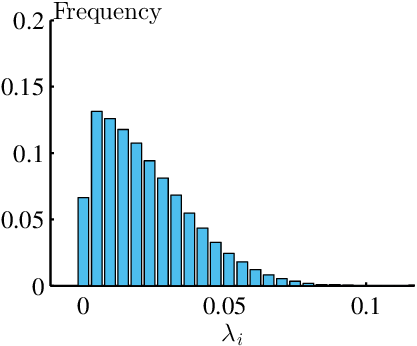}
\par\end{centering}
}
\hspace{0.0cm}
\subfloat[$\delta=0.45$\label{subfig:hist_opt_deltadot45}]{\begin{centering}
\includegraphics[clip,scale=0.5]{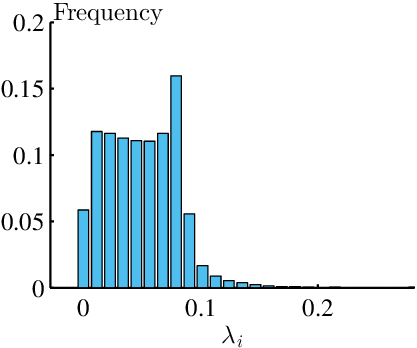}
\par\end{centering}
}
\hspace{0.0cm}
\subfloat[$\delta=1$\label{subfig:hist_opt_delta1}]{\begin{centering}
\includegraphics[clip,scale=0.5]{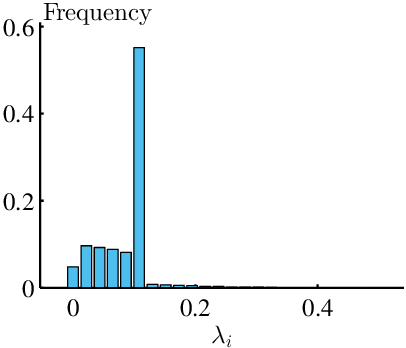}
\par\end{centering}
}
\par\end{centering}
\caption{(a): Theoretical predictions (solid lines) v.s. empirical results. Here,
$\protect\sgli_{i}$ are i.i.d. Bernoulli random variables with $\P(\sgli_{i}=1) = 0.2$
and $\noisei_{i}\protect\iid\mathcal{N}(0,\,0.04)$.
In our simulation, we choose
$p=2048$ and the empirical results are averaged over 20 independent trials.
(b)-(d): Empirical distributions of optimal regularizing sequences under 3 different sampling ratios.
\label{fig:Comparison-of-MSE-0}}
\end{figure}

An illustration of asymptotic characterization and optimality results stated above are presented in Fig. \ref{fig:Comparison-of-MSE-0}. We consider three different regularizing sequences: LASSO, BHq sequence proposed in \cite{bogdan2013statistical} and the optimal sequence given by Proposition \ref{prop:min_MSE} below. In Fig. \ref{subfig:MSE_delta}, we plot the empirical MSEs  and compare them with the theoretical results. We can see they match well under all settings. Moreover, all the recorded MSE values are lower bounded by the fundamental limits predicted by our theory (red curve in the figure) and they can be achieved by the optimally designed sequences (red circles in the figure). For comparison, we also enclose the curve of minimum mean square error (MMSE) of linear Gaussian model, which was derived in \cite{barbier2016mutual,reeves2016replica}. Finally, to help the readers get a sense of what the optimal regularizing sequences look like, in Fig. \ref{subfig:hist_opt_deltadot1}-\ref{subfig:hist_opt_delta1} we
plot their empirical distributions under 3 different sampling ratios $\delta$. Interestingly, we can find they exhibit very different distributions as we change $\delta$.
%
\subsection{Related Work}
\subsubsection{Exact asymptotic characterization}
There has been a growing body of works studying the exact asymptotics in high-dimensional statistical problems under random design assumptions. A partial list of these works includes \cite{donoho2009observed,donoho2009message,bayati2011dynamics,chandrasekaran2012convex,bayati2012lasso,krzakala2012statistical,javanmard2013state,el2013robust,amelunxen2014living,thrampoulidis2015regularized,donoho2016high,dhifallah2017phase,thrampoulidis2018precise,sur2019modern}.
One distinct feature of these  type of results is that they provide sharp performance guarantee that does not contain loose constants. From a technical viewpoint, these works are built on powerful tools including statistical physics \cite{tanaka2002statistical,kabashima2009typical}, approximate message passing (AMP) \cite{donoho2009message,bayati2011dynamics}, Gaussian width or statistical dimensions \cite{chandrasekaran2012convex,amelunxen2014living}, leave-one-out analysis \cite{el2013robust,el2018impact}, Gordon's Gaussian comparison lemma \cite{gordon1985some}, etc. Our main asymptotic characterization is proved based on convex min-max Gaussian theorem (CGMT) \cite{stojnic2013framework,oymak2013squared,thrampoulidis2015regularized}, which is a tight version of Gordon's comparison lemma in the convex setting. This framework was developed through a series of works \cite{stojnic2013framework,oymak2013squared,thrampoulidis2015regularized,thrampoulidis2018precise} and have now been successfully applied in a variety of problems such as binary detection \cite{thrampoulidis2016ber}, regularized M-estimation \cite{thrampoulidis2015regularized,thrampoulidis2018precise}, phase retrieval \cite{dhifallah2017phase,salehi2018precise} and high-dimensional classification \cite{deng2019model,montanari2019generalization,kini2020analytic}.
\subsubsection{Optimal M estimation in high dimensions}
The optimality part of this work falls within the line of research pursuing the optimal M-estimation in high-dimensional regression. The general form of M-estimator
is as follows:
\begin{equation}
\label{eq:gene_M_estimator}
\sol \in \argmin{\est} \sum_{i=1}^{n}\ell(y_i,\va_i^{\T} \est)+r(\est)
\end{equation}
and the question is what is the optimal statistical performance achievable by \eqref{eq:gene_M_estimator} and how to optimally design the loss function $\ell$ and the regularizer $r$.
The exact asymptotic characterizations open up the possibility of obtaining a precise answer to the above question.
This line of research is initiated by the papers \cite{bean2013optimal} and \cite{donoho2016high}, where the authors study the fundamental limits of the unregularized M-estimator (\emph{i.e.}, the case when $r=0$) in the linear model. In particular, a computational feasible recipe is provided in \cite{bean2013optimal} for constructing the optimal loss function $\ell$ that minimizes the estimation errors.
Similar types of results are also recently established for the binary models \cite{taheri2021sharp}. When a regularizer is included, the optimal performance of \eqref{eq:gene_M_estimator} in the linear model is studied in \cite{advani2016statistical} and recently extended to binary model for the special case of quadratic regularization \cite{aubin2020generalization,taheri2021fundamental}. In the meantime, a series of papers study the optimal $\ell_q$-norm regularized least square regression \cite{zheng2017does,weng2018overcoming,wang2020bridge}. In some limiting regimes, explicit answers are provided regarding the optimal choice of $q$.
Note that all the aforementioned works consider the separable regularizer: $r(\est)=\sum_{i=1}r_i(b_i)$, while SLOPE regularizer considered in this paper is not separable.

Closely related with current work is the paper by Celentano and Montanari \cite{celentano2019fundamental}. One of their main results is on the optimal estimation performance achievable by quadratic loss regularized by any lower semi-continuous, proper, convex and symmetric \footnote{Symmetric means $r(\est)$ is permutation invariant to coordinates of $\est$.} function. It is not hard to check that SLOPE norm belongs to this family of functions. In fact, the optimality results presented in their paper and ours share a very similar form. We will elaborate more on this in Sec. \ref{sec:mse}.

\subsubsection{Three Parallel works}
Finally, we mention three parallel works that also study the limiting behavior of SLOPE under the same asymptotic setting.
\begin{enumerate}
  \item
  From an algorithmic perspective, \cite{bu2020algorithmic} consider solving the SLOPE minimization problem \eqref{eq:slope_opt} using the AMP algorithm. By relating the stationary point of AMP iterations to SLOPE estimator, they also establish the same characterization (as shown in Theorem \ref{thm:asymp_char} below). In the proof, they also utilize the asymptotic separability property proved in Proposition \ref{prop:prox}.
  \item
  The CGMT framework is also applied in \cite{wang2019does} to obtain the limiting mean square errors (MSE) of SLOPE, together with a finite-sample concentration bound. The authors quantitatively compares the MSEs of different regularizing sequences in some limiting regimes. In particular, it is shown that in the high SNR regimes, LASSO regularization is optimal. A major difference from our work is that they do not exploit the asymptotic separability of SLOPE and the optimal performance in the general regime is not addressed.
  \item
  In \cite{celentano2019approximate}, the asymptotic separability properties is further extended to all lsc, proper, convex and symmetric regularizers using an elegant lifting and embedding idea. A finite-sample concentration bound is also given. Using the general asymptotic separability results, the author proves a conjecture in \cite{celentano2019fundamental}: the MSE lower bound achievable by \emph{non-separable} convex symmetric regularizers will be the same if we are restricted to the \emph{separable} convex regularizers. However, the performance of variable selection is not addressed.
\end{enumerate}

\subsection{Notations\label{subsec:DefNota}}
For a vector $\vx\in\R^{p}$ and a scalar function $f(\cdot):\R\to\R$, $f(\vx)$ means $f(\cdot)$ is applied to vector $\vx$ coordinate-wise. $\|\vx\|$ denotes the $\ell_2$ norm, $x_i$ (or $[\vx]_i$) denotes the $i$th coordinate of $\vx$ and $|x|_{(i)}$ (or $|\vx|_{(i)}$) denotes the $i$th largest coordinate of $|\vx|$.
The Euclidean ball in $\R^{p}$ centered on $\va$ with radius $r\geq 0$ is denoted as: $\bdball_{r}(\va):=\{\vv:\|\vv-\va\|\leq r\}$ and $\bdball_{r}\bydef\bdball_{r}(\boldsymbol{0})$. Also we define $\outbdball_{r}(\va)\bydef \{\vv:\|\vv-\va\|\geq r\}$.

For a probability measure $\mealim$, we denote $\text{Supp}(\mealim)$ as its support. For random variables $X,Y$, we denote $\meajoint X Y$ and $\meamarginal X$, $\meamarginal Y$ as their joint and marginal measures and $F_{X}$, $F_{Y}$ as the corresponding (marginal) cumulative distribution function (CDF). The quantile function of random variable $X$ is denoted as $F_{X}^{-1}(p)$, where $F_{X}^{-1}(p) \bydef \inf\{x\in\R: F_{X}(x)\geq p\}$.
Specifically, we use $\Phi$ and $\Phi^{-1}$ to denote the CDF and quantile function of standard Gaussian.
For vectors $\vx,\vy\in\R^{p}$, we denote $\meajoint \vx \vy$ and $\meamarginal \vx$, $\meamarginal \vy$ as their joint and marginal empirical measures and
$F_{\vx}$, $F_{\vy}$ as the corresponding (marginal) empirical CDF.
Also we denote the empirical quantile function of $\vx$ as $F_{\vx}^{-1}$.

We denote $\mathcal{P}_{q}(\R^{k})$, for some $q\geq 1$ and $k\in\mathbb{Z}^{+}$, as the space of all probability measures on $\R^{k}$ with bounded moments of order $q$, \emph{i.e.}, for any $\mu\in\mathcal{P}_{q}(\R^{k})$,  it holds that $\E_{\mu}(\|X\|^q)<\infty$.
For two measures $\mu,\nu\in \mathcal{P}_{q}(\R^{k})$, their Wasserstein-$q$ distance is defined as:
\begin{equation*}
W_q(\mu,\nu) \bydef \Big(\inf_{\pi\in\Pi(\mu,\nu)} \E \|X-Y\|_2^q \Big)^{1/q},
\end{equation*}
where $(X, Y)\sim \pi$ and $\Pi(\mu,\nu)$ is the set of all couplings of $\mu$ and $\nu$.


\subsection{Asymptotic Setting}
There are four main objects in the description of our model and algorithm: (1) the unknown vector $\sgl$; (2) the design matrix $\dmtx$;
(3) the noise vector $\noise$; and (4) the regularizing sequence $\vlambda$. Since we study the asymptotic limit (with $p \to \infty$), we will consider a sequence of instances $\big\{ \sgl^{(p)},\,\dmtx^{(p)},\,\vw^{(p)},\,\vlambda^{(p)}\big\} _{p\in\mathbb{N}}$
with increasing dimensions $p$, where $\sgl^{(p)},\,\vlambda^{(p)}\in\R^{p}$,
$\dmtx^{(p)}\in\R^{n\times p}$ and $\noise^{(p)}\in\R^{n}$.
A sequence of vectors $\{\boldsymbol{x}^{(p)}\}_{p\in\mathbb{Z}}$ (or $\{ \boldsymbol{x}^{(p)}, \boldsymbol{y}^{(p)}\}_{p\in\mathbb{Z}}$), with $p$ indexing the growing dimensions, is called a \emph{converging sequence}, if its empirical measure $\meamarginal{\vx^{(p)}}$ (or $\meajoint{\vx^{(p)}}{\vy^{(p)}}$)
converges in Wasserstein-2 distance to a probability measure $\meamarginal{X}$ (or $\meajoint{X}{Y}$) as $p\to\infty$.
For notational brevity, we will omit the superscript ``$(p)$'' when it is clear from the context.
\subsection{Paper Outline}
The rest of the paper is organized as follows. In Sec. \ref{sec:Asymptotic-Separability}, we first prove the asymptotic separability of the proximal operator associated with $\regu(\vx)$. This property allows us to derive our asymptotic
characterization of SLOPE in Sec. \ref{sec:Asymptotic-Characterization}. Based on this analysis, we derive the fundamental limit and present the optimal design of the regularizing sequence in Sec. \ref{sec:Oracle-Optimality}. Numerical simulations are provided to verify our asymptotic characterizations. They also demonstrate the superiority of our optimal regularization over LASSO and BHq sequence in \cite{su2016slope}. In Sec. \ref{sec:Proof-of-Main-Results}, we provide the proof of all our main results. We conclude the paper in Sec. \ref{sec:Conclusions} and discuss some possible directions for future work. 

\section{Proximal Problem and Asymptotic Separability\label{sec:Asymptotic-Separability}}

We start by studying the following proximal problem:
\begin{equation}
\mathcal{M}_{\vlambda}(\vy;\tau) \bydef\min_{\vx} \frac{1}{2\tau}\|\vy-\vx\|_{2}^{2}+\regu(\vx),\label{eq:slope_prox}
\end{equation}
where $\tau>0$, $\vy\in\R^{p}$ and $\regu(\vx)=\sum_{i=1}^{p}\lambda_{i}|x|_{(i)}$, with $0\leq\lambda_{1}\leq\lambda_{2}\leq\cdots\leq\lambda_{p}$.
$\mathcal{M}_{\vlambda}(\vy;\tau)$ in \eqref{eq:slope_prox} is known as the \emph{Moreau envelope} of $\regu(\vx)$ evaluated at $\vy$ and $\tau$ is the smoothing parameter. The unique minimizer of \eqref{eq:slope_prox} is the \emph{proximal operator} associated with $\regu(\vx)$ under parameter $\tau$. From \eqref{eq:slope_prox}, we know the proximal operator of $\regu(\vx)$ is fully determined by $\vy$ and $\tau\vlambda$, so we simply denote it as $\text{Prox}_{\tau\boldsymbol{\lambda}}(\vy)$.
It turns out that the asymptotics of the original problem \eqref{eq:slope_opt} is closely related to \eqref{eq:slope_prox}. Thus, as a preliminary step, we will first analyze its limiting properties.
To state our result, we introduce the following functional optimization problem. For $\mealim_{Y},\mealim_{\Lambda}\in \Wspace(\R)$, with $\P(\Lambda\geq 0 )= 1$,  define
\begin{align}
\mathcal{M}_{\mealim_\Lambda}(\mealim_Y;\tau) \bydef \min_{g\in\Lipndset}\frac{1}{2\tau}\E_{\mealim_{Y}}[Y-g(Y)]^{2}+\int_{0}^{1}F_{\Lambda}^{-1}(u)F_{|g(Y)|}^{-1}(u)du, \label{eq:infinite_SLOPE}
\end{align}
where
\begin{equation}
\Lipndset\bydef\{g(y) \mid g(y) \text{ is odd, non-decreasing and 1-Lipschitz}\}.
\end{equation}
Also we denote $\prox(\cdot;\mealim_Y,\mealim_{\tau\Lambda})$ as the optimal solution of \eqref{eq:infinite_SLOPE}. Comparing \eqref{eq:infinite_SLOPE} with \eqref{eq:slope_prox}, we can intuitively interpret $\mathcal{M}_{\mealim_\Lambda}(\mealim_Y;\tau)$ and $\prox(\cdot;\mealim_Y,\mealim_{\tau\Lambda})$ as the functional-form Moreau envelope and proximal operator.

We are now ready to state our main result on the asymptotics of the proximal problem \eqref{eq:slope_prox}.
\begin{prop}
\label{prop:prox}
Let $\{ \vy\}_{p\in \mathbb{N}}$ and $\{\vlambda\}_{p\in \mathbb{N}}$ be two converging sequences, with limiting measures $\mealim_Y$ and $\mealim_\Lambda$ satisfying $\P(\Lambda\geq 0 )= 1$. It holds that for any $\tau > 0$,
\begin{equation}
\label{eq:Moreau_conv}
\frac{1}{p} \mathcal{M}_{\vlambda}(\vy;\tau)\to \mathcal{M}_{\mealim_\Lambda}(\mealim_Y;\tau)
\end{equation}
and
\begin{equation}
\label{eq:prox_sep}
\frac{1}{p}\,\|\text{Prox}_{{\tau\vlambda}}(\vy)-\prox(\vy; \mealim_Y, \mealim_{\tau\Lambda})\|^{2} \to 0,
\end{equation}
where $\mathcal{M}_{\mealim_\Lambda}(\mealim_Y;\tau)$ and $\prox(\cdot;\mealim_Y,\mealim_{\tau\Lambda})$ are the optimal value and the unique (up to a set of measure zero with respect to $\mealim_{Y}$) optimal solution of \eqref{eq:infinite_SLOPE}.
\end{prop}


The proof of Proposition \ref{prop:prox} will be provided in Appendix \ref{subsec:Proof-of-Asymp-sepa}. We will also see that the limiting characterization of $\mathcal{M}_{\vlambda}(\vy;\tau)$ in \eqref{eq:infinite_SLOPE} and the asymptotic separability of $\text{Prox}_{\tau\boldsymbol{\lambda}}(\cdot)$ in \eqref{eq:prox_sep} greatly facilitates our asymptotic analysis
and the optimal design of $\vlambda$, since this allows us to reduce the original high-dimensional
problem to an equivalent one-dimensional problem, as in the LASSO case. Indeed, $\prox(\cdot; \meamarginal{Y}, \meamarginal{\tau\Lambda})$ in \eqref{eq:prox_sep} is exactly the {limiting scalar function} $\prox(\cdot)$ shown earlier in Fig. \ref{fig:prox_dist}. We will still sometimes adopt the lighter notation $\prox(\cdot)$, when doing so causes no confusion.

Note that \eqref{eq:infinite_SLOPE} is involved with an infinite-dimensional optimization, which typically permits no simple analytical solutions. To gain more intuition, before moving on, let us consider two examples where closed-form solutions do exist.
\begin{example}[LASSO]
\label{ex:lasso}
The LASSO case corresponds to $\P(\Lambda=\lambda)=1$. When $\tau=1$, optimization in \eqref{eq:infinite_SLOPE} then reduces to
\begin{align}
\min_{g\in\Lipndset}\frac{1}{2}\E\big[\underbrace{|Y|-\lambda}_{:=f(|Y|)}-g(|Y|)\big]^{2} + \text{constant}\label{eq:LASSO_fy}.
\end{align}
Note that the function $f(y)=y-\lambda$ in \eqref{eq:LASSO_fy} is non-decreasing and 1-Lipschitz on $\R_{\geq 0}$ and $f(0)\leq 0$. It is not hard to show in this case, the optimal solution of \eqref{eq:LASSO_fy} equals to
\begin{align}
\prox(y; \meamarginal{Y}, \meamarginal{\Lambda}) &=\sgn(y)\max\big(f(|y|),0\big)\nonumber\\
&= \sgn(y)\max(|y|-\lambda,0)\nonumber,
\end{align}
which is exactly the soft-thresholding function.
\end{example}
\begin{example}[BHq \cite{bogdan2013statistical}]
\label{exa:BHq}
The BHq regularization corresponds to $\Lambda\sim\Phi^{-1}(1-\frac{q}{2}+\frac{q}{2}U)$, where $q\in(0,1]$ and $U$ is uniformly distributed over $[0,1]$. Then we have $F_{\Lambda}^{-1}(u) = \Phi^{-1}(1-\frac{q}{2}+\frac{q}{2}u)$. Further, we consider $Y\sim\mathcal{N}(0,1)$. It holds that $F_{|Y|}(y)=2\Phi(y)-1$ and
$F_{|g(Y)|}^{-1}\big(F_{|Y|}(y)\big) = g(y)$, for $y\geq 0$. Therefore,
\begin{equation}
\label{eq:lambda_y_1}
\begin{aligned}
\int_{0}^{1}F_{\Lambda}^{-1}(u)F_{|g(Y)|}^{-1}(u)du &=
\int_{0}^{\infty} \underbrace{\Phi^{-1}\big(1-{q}+{q}\cdot\Phi(y)\big)}_{:=\lambda(y)} g(y) dF_{|Y|}(y),
\end{aligned}
\end{equation}
where we apply a change of variable $u=F_{|Y|}(y)$.
In this case, \eqref{eq:infinite_SLOPE} becomes
\begin{equation*}
\begin{aligned}
&\min_{g\in\Lipndset}\frac{1}{2}\E\big[|Y|-g(|Y|)\big]^{2}+ \E\big[\lambda(|Y|) g(|Y|)\big]\\
=& \min_{g\in\Lipndset}\frac{1}{2}\E\big[|Y|-\lambda(|Y|)-g(|Y|)\big]^{2} + \text{constant}.
\end{aligned}
\end{equation*}
On the other hand, by direct differentiation of $\lambda(y)$ in \eqref{eq:lambda_y_1}, we can get $\lambda'(y)=\frac{q\phi(y)}{\phi(\lambda(y))}$, where $\phi$ is the density function of standard Gaussian. It is not hard to verify $\lambda'(y)\in (0,1]$ when $y\geq 0$. Therefore, $y\mapsto y-\lambda(y)$ is non-decreasing and 1-Lipschitz on $\R_{\geq 0}$. On the other hand, $\lambda(0) = \Phi^{-1}(1-\frac{q}{2})\geq 0$. Then following the same argument in Example \ref{ex:lasso}, we get
$\prox(y; \meamarginal{Y}, \meamarginal{\Lambda})
= \sgn(y)\max\big(|y|-\lambda(|y|),0\big).$
\end{example}
\begin{rem}
More generally, we can show when $Y$ has a density supported on $\R$ and $y\mapsto y - F_{\Lambda}^{-1}(F_{|Y|}(y))$ is non-decreasing and 1-Lipschitz on $\R_{\geq 0}$, then $\prox(y; \meamarginal{Y}, \meamarginal{\Lambda})
= \sgn(y)\max\big(|y|-F_{\Lambda}^{-1}\big(F_{|Y|}(|y|)\big),0\big).$ In some sense, $F_{\Lambda}^{-1}\big(F_{|Y|}(|y|)\big)$ can be viewed as the equivalent regularization function. This equivalent regularization is adaptive to $y$. As a comparison, the regularization is a constant $\lambda$ in the LASSO case.
\end{rem}

\section{Asymptotic Characterization of SLOPE\label{sec:Asymptotic-Characterization}}
Based on the asymptotic separability properties established in the last section, we are now ready to tackle the original optimization problem (\ref{eq:slope_opt}). We are going to obtain the precise characterizations of SLOPE in both estimation and variable selection problems.

\subsection{Technical Assumptions\label{subsec:DefandAssump}}


Our results are proved under the following assumptions:

\begin{enumerate}[label={(A.\arabic*)}]
\item \label{a:sampling_ratio} The number of observations grows in proportion to $p$: $n^{(p)}/p\to\delta\in(0,\infty)$.
\item \label{a:sparsity_ratio}The number of nonzero elements in $\sgl^{(p)}$ grows in proportion to
$p$: $r_{0}^{(p)}/p\to\rho\in[0,1]$.
\item The elements of $\dmtx^{(p)}$ are i.i.d. Gaussian distribution: $\dmtxi_{ij}^{(p)}\iid\mathcal{N}(0,\,\frac{1}{n})$.
\item \label{a:converging} $\{\sgl^{(p)}\}_{p\in \mathbb{N}}$, $\{\vw^{(p)}\}_{p\in \mathbb{N}}$ and
$\{\vlambda^{(p)}\}_{p\in \mathbb{N}}$ are converging sequences. The limiting measures are denoted by $\meamarginal{B}$, $\meamarginal{W}$ and $\meamarginal{\Lambda}$, respectively. In addition, $\P(B\neq 0)=\rho$, 
$\sigma_{\noisei}^{2} = \E[W^{2}]>0$ and $\P(\Lambda\neq0)>0$ when $\delta\leq 1$,
where the probability $\P(\cdot)$ and the expectations $\E[\cdot]$ are all computed with respect to the limiting measures.
\end{enumerate}

\subsection{Asymptotic Performance of Estimation}
The main goal of this section is to derive the limiting MSE of SLOPE: $\lim_{p\to\infty}\frac{1}{p}\|\sol-\sgl\|^2$.
As in \cite{bayati2012lasso}, we are going to prove a more general result,
which characterizes the joint empirical measure of $(\sol,\,\sgl)$ through its action on \emph{pseudo-Lipschiz} functions.
\begin{defn}[Pseudo-Lipschiz function]
A function $\psi:\,\R^{2}\to\R$
is called {pseudo-Lipschiz} if $|\psi(\vx)-\psi(\vy)|\leq L(1+\|\vx\|+\|\vy\|)\|\vx-\vy\|$ for all $\vx,\,\vy\in\R^{2}$, where $L$ is a positive constant.
\end{defn}
To compute the limiting MSE, we just need to let $\psi(\vx) = (x_1-x_2)^2$, which is a pseudo-Lipschiz function by the above definition.
The general theorem is as follows, whose proof is deferred to Sec. \ref{subsec:Asymptotic-Estimation-Performanc-proof}.
\begin{thm}
\label{thm:asymp_char} Assume \ref{a:sampling_ratio} -- \ref{a:converging} hold. For any pseudo-Lipschiz function $\psi$, we have
\begin{equation}
\frac{1}{p}\sum_{i=1}^{p}\psi(\soli_{i},\,\sgli_{i})
\pconv\E[\psi(\prox(\YOpt;\,\mealim_{\YOpt}, \mealim_{\tauOpt \Lambda}),B)],\label{eq:weak_convergence}
\end{equation}
where $\YOpt=B+\sigOpt H$ with $B \sim \mealim_{B}$, $H \sim\mathcal{N}(0,1)$ independent and $\prox$ is the limiting scalar function defined in Proposition \ref{prop:prox}.  In the above, the scalar pair $(\sigOpt,\,\tauOpt)$ is the unique solution of the following equations:
\begin{align}
\sigma^{2} & =\sigma_{\noisei}^{2}+\frac{1}{\delta}\E\big[\big(\prox(B+\sigma H;\meamarginal{B+\sigma H}, \meamarginal{\tau \Lambda})-B\big)^{2}\big]\label{eq:sigmatau_1}\\
1 & =\tau\Big[1-\frac{1}{\delta}\E\prox'(B+\sigma H;\meamarginal{B+\sigma H}, \meamarginal{\tau \Lambda})\Big].\label{eq:sigmatau_2}
\end{align}
\end{thm}
Theorem \ref{thm:asymp_char} essentially says that the joint empirical measure of $(\sol^{(p)},\,\sgl^{(p)})$ converges to the law of $(\prox(\YOpt;\,\mealim_{\YOpt}, \mealim_{\tauOpt \Lambda}),B)$.
This means that although the original problem (\ref{eq:slope_opt})
is high-dimensional, its asymptotic performance can be succinctly captured
by merely two scalars random variables. From \eqref{eq:weak_convergence} and \eqref{eq:sigmatau_1}, we know the limiting MSE equals to
\begin{equation}
\label{eq:limit_MSE}
\lim_{p\to\infty} \frac{1}{p} \|\sol-\sgl\|^2 = \delta(\sigOpt^2 - \sigma_{\noisei}).
\end{equation}

Readers familiar with the asymptotic analysis of LASSO will recognize that the forms
of (\ref{eq:sigmatau_1}) and (\ref{eq:sigmatau_2}) look identical to the results of LASSO obtained in \cite{bayati2012lasso,miolane2018distribution}. Indeed, the proof of Theorem \ref{thm:asymp_char} directly applies the framework of analyzing
LASSO asymptotics using convex Gaussian min-max
theorem (CMGT) \cite{thrampoulidis2015regularized, thrampoulidis2018precise, miolane2018distribution}. In a nutshell, the CGMT framework builds a connection between the asymptotics of the original high-dimensional problem \eqref{eq:slope_opt} and the optimal solution of the following two-dimensional minimax problem:
\begin{equation}
\label{eq:minimax_1}
\min_{\sigma\geq\sigma_{\noisei}}\max_{\theta\geq0}\frac{\theta}{2}\Big(\frac{\sigma_{\noisei}^{2}}{\sigma}+\sigma\Big)-\frac{\theta^{2}}{2}+\frac{1}{\delta}
\Big[\underbrace{\lim_{p\to\infty}\frac{1}{p}\mathcal{M}_{\vlambda}\big(\sgl+\sigma\vh;\frac{\sigma}{\theta}\big)}_{:=\Moreau(\sigma,\theta)}-\frac{\theta\sigma}{2}\Big],
\end{equation}
where $\sgl$ is the true signal vector in \eqref{eq:model}, $\vh\sim\mathcal{N}(\boldsymbol{0},\mI_p)$ and $\mathcal{M}_{\vlambda}(\cdot;\cdot)$ is the Moreau envelope defined in \eqref{eq:slope_prox}. In fact, equation \eqref{eq:sigmatau_1} and \eqref{eq:sigmatau_2} corresponds to the first-order optimality condition of \eqref{eq:minimax_1}.
Proposition \ref{prop:prox} enables us to  justify and explicitly compute the limit in \eqref{eq:minimax_1}, as well as the first-order derivatives $\frac{\partial\Moreau(\sigma,\theta)}{\partial\sigma}$ and $\frac{\partial\Moreau(\sigma,\theta)}{\partial\theta}$, which are crucial in obtaining the optimal point of \eqref{eq:minimax_1}.
\subsection{Asymptotic Performance of Variable Selection\label{subsec:Testing-Asymptotics}}

Next we study the asymptotic performance of SLOPE, when it is used as a variable selection methodology. Under this setting, the goal is to accurately select all the non-zero coordinates of $\sgl$. Based on SLOPE estimate,
we select the non-zero coordinates of estimate $\sol$. Ideally, we hope that the selected set includes the non-zero coordinates of $\sgl$, while do not contain zero coordinates of $\sgl$.
The usual performance metrics for this task include Type-I error, power, false discovery rate (FDR), etc.
Most of these performance metrics can be expressed
as a function of the spasiry level $r_{0}^{(p)}$ and the following quantities
\begin{equation}
\begin{aligned}R_{0}^{(p)}= & \frac{1}{p}\sum_{i=1}^{p}\indicatorfn_{\soli_{i}=0}, & V^{(p)}= & \frac{1}{p}\sum_{i=1}^{p}\indicatorfn_{\soli_{i}\neq0,\sgli_{i}=0},\end{aligned}
\label{eq:three_testing_related modules}
\end{equation}
where $R_{0}^{(p)}$ and $V^{(p)}$ are the proportions of discoveries
and false discoveries.
In the following, we will adopt Type-I error and power as our performance metrics, which can be
written as
\begin{equation}
\begin{aligned}\text{Type-I error}= & \frac{V^{(p)}}{\max\{1-r_{0}^{(p)},1/p\}}, & \text{Power}= & \frac{1-R_{0}^{(p)}-V^{(p)}}{\max\{r_{0}^{(p)},1/p\}}.\end{aligned}
\label{eq:type-I_power_formula}
\end{equation}
In order to study the asymptotics of these testing statistics,
we need to obtain the limits of $R_{0}^{(p)}$ and $V^{(p)}$ in (\ref{eq:three_testing_related modules}).

Note that the test functions involved in (\ref{eq:three_testing_related modules})
($\indicatorfn_{x=0}$ and $\indicatorfn_{x\neq0,y=0}$) are discontinuous,
so we can not directly apply (\ref{eq:weak_convergence}) in Theorem
\ref{thm:asymp_char} to compute $\lim_{p\to\infty}R_{0}^{(p)}$ and
$\lim_{p\to\infty}V^{(p)}$. Further justifications are needed to
obtain companion results for the testing-related statistics in \eqref{eq:three_testing_related modules}. Before
delving into techincal descriptions, we first show that counter examples
do exist where the quantities in (\ref{eq:three_testing_related modules})
fail to converge, while the assumptions in Theorem \ref{thm:asymp_char}
are still satisfied. This is different from the LASSO case, where the prediction \eqref{eq:weak_convergence} is shown to be still correct for the above non-smooth indicator functions \cite{bogdan2013statistical}.

\begin{example}[A counter example]
\label{exa:counter_unstable_2}Consider $\mealim_{B}$ being a spike-and-slab
distribution: $\mealim_{B}=0.5\cdot\delta_{0}+0.5\cdot\mathcal{N}(1,0.5^{2})$
and $\{\sgli_{i}\}_{i\in[p]}$ are i.i.d. generated from $\mealim_{B}$.
Let $(\sigOpt,\tauOpt)$ be the solution of (\ref{eq:sigmatau_1})-(\ref{eq:sigmatau_2})
in the LASSO case, where $\P(\Lambda=1)=1$.
Then we construct
the following class of distribution of $\Lambda$, parameterized by
$\vartheta\in[0,1]$:
\begin{equation}
\Lambda_{\vartheta}=\begin{cases}
\vartheta & |\YOpt|<\vartheta\tauOpt,\\
\frac{|\YOpt|}{\tauOpt} & \vartheta\tauOpt\leq|\YOpt|<\tauOpt,\\
1 & |\YOpt|\geq\tauOpt,
\end{cases}\label{eq:unstable_Lambda}
\end{equation}
where $\YOpt=B+\sigOpt H$. Here $\vartheta$ is a tuning parameter
and $\vartheta=1$ corresponding to the LASSO regularization. In Fig.
\ref{fig:Counter-example-R0}, we plot the empirical $R_{0}^{(p)}$
and MSEs under different values of $\vartheta$. It can be seen from
Fig. \ref{subfig:Counter-example-R0-MSE} that for different values
of $\vartheta$, %
{} the empirical MSEs all concentrate around the predicted values from
Theorem \ref{thm:asymp_char}, when $\Lambda=1$. On the contrary,
from Fig. \ref{subfig:Counter-example-R0-R0} we can find when $\vartheta<1$,
$R_{0}^{(p)}$ does not converge to $\P\big(\prox(\YOpt)=0\big)$, which is the limit indicated by Theorem \ref{thm:asymp_char}.
Moreover, as $\vartheta$ becomes smaller, the SLOPE estimator becomes
less conservative and the variances of $R_{0}^{(p)}$ become increasingly
notable. Also we can see $R_{0}^{(p)}$ does converge to $\P\big(\prox(\YOpt)=0\big)$,
when $\vartheta=1$.

\begin{figure}
\begin{centering}
\subfloat[\label{subfig:Counter-example-R0-R0}]{\begin{centering}
\includegraphics[scale=0.49]{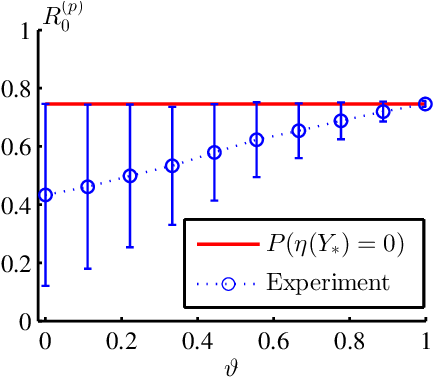}
\par\end{centering}
}
\subfloat[\label{subfig:Counter-example-R0-MSE}]{\begin{centering}
\includegraphics[scale=0.49]{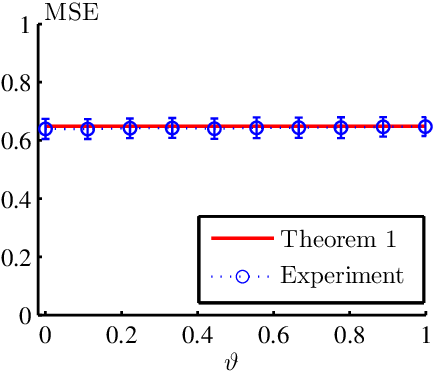}
\par\end{centering}
}
\subfloat[\label{subfig:Counter-example-R0_hist1}]{\begin{centering}
\includegraphics[scale=0.49]{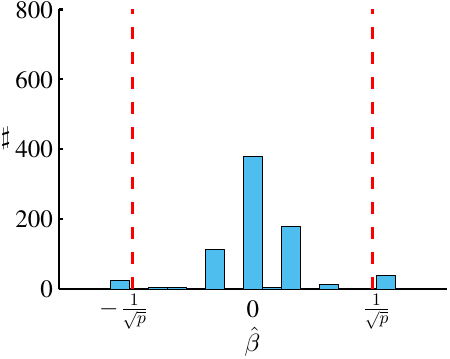}
\par\end{centering}
}
\subfloat[\label{subfig:Counter-example-R0_hist2}]{\begin{centering}
\includegraphics[scale=0.49]{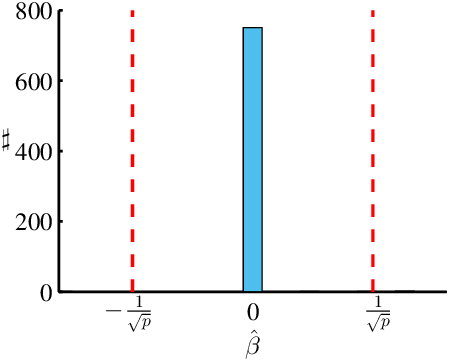}
\par\end{centering}
}
\end{centering}
\caption{Counter example when $R_{0}^{(p)}\protect\not\to\protect\P\big(\protect\prox(\protect\YOpt)=0\big).$
In the experiment, $p=1024$. The regularizing sequence $\protect\vlambda$
is generated by reordering $p$ i.i.d. samples of $\Lambda_{\vartheta}$ defined in \eqref{eq:unstable_Lambda}.
(a) and (b): empirical $R_{0}^{(p)}$ and $\text{MSE}$ v.s. theoretical predictions based on Theorem \ref{thm:asymp_char}, under different values of $\vartheta$. The error bars in (a) and (b) are
plotted using 1000 independent runs.
(c) and (d): the histograms
of $\protect\sol$ near zero when $\vartheta=0$ and $\vartheta=1$.
When $\vartheta=0$, it can be observed that two clumps of pseudo-zero
entries appear within the tiny interval $\big[-\frac{1}{\sqrt{p}},\frac{1}{\sqrt{p}}\big]$,
while when $\vartheta=1$, there is no pseudo-zero cluster. \label{fig:Counter-example-R0}}
\end{figure}
\end{example}

We will explain the logic behind the construction of $\Lambda_{\vartheta}$
in Remark \ref{rem:counter_construct} below. The counter example
above suggests that some additional constraints are needed, so that the testing statistics in \eqref{eq:three_testing_related modules} have
well-defined limits and
\eqref{eq:weak_convergence} can be used to compute these limits. %
It turns out that we just need one more condition to guarantee their
convergence. 

\begin{prop}
\label{prop:valid_testing_2}
Under the same settings as Theorem
\ref{thm:asymp_char},
define $\probthresh\bydef\P(\prox(\YOpt)=0).$
If the following condition holds:
\begin{enumerate}[label={(R.\arabic*)}]
\item \label{enu:R_strict_majorization}$\probthresh=0$ or for any $t\in[0,\probthresh)$,
$\int_{t}^{\probthresh}F_{|\YOpt|}^{-1}(u)du<\int_{t}^{\probthresh}F_{\tauOpt\Lambda}^{-1}(u)du$,
\end{enumerate}
then we have
\begin{equation}
\begin{aligned} & R_{0}^{(p)}\pconv\P\big(\prox(\YOpt)=0\big)\text{ and }V^{(p)}\pconv\P\big(\prox(\YOpt)\neq0,B=0\big),\end{aligned}
\label{eq:R_V_valid_limit}
\end{equation}
where $R_{0}^{(p)}$ and $V^{(p)}$ are defined in (\ref{eq:three_testing_related modules}).
\end{prop}
The proof of Proposition \ref{prop:valid_testing_2} will be provided
in Appendix \ref{subsec:Proof-of-Testing}, along with some explanations
for condition \ref{enu:R_strict_majorization} (see Remark \ref{rem:R1_explanation}).
\begin{rem}
\label{rem:counter_construct}In fact, $\Lambda_{\vartheta}$ in (\ref{eq:unstable_Lambda})
is constructed so that condition \ref{enu:R_strict_majorization}
is violated for all $\vartheta<1$. One can easily check that under
the setting of Example \ref{exa:counter_unstable_2}, we have $\probthresh=F_{|\YOpt|}(\tauOpt)>0$.
From (\ref{eq:unstable_Lambda}) we can get $F_{\tauOpt\Lambda_{\vartheta}}^{-1}(u)=F_{|\YOpt|}^{-1}(u)$,
for all $u\in[F_{|\YOpt|}(\vartheta\tauOpt),F_{|\YOpt|}(\tauOpt)]$.
Also due to the fact that $\YOpt$ is supported on $\R$, we have
$F_{|\YOpt|}(\vartheta\tauOpt)<F_{|\YOpt|}(\tauOpt)$, when $\vartheta<1$.
Therefore, $\int_{t}^{\probthresh}F_{|\YOpt|}^{-1}(u)du=\int_{t}^{\probthresh}F_{\tauOpt\Lambda_{\vartheta}}^{-1}(u)du$
for any $t\in[F_{|\YOpt|}(\vartheta\tauOpt),\probthresh)$, where
we have used $\probthresh=F_{|\YOpt|}(\tauOpt)$. This violates condition
\ref{enu:R_strict_majorization}. On the other hand, we can also check
when $\vartheta=1$,\emph{ i.e.}, in the LASSO case, condition \ref{enu:R_strict_majorization}
is satisfied. Indeed, in this case $\Lambda_{\vartheta}=1$ and $F_{\tauOpt\Lambda_{\vartheta}}^{-1}(u)=F_{\tauOpt}^{-1}(u)=\tauOpt$
for any $u\in[0,\probthresh]$. Besides, since $\probthresh=F_{|\YOpt|}(\tauOpt)>0$
and $\YOpt$ is supported on $\R$, we get $F_{|\YOpt|}^{-1}(u)<\tauOpt$
for any $u\in[0,\probthresh)$. Therefore, $\int_{t}^{\probthresh}F_{|\YOpt|}^{-1}(u)du<\int_{t}^{\probthresh}F_{\tauOpt\Lambda_{\vartheta}}^{-1}(u)du$
for any $t\in[0,\probthresh)$.
\end{rem}
\begin{rem}
In Example \ref{exa:counter_unstable_2}, a superficial reason for $R_{0}^{(p)}\not\to\P\big(\prox(\YOpt)=0\big)$
when $\vartheta<1$ is that $\vlambda$ generated from such $\Lambda$
will lead to many pseudo-zero entries in $\sol$, \emph{i.e.}, entries
that are very closed to 0, but not strictly 0. This is illustrated
in Fig. \ref{subfig:Counter-example-R0_hist1} and \ref{subfig:Counter-example-R0_hist2}.
In practice, the pseudo-zero effects
can be mitigated by employing post-screening to
$\sol$. This is done by first specifying a threshold $h>0$ and then
setting all the entries in $\sol$ with $|\soli_{i}|<h$ to be zero.
However, this creates a new problem of choosing the appropriate $h$.
Our claim is that this problem can be completely avoided by adding an extra constraint on the regularizing sequence. Moreover,
as will be clarified in Sec. \ref{sec:max_power}, this additional constraint will not harm the diversity of our design choices.
\end{rem}
Based on Proposition \ref{prop:valid_testing_2}, we can now compute
the limiting Type-I error and power of SLOPE. %

\begin{cor}
\label{cor:lim_error_power}
When $\P(B=0)\in(0,1)$, we have
\begin{align}
\lim_{p\to\infty}\text{Type-I error} & =\P(|\sigOpt H|\geq\ythresh)\label{eq:type-I_error_limit}
\end{align}
and
\begin{align}
\lim_{p\to\infty}\text{Power}= & \P(|B+\sigOpt H|\geq\ythresh\mid B\neq0),\label{eq:power_limit}
\end{align}
where $\ythresh= \sup_{y\geq0}\{y\mid\prox(y;\mealim_{\YOpt},\mealim_{\tauOpt\Lambda})=0\}$.
\end{cor}
The proof of Corollary \ref{cor:lim_error_power} directly follows
from (\ref{eq:type-I_power_formula}), (\ref{eq:R_V_valid_limit}) and the assumption that $r_{0}^{(p)}/p \to \P(B\neq 0)$.
Formulas (\ref{eq:type-I_error_limit}) and (\ref{eq:power_limit})
will be useful in Sec. \ref{sec:max_power}, where we analyze the optimal performance of SLOPE for variable selection.
\begin{rem}
In Corollary \ref{cor:lim_error_power}, we require that $\P(B=0)\in(0,1)$.
This means asymptotically, the proportions of zero and non-zero entries
of $\sgl$ are both non-vanishing. We need this assumption on the
distribution of $B$, because otherwise the limiting formula of Type-I error and power will
involve with $\frac{0}{0}$ term, when we apply (\ref{eq:type-I_power_formula}).
This is beyond the scope of asymptotic setting considered in this paper.
\end{rem} 

\section{Fundamental Limits and Optimal Regularization\label{sec:Oracle-Optimality}}

Armed with the asymptotic characterizations in Theorem \ref{thm:asymp_char} and Proposition \ref{prop:valid_testing_2}, we are now ready to analyze the optimal performance of SLOPE in both estimation and variable selection setting.
\subsection{Estimation with Minimum MSE}
\label{sec:mse}

We first turn to the problem of finding the minimum MSE achievable by SLOPE estimator and the corresponding optimal regularization. In the current asymptotic setting, this can be formulated as follows:
\begin{equation}
\label{eq:optdesign_formulation}
\begin{aligned}
\inf_{\mealim_{\Lambda}\in\Lamspace} \lim_{p\to\infty}\frac{1}{p}\|\sol-\sgl\|_{2}^{2}
\end{aligned}
\end{equation}
where $\Lamspace\bydef\{\mealim_{\Lambda}\mid\mealim_{\Lambda}\in\Wspace(\R)\text{ and }\P(\Lambda\neq 0)>0, \text{ when }\delta\leq 1 \}$ is the admissible set of $\mealim_{\Lambda}$, under which the asymptotic characterization in Theorem \ref{thm:asymp_char} is valid.
By \eqref{eq:limit_MSE},
solving \eqref{eq:optdesign_formulation} is equivalent to solving
\begin{equation}
\label{eq:inf_sigmastar}
\inf_{\mealim_{\Lambda}\in\Lamspace}\sigOpt.
\end{equation}
In the current context, $\sigOpt$ should be understood as a function of $\mealim_{\Lambda}$, but for notational simplicity, we will drop its dependency on $\mealim_{\Lambda}$, when doing so causes no confusion.

Note that $\sigOpt$
is determined by $\mealim_{\Lambda}$ implicitly through the nonlinear fixed point equation (\ref{eq:sigmatau_1})-(\ref{eq:sigmatau_2}), so a
direct optimization over $\mealim_{\Lambda}$ as in \eqref{eq:inf_sigmastar} is not viable.
To proceed, a key observation from (\ref{eq:sigmatau_1})-(\ref{eq:sigmatau_2}) is that the influence of $\mealim_{\Lambda}$ is exerted only through the limiting scalar function $\prox$. In light of this, \eqref{eq:inf_sigmastar} can be alternatively solved via the following two-step scheme:
\begin{enumerate}
  \item [Step 1.]
  Search over all \emph{realizable} $\prox$ such that there exists $\sigma, \tau>0$ satisfying
    \begin{align}
    \sigma^{2} & =\sigma_{\noisei}^{2}+\frac{1}{\delta}\E[(\prox(B+\sigma H)-B)^{2}]\label{eq:sigmatau_12}\\
    1 & =\tau\big(1-\frac{1}{\delta}\E[\prox'(B+\sigma H)]\big)\label{eq:sigmatau_22}
    \end{align}
  and find optimal $\prox^{\star}$ that yields the minimum feasible $\sigma$. Denote the corresponding solution of \eqref{eq:sigmatau_12}-\eqref{eq:sigmatau_22} as $(\sigma^{\star},\tau^{\star})$.
  \item [Step 2.]
  Find corresponding $\mealim_{\Lambda}$ such that $\prox(y;\mealim_{B+\sigma^{\star} H},\mealim_{\tau^{\star}\Lambda})=\prox^{\star}(y)$.
\end{enumerate}
Note that in Step 1, $\prox$ is treated as an optimization variable that do not depend on other parameters,
which greatly simplifies the original formulation \eqref{eq:inf_sigmastar}.
However, to implement this scheme, we still need to guarantee two things. First, the \emph{realizable} set of $\prox$ (as required in Step 1) needs to be decided. Second, for any realizable $\prox$, the corresponding $\Lambda$ can be efficiently computed. These are both addressed in the following result.
\begin{prop}
\label{prop:function_space}
For a probability measure $\mealim_{Y}\in\Wspace(\R)$, define
\begin{equation}
\mathcal{M}_{\mealim_{Y}}\bydef\left\{\prox(\cdot\,; \mealim_Y, \mealim_\Lambda)\mid \mealim_{\Lambda}\in\Wspace(\R)\right\},
\end{equation}
where $\prox(\cdot\,; \mealim_Y, \mealim_\Lambda)$ is the limiting scalar function in Proposition \ref{prop:prox}. Then for any $\mealim_{Y}\in\Wspace(\R)$, we have $\mathcal{M}_{\mealim_{Y}}=\Lipndset$.
Correspondingly, for any $\proxfree(y)\in \Lipndset$, we can take $\Lambda\sim|Y|-\proxfree(|Y|)$, with $Y\sim\mealim_{Y}$, so that $\prox(y\,; \mealim_Y, \mealim_\Lambda)=\proxfree(y)$.
\end{prop}

The proof of Proposition \ref{prop:function_space} will be presented in Appendix \ref{subsec:Proof-of-Proposition-functionspace}. It is the key ingredient in proving our optimality results. It shows that, with different choices of $\mealim_{\Lambda}$, one can reach any non-decreasing and odd
function that is Lipschitz continuous with constant 1. Clearly, the soft-thresholding functions associated with LASSO belongs to $\mathcal{M}_{\mealim_{Y}}$, but the set $\mathcal{M}_{\mealim_{Y}}$ is much richer. This is how SLOPE generalizes LASSO: it allows for more degrees of freedom in the regularization.


Based on Proposition \ref{prop:function_space}, we are now ready to show the two-step scheme sketched above indeed yield a computationally feasible procedure to obtain the minimum MSE and the optimal $\mealim_{\Lambda}$. Before that, we first introduce the following function:
\begin{equation}
\label{eq:opt_esti_prox}
\begin{aligned}
\mathcal{L}(\sigma)\bydef& \inf_{\proxfree\in\Lipndset} \E[\proxfree(B+\sigma H)-B]^{2}\\
& \text{ s.t. } \delta^{-1}\E[\proxfree'(B+\sigma H)]\leq 1.
\end{aligned}
\end{equation}
We will see for any $\sigma>0$, problem \eqref{eq:opt_esti_prox} is convex and there exists a unique optimal solution.
Given $\mathcal{L}(\sigma)$, we then introduce the following equation on $\sigma$:
\begin{equation}
\mathcal{L}(\sigma)=\delta(\sigma^2 - \sigma_{\noisei}^2).
\label{eq:sigma_min_equation}
\end{equation}
As is shown in Proposition \ref{prop:min_MSE} below, the minimum limiting MSE is closely related with the minimum solution of equation \eqref{eq:sigma_min_equation}.
\begin{prop}
\label{prop:min_MSE}
Under the same setting as Theorem \ref{thm:asymp_char}, we have
\begin{itemize}
  \item [(a)]  For any $\sigma>0$, problem \eqref{eq:opt_esti_prox} is convex and there exists a unique optimal solution $\proxfree_{\sigma}\in\Lipndset$.
  \item [(b)] $\mathcal{L}(\sigma)$ defined in \eqref{eq:opt_esti_prox} is continuous on $\R_{>0}$ and equation \eqref{eq:sigma_min_equation} always has a solution. The minimum solution
  $\minsigsol\in \big[\sigma_{\noisei},\,\sqrt{\sigma_{\noisei}^{2}+\delta^{-1}\E B^{2}}\big].$
  \item [(c)]
    Define $\optY:=B+\minsigsol H$ and $\mintausol := \left[1-{\delta^{-1}}{\E \proxfree_{\minsigsol}'(\optY)}\right]^{-1}.$ It always holds that
    \begin{equation}
    \label{eq:MSElbd}
    \lim_{p\to\infty}\frac{1}{p}\|\sol-\sgl\|_{2}^{2}\geq \delta(\minsigsol^2 - \sigma_{\noisei}^2).
    \end{equation}
    Moreover, if $\delta^{-1}\E\big[\proxfree_{\minsigsol}'(\optY)\big]<1$, the equality in \eqref{eq:MSElbd} can be attained, when $\mealim_{\Lambda}$ is the law of
    \begin{equation}
    \label{eq:optimal_estimation_dist}
    \frac{1}{\mintausol}\big[|\optY|-\proxfree_{\minsigsol}(|\optY|)\big].
    \end{equation}
\end{itemize}
\end{prop}
The proof of Proposition \ref{prop:min_MSE} is deferred to Appendix \ref{subsec:min_MSE_proof}. To solve the infinite-dimensional optimization problem \eqref{eq:opt_esti_prox} in practice, we can discretize over $\R$ and obtain a finite-dimensional approximation. Naturally, this finite-dimensional problem is still convex. In our simulation, we use an approximation with 2048 grids.

We have a couple of comments regarding Proposition \ref{prop:min_MSE} as follows.
\begin{rem}[Interpretation of $\mathcal{L}(\sigma)$]
Consider the optimization in \eqref{eq:opt_esti_prox}:
\begin{equation}
\label{eq:gseq_minMSE}
\inf_{\proxfree\in\Lipndset} \E[\proxfree(B+\sigma H)-B]^{2}
\end{equation}
and we neglect the constraint $\delta^{-1}\E[\proxfree'(B+\sigma H)]\leq 1$ for a moment. From Proposition \ref{prop:prox} and Proposition \ref{prop:function_space} we know minimization in \eqref{eq:gseq_minMSE} is equivalent to
\[
\inf_{\mealim_{\Lambda}\in\Wspace(\R)}\lim_{p\to\infty}\frac{1}{p}\,\|\text{Prox}_{{\vlambda}}(\sgl + \sigma \vh)-\sgl\|^{2},
\]
where $\vh\sim\mathcal{N}(\boldsymbol{0},\mI_{p})$ and $\mealim_{\Lambda}=\lim_{p\to\infty}\mealim_{\vlambda}$. In other words, we are estimating $\sgl$ from the noisy observation: $\vy = \sgl + \sigma \vh$ using SLOPE and we want to find the optimal regularization (specified by its limiting distribution $\mealim_{\Lambda}$) such that the estimation error of $\sgl$ is minimized. Then $\mathcal{L}(\sigma)$ can be understood as the minimum MSE we can achieve, if we put an additional constraint on the average slope of limiting scalar function. On the other hand, if at the optimal solution $\proxfree_{\sigma}$, the constraint is inactive, \emph{i.e.}, $\delta^{-1}\E[\proxfree_{\sigma}'(B+\sigma H)]<1$, then $\mathcal{L}(\sigma) = \inf_{\proxfree\in\Lipndset} \E[\proxfree(B+\sigma H)-B]^{2}$. This can be easily verified as follows. Assume there exists $\proxfree_{\star}\in \Lipndset$ such that $\E[\proxfree_{\star}(B+\sigma H)-B]^{2} < \mathcal{L}(\sigma)$. Then consider the convex combination $\proxfree_{t}:=t \proxfree_{\star} + (1-t)\proxfree_{\sigma}$, for $t\in(0,1)$. Clearly, $\proxfree_{t}\in\Lipndset$ and it is not hard to check for small enough $t$, $\delta^{-1}\E[\proxfree_{t}'(B+\sigma H)]\leq 1$. However, due to the convexity of objective function in  \eqref{eq:opt_esti_prox},
\begin{align*}
\E[\proxfree_{t}(B+\sigma H)-B]^{2} &\leq t \underbrace{\E[\proxfree_{\star}(B+\sigma H)-B]^{2}}_{<\mathcal{L}(\sigma)} + (1-t)
\underbrace{\E[\proxfree_{\sigma}(B+\sigma H)-B]^{2}}_{=\mathcal{L}(\sigma)} < \mathcal{L}(\sigma),
\end{align*}
which leads to a contradiction.
\end{rem}
\begin{rem}[Tightness of lower bound \eqref{eq:MSElbd}]
We require $\delta^{-1}\E\proxfree_{\sigma_{0}}'(B+\sigma_{0} H)<1$ so that the lower bound \eqref{eq:MSElbd} is tight. The question is whether it is possible that $\delta^{-1}\E\proxfree_{\sigma_{0}}'(B+\sigma_{0} H)=1$. This will not happen when $\delta>1$, since $\proxfree_{\sigma_{0}}'\leq 1$. When $\delta\leq 1$, we do not have a rigorous proof yet. Numerically, this never happens either.
Here we provide an intuitive argument. Suppose for certain configurations of $(\delta,\rho,\sigma_{\noisei}, \mealim_{B})$, we do have $\delta^{-1}\E\proxfree_{\sigma_{0}}'(B+\sigma_{0} H)=1$. Under this scenario, let us consider the following approximation of \eqref{eq:opt_esti_prox} and \eqref{eq:sigma_min_equation}, parameterized by $\veps>0$:
\begin{equation}
\label{eq:opt_esti_prox_eps}
\begin{aligned}
\mathcal{L}_{\veps}(\sigma)\bydef& \inf_{\proxfree\in\Lipndset} \E[\proxfree(B+\sigma H)-B]^{2}\\
& \text{ s.t. } \delta^{-1}\E[\proxfree'(B+\sigma H)]\leq 1-\veps.
\end{aligned}
\end{equation}
and
\begin{equation}
\mathcal{L}_{\veps}(\sigma)=\delta(\sigma^2 - \sigma_{\noisei}^2).
\label{eq:sigma_min_equation_eps}
\end{equation}
Denote $\sigma_{0,\veps}$ as the minimum solution of equation \eqref{eq:sigma_min_equation_eps} and $\proxfree_{\veps}$ as the optimal solution of \eqref{eq:opt_esti_prox_eps}, when $\sigma=\sigma_{0,\veps}$. If we take $\mealim_{\Lambda}$ to be the law of
\begin{equation}
\label{eq:meaLambda_eps}
\frac{1}{\tau_{0,\veps}}[|Y_{0,\veps}|-\proxfree_{\veps}(|Y_{0,\veps}|)],
\end{equation}
where $Y_{0,\veps} = B+\sigma_{0,\veps}H$ and $\tau_{0,\veps} = \big[1-{\delta^{-1}}{\E \proxfree_{\veps}'(Y_{0,\veps})}\big]^{-1}<\infty$. Then similar as Proposition \ref{prop:min_MSE}, it is not hard to show
$\lim_{p\to\infty}\frac{1}{p}\|\sol_{\veps}-\sgl\|_{2}^{2}= \delta(\sigma_{0,\veps}^2 - \sigma_{\noisei}^2)$, where $\sol_{\veps}$ denotes the corresponding estimator.
Intuitively, we could also expect $\sigma_{0,\veps}\to\sigma_{0}$ and $\tau_{0,\veps}\to\infty$, as $\veps\to 0$. This implies the MSE can be made arbitrarily close to the lower bound \eqref{eq:MSElbd} using a sequence $\{\mealim_{\Lambda,\veps}\}_{\veps>0}$ which converges to the probability mass at 0 as $\veps\to 0$.
Recall that we have assumed $\sigma_{\noisei}>0$, so this means the optimal regularization in a noisy overparameterized linear model should be vanishingly small, which is not likely the case.
\end{rem}
\begin{rem}[Comparison with \cite{celentano2019fundamental,celentano2019approximate}]
In \cite{celentano2019fundamental,celentano2019approximate}, the authors also analyze the problem of optimal estimation in the linear model \eqref{eq:model} with i.i.d. Gaussian design. For the convenience of comparison, here we rephrase their results in our notations. The optimality they consider is with respect to the following class of estimator:
\begin{equation}
\label{eq:symtric_regu_estimator}
\Big\{\sol~\Big|~\sol \in \argmin{\est} \frac{1}{2}\|\vy-\dmtx\est\|_{2}^{2}+r_p(\est), r_p\in\mathcal{C}_{p}\Big\}
\end{equation}
where
$$
\mathcal{C}_{p}\bydef \{r_p:\R^{p}\to\bar{\R}\mid r_p \text{ is lsc, proper, convex and symmetric} \}.
$$
The optimal estimation within the class of $\mathcal{C}_{p}$ is formulated as:
\begin{equation}
\label{eq:min_MSE_cvx}
\text{MSE}_{\text{ cvx}}\bydef\inf_{\substack{\forall p}, r_p\in \mathcal{C}_{p}\cap\mathcal{W}_{p}} \liminf_{p\to\infty}\frac{1}{p}\|\sol-\sgl\|_{2}^{2},
\end{equation}
where $\mathcal{W}_{p}$ is some set that ensures $\sol$ is unique
\footnote{In fact, $\mathcal{W}_{p}$ corresponds to the tightness condition $\delta^{-1}\E\big[\proxfree_{\sigma_{0}}'(B+\sigma_{0}H)\big]<1$ in Proposition \ref{prop:min_MSE} (c).}.
One of their main results states that under certain conditions, the minimum achievable limiting MSE defined in \eqref{eq:min_MSE_cvx} satisfies: $\text{MSE}_{\text{ cvx}} \geq \delta(\sigma_{\text{cvx}}^2 - \sigma_{\noisei}^2)$, where
\begin{equation}
\label{eq:sigma_cvx_def}
\sigma_{\text{cvx}}^2 = \sup\{\sigma^2\mid\delta(\sigma^2-\sigma_{\noisei}^2) < \textstyle\inf_{\proxfree\in\mathcal{J}} \E[\proxfree(B+\sigma H)-B]^{2} \},
\end{equation}
with $\mathcal{J}\bydef\{f: \R\to\R \mid f \text{ is non-decreasing and 1-Lipschitz continuous}\}$. Comparing their results with ours, we can find the lower bounds in both settings follow the same type of characterization.
Specifically, lying at the heart of this characterization is an optimization problem: $\inf_{\proxfree\in\mathcal{F}} \E[\proxfree(B+\sigma H)-B]^{2}$, which aims at finding the optimal estimator $\proxfree$ of $B$ under the noisy observation $B+\sigma H$.
The only difference is on the feasible set $\mathcal{F}$: $\mathcal{F}=\mathcal{J}$ in \eqref{eq:sigma_cvx_def}, while $\mathcal{F}=\Lipndset\subset\mathcal{J}$ in \eqref{eq:opt_esti_prox}.
This agreement is not a coincidence, but related with the fact that the proximal operator of all functions in $\mathcal{C}_{p}$ is asymptotically separable as proved in \cite{celentano2019approximate}.
In fact, $\proxfree$ corresponds to the limiting proximal operator of the regularizer $r_p$.
In our settings, $r_p$ is chosen from the set of all possible sorted $\ell_1$ norms (denoted by $\mathcal{S}_p$), while in their settings,
it is chosen from the set $\mathcal{C}_p$. Correspondingly, $\Lipndset$ is the set of all limiting proximal operators associated with $\mathcal{S}_p$ and $\mathcal{J}$ is the one associated with $\mathcal{C}_p$. It is not hard to check $\mathcal{S}_p\subset \mathcal{C}_p$ and consequently, we have $\Lipndset\subset \mathcal{J}$.
\end{rem}
\begin{figure}[t]
\begin{centering}
\subfloat[$\text{SNR}=4$\label{subfig:MSE_sparsity}]{\begin{centering}
\includegraphics[clip,scale=0.45]{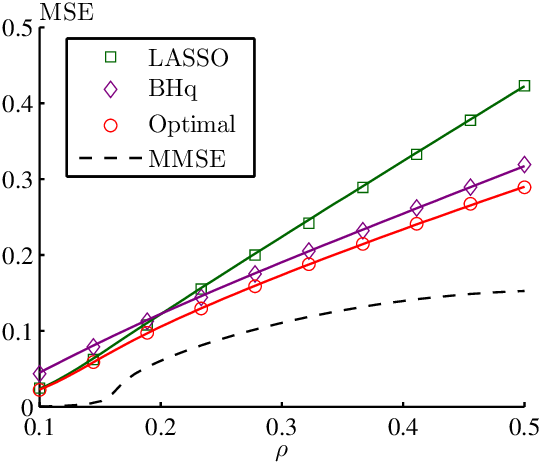}
\par\end{centering}
}
\hspace{0.0cm}
\subfloat[$\rho=0.1$\label{subfig:MSE_SNR1}]{\begin{centering}
\includegraphics[clip,scale=0.45]{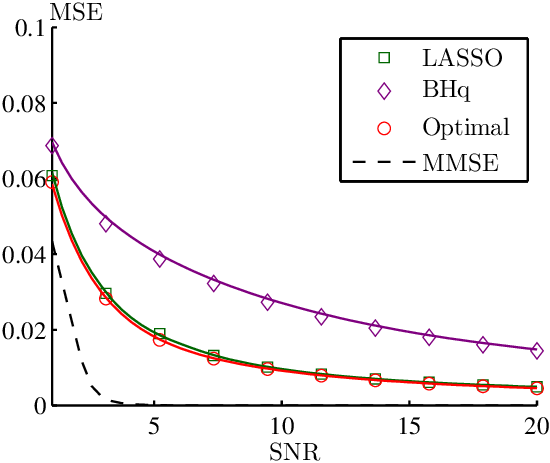}
\par\end{centering}
}
\hspace{0.0cm}
\subfloat[$\rho=0.2$\label{subfig:MSE_SNR2}]{\begin{centering}
\includegraphics[clip,scale=0.45]{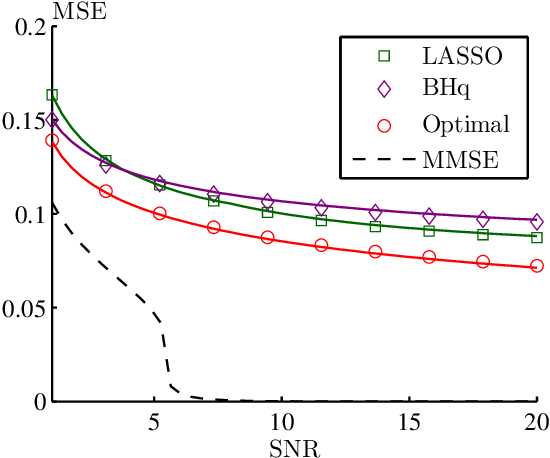}
\par\end{centering}
}
\par\end{centering}
\caption{Comparison of MSEs obtained by three regularization sequences: LASSO,
BHq and the oracle optimal design. Here,
$\protect\sgli_{i}$ are i.i.d. Bernoulli random variables, with $\P(\sgli_{i}=1)=\rho$
and $w_{i}\protect\iid\mathcal{N}(0,\,\sigma_w^{2})$, with $\sigma_w^{2} = \rho/\text{SNR}$
and the BHq sequences are generated by reordering i.i.d. samples of: $\sigma_{\noisei}\Phi^{-1}(1-\frac{q}{2}+\frac{q}{2}U)$, where $q\in(0,1]$ and $U$ follows the uniform distribution over $[0,1]$.
In our simulation, we fix
$p=2048$, $\delta=0.5$ and the empirical results are averaged over 20 independent trials. The dash curves correspond to the information-theoretic limit obtained in \cite{barbier2016mutual,reeves2016replica}.
\label{fig:Comparison-of-MSEs}}
\end{figure}
In Fig. \ref{fig:Comparison-of-MSEs}, we compare the MSEs achieved by different regularizing sequences (LASSO, BHq and oracle optimal design), at different SNR and sparsity
levels.
Since we are concerned with oracle optimality, for fair comparison, we search through the parameters of the BHq and LASSO sequences (in particular, $q$ for BHq and $\lambda$ for LASSO) and report the minimum MSEs that can be achieved.
The solid curves correspond to the theoretical MSEs predicted by Theorem \ref{thm:asymp_char} and Proposition \ref{prop:min_MSE}.
Note that the empirical MSEs match
well with theoretical predictions
\footnote{Here, the MSEs of LASSO and BHq are obtained by optimizing over the parameters $\lambda$ and $q$, so strictly speaking, the theoretical curves are valid only if a stronger uniform convergence result holds. The uniform convergence for LASSO case is proved in \cite{mousavi2018consistent,miolane2018distribution} and we conjecture that it also holds true for BHq sequences.}.
It is also observed that under each setting,
the MSEs of different regularizing sequences are all above the lower bound obtained in \eqref{eq:MSElbd} (red curve in the figure).
Also we can see this lower bound can be attained when the limiting empirical distribution of $\vlambda$ follows prescribed optimal distribution \eqref{eq:optimal_estimation_dist}.
We also have the following findings:
\begin{enumerate}
  \item
  As can be seen from Fig. \ref{subfig:MSE_sparsity} and Fig. \ref{subfig:MSE_SNR1}, when $\rho$ is small, LASSO performs well and the corresponding MSEs almost match the theoretical lower bound, across different values of SNR. However, its performance degrades faster than the other two sequences, as $\rho$ grows. This is because LASSO's penalization is not adaptive to the underlying sparsity levels and it incurs higher bias under larger $\rho$ \cite{su2016slope}.
  \item
  From Fig. \ref{subfig:MSE_SNR1} and Fig. \ref{subfig:MSE_SNR2}, we can find that at low SNR regimes, the BHq sequence can lead to comparable performance
    as the optimal design. However, at higher SNR regimes, the optimal design notably outperforms
    the BHq sequence. To explain this phenomenon, we plot in Fig. \ref{fig:Comparison-of-distributions} the empirical distributions of the $\vlambda$-sequences associated with the optimal design and the BHq design, respectively. It turns out that, in the low SNR case, the optimal design and BHq have similar distributions, while at higher SNRs, the distribution of the optimal design is close to a mixture of a delta mass and uniform distribution.
\end{enumerate}



\begin{figure}
\begin{centering}
\subfloat[Optimal, SNR=1]{\begin{centering}
\includegraphics[clip,scale=0.49]{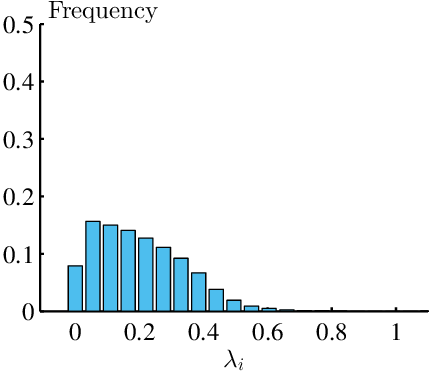}
\par\end{centering}
}
\subfloat[BHq, SNR=1]{\begin{centering}
\includegraphics[clip,scale=0.49]{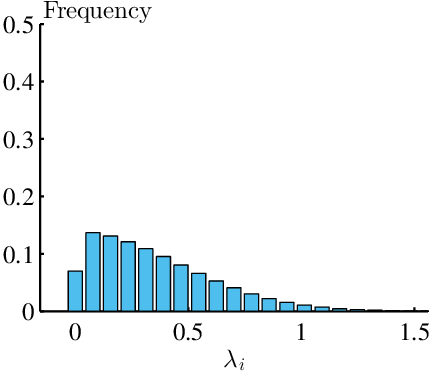}
\par\end{centering}
}
\subfloat[Optimal, SNR=20]{\begin{centering}
\includegraphics[clip,scale=0.49]{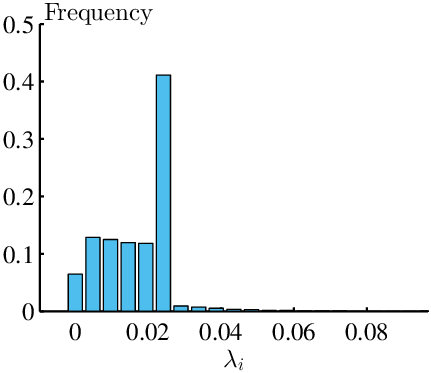}
\par\end{centering}
}
\subfloat[BHq, SNR=20]{\begin{centering}
\includegraphics[clip,scale=0.49]{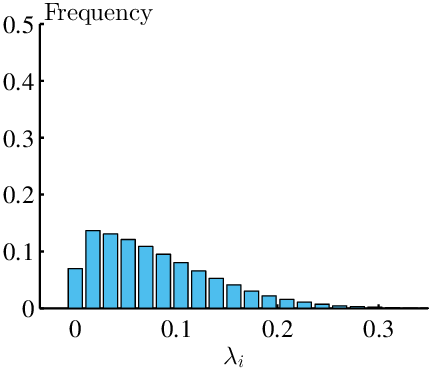}
\par\end{centering}
}
\par\end{centering}
\caption{Comparison of empirical distributions of two regularizing sequences ``BHq'' and ``Optimal'' in Fig.
\ref{subfig:MSE_SNR2}.
\label{fig:Comparison-of-distributions}
}
\end{figure}

\subsection{Variable Selection with Maximum Power}
\label{sec:max_power}
Next we consider using SLOPE
for variable selection. Our goal is
to find the optimal regularizing sequence to achieve the highest possible power, under a given level of type-I error $\alpha$, formulated as:
\begin{equation}
\label{eq:slope_test_opt_0}
\begin{aligned}
\limpower(\alpha)\bydef\sup_{{\Lambda}\in{\tLamspace}} \hspace{0.5em} & \lim_{p\to\infty}\text{Power} \\
\text{s.t.}\hspace{0.5em} & \lim_{p\to\infty}\text{Type-I error} \leq \alpha,
\end{aligned}
\end{equation}
where ${\tLamspace} \bydef \{\Lambda\in \Lamspace:  \text{\ref{enu:R_strict_majorization} is satisfied}\}$ is the admissible set of $\mealim_{\Lambda}$, with which the limits in \eqref{eq:slope_test_opt_0} exist.
In light of \eqref{eq:type-I_error_limit} and \eqref{eq:power_limit}, if $\P(B=0)\in(0,1)$,
optimization problem \eqref{eq:slope_test_opt_0} is equivalent to:
\begin{equation}
\label{eq:slope_test_opt}
\begin{aligned}
\limpower(\alpha)=\sup_{{\Lambda}\in{\tLamspace}} \hspace{0.5em} & \P(|B+\sigOpt H|\geq\ythresh\mid B\neq 0) \\
\text{s.t.}\hspace{0.5em} & \P(|\sigOpt H|\geq\ythresh) \leq \alpha,
\end{aligned}
\end{equation}
where $\ythresh=\sup_{y\geq0}\{y\mid\prox(y;\mealim_{\YOpt},\mealim_{\tauOpt\Lambda})=0\}$. Comparing the admissible set $\tLamspace$ with $\Lamspace$ in \eqref{eq:inf_sigmastar}, it can be seen the only difference is that here we need an additional condition \ref{enu:R_strict_majorization} to ensure the limits of $\text{Type-I error}$ and $\text{Power}$ both exist (see Proposition \ref{prop:valid_testing_2}).
%

To state our results, we first introduce the following function, which is the counterpart of \eqref{eq:opt_esti_prox}.
\begin{equation}
\label{eq:opt_testing_prox}
\begin{aligned}
\mathcal{L}_{\alpha}(\sigma)\bydef\inf_{\proxfree\in\Lipndset\cap\softfunset}\hspace{0.05em} & \E[\proxfree(B+\sigma H)-B]^{2}\\
\text{s.t.}\hspace{0.5em} & \delta^{-1}\E[\proxfree'(B+\sigma H)]\leq 1
\end{aligned}
\end{equation}
where $\alpha\in[0,1]$ is a prescribed Type-I error level and $\softfunset\bydef\big\{\proxfree(y):\proxfree(y)=0\text{ for }|y|\leq\Phi^{-1}(1-\tfrac{\alpha}{2})\sigma\big\}$.
Similar as Proposition \ref{prop:min_MSE}, we will see that the maximum power achievable by SLOPE under Type-I error level $\alpha$ is related with the following equation:
\begin{equation}
\mathcal{L}_{\alpha}(\sigma)=\delta(\sigma^2 - \sigma_{\noisei}^2),
\label{eq:sigma_min_equation1}
\end{equation}
where $\mathcal{L}_{\alpha}(\sigma)$ is the function defined in \eqref{eq:opt_testing_prox}.

\begin{figure}
%
\begin{centering}
\subfloat[\label{fig:ROC1}]{\begin{centering}
\includegraphics[scale=0.5]{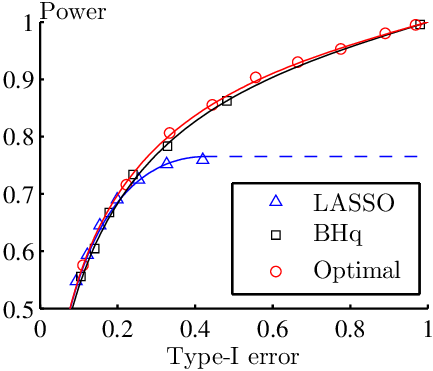}
\par\end{centering}
}
\hspace{0.1cm}
\subfloat[\label{fig:ROC2}]{\begin{centering}
\includegraphics[scale=0.5]{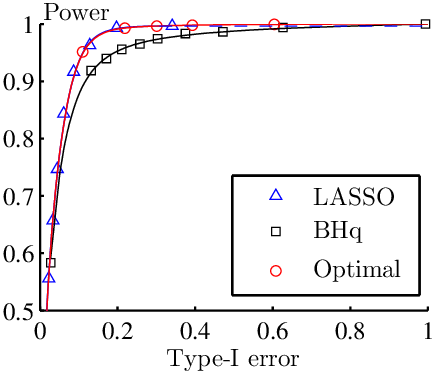}
\par\end{centering}
}
\hspace{0.1cm}
\subfloat[\label{fig:ROC3}]{\begin{centering}
\includegraphics[scale=0.5]{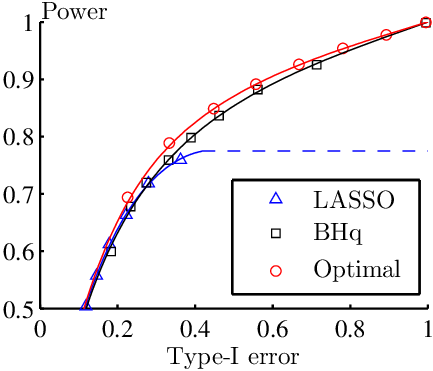}
\par\end{centering}
}
\par\end{centering}
\caption{Theoretical predictions v.s. empirical results of testing performance using LASSO, BHq and oracle optimal sequences. Here,
$\protect\sgli_{i}$ are i.i.d. Bernoulli random variables, with $\P(\sgli_{i}=1)=\rho$
and $w_{i}\protect\iid\mathcal{N}(0,\,\sigma_w^{2})$, with $\sigma_w^{2} = \rho/\text{SNR}$. The empirical results are generated under $p=2048$ and  $\delta=0.5$ and we choose different $\rho$ and SNR values:
(a) $\rho=0.1$, $\text{SNR}=0.6$, (b) $\rho=0.1$, $\text{SNR}=4$, (c) $\rho=0.2$, $\text{SNR}=4$. Dash curves correspond to the observed upper bound of power achieved by LASSO.
\label{fig:Optimal-hypothesis-testing}}
\end{figure}

We are now ready to state our main optimality results for variable selection.
\begin{prop}
\label{prop:max_power}
Under the same setting as Proposition \ref{prop:valid_testing_2}, assume $\P(B=0)\in(0,1)$. Then we have
\begin{itemize}
  \item [(a)]  For any $\alpha\in[0,1]$ and $\sigma>0$, problem \eqref{eq:opt_testing_prox} is convex and there exists a unique optimal solution $\proxfree_{\alpha,\sigma}\in\Lipndset$.
  \item [(b)]  For any $\alpha\in[0,1]$, $\mathcal{L}_{\alpha}(\sigma)$ is continuous on $\R_{>0}$ and equation \eqref{eq:sigma_min_equation1} always has a solution. The minimum solution
  $\tminsigsol\in \big[\sigma_{\noisei},\,\sqrt{\sigma_{\noisei}^{2}+\delta^{-1}\E B^{2}}\big].$
  \item [(c)]
    Let $\toptY := B+\tminsigsol H$ and $\proxfree_{\alpha}:=\proxfree_{\alpha,\tminsigsol}$. If $\lim_{p\to\infty}\text{Type-I error} \leq \alpha$, then
    \begin{equation}
    \label{eq:powerubd}
    \lim_{p\to\infty}\text{Power}\leq\P\big(|\toptY|\geq\Phi^{-1}(1-\tfrac{\alpha}{2})\tminsigsol\mid B\neq0\big).
    \end{equation}
    Moreover,
    if $\delta^{-1}\E\big[\proxfree_{\alpha}'(\toptY)\big]< 1$ and $\tythreshzero=\Phi^{-1}(1-\frac{\alpha}{2})\tminsigsol$,
    the upper bound in \eqref{eq:powerubd} can be attained by $\mealim_{\Lambda}=\mealim_{\text{opt},\alpha}$, with $\mealim_{\text{opt},\alpha}$ being the law of
    \begin{equation}
    \label{eq:optimal_testing_dist}
    \frac{1}{\tmintausol}\max\big\{\tythreshzero,|\toptY|-\proxfree_{\alpha}(|\toptY|)\big\}.
    \end{equation}
    Here, $\tythreshzero=\sup_{y\geq0}\{y\mid\proxfree_{\alpha}(y)=0\}$ and $\tmintausol=\big[1-\delta^{-1}\E\proxfree_{\alpha}'(\toptY)\big]^{-1}$.
\end{itemize}
\end{prop}
The proof of Proposition \ref{prop:max_power}, which is similar to that of Proposition \ref{prop:min_MSE}, will be given in Sec. \ref{subsec:max_power_proof}. A key step is to show the realizable set of $\prox$ in the variable selection setting is still equal to $\Lipndset$ (see Lemma \ref{lem:function_space1} in Appendix \ref{subsec:Auxiliary-Results-for-testing}), although the admissible set of $\mealim_{\Lambda}$ is replaced by $\tLamspace$, which is a subset of $\Lamspace$ in the estimation setting.
\begin{rem}
Comparing the results in Proposition \ref{prop:min_MSE} and Proposition \ref{prop:max_power}, we can find that although at the beginning, we are dealing with two different problems (the objective of the first one is minimizing the MSE, while the other is on maximizing the power under a given Type-I error), we end up with two procedures of very similar natures. Both problems can finally be converted into a formulation involving finding the optimal estimation of $\sgl$ that can be achieved by SLOPE under the observation $\vy=\sgl+\sigma\vh$, with $\vh\sim\mathcal{N}(\boldsymbol{0},\mI)$. The only difference is that in the second problem, we need to enforce an additional restriction on the regularization sequence $\vlambda$ to ensure the Type-I error is below certain threshold $\alpha$.
\end{rem}
\begin{rem}[Tightness of upper bound \eqref{eq:powerubd}]
The tightness of the upper bound for power relies on the conditions: $\delta^{-1}\E\big[\proxfree_{\alpha}'(\toptY)\big]< 1$ and $\tythreshzero=\Phi^{-1}(1-\frac{\alpha}{2})\tminsigsol$.
Numerically they hold under all the settings considered. We conjecture that within our assumptions, this condition always hold and the upper bound \eqref{eq:powerubd} is tight.
\end{rem}
In Fig.\ref{fig:Optimal-hypothesis-testing}, we compare the variable selection performance achieved by the optimal regularization with that of LASSO and BHq sequences.
We show both theoretical ROC curves and the empirical power under given Type-I error levels. Here each empirical $(\text{Type-I error}, \text{power})$ pair is generated by first fixing all the parameters (including the tuning parameters such as $\lambda$ and $q$) and then averaging over 20 independent trials. It can seen that the empirical results match well with the theoretical predictions (solid curves in the figures) and the optimal design of regularization dominates the other two regularizing sequences. We also have the following observations:
\begin{enumerate}
  \item
  In all cases, the theoretical upper bounds on power \eqref{eq:powerubd} can be achieved by choosing $\mealim_{\Lambda}$ to be the law of \eqref{eq:optimal_testing_dist}.
  \item
  The performance of LASSO is closed to the fundamental limit at low sparsity and high SNR regimes, while its performance is significantly degraded as sparsity grows higher or SNR grows lower. In particular, we can find in such cases, the maximum power achievable by LASSO is less than 1. This phenomenon is also discussed in \cite{bogdan2015slope,su2016slope,su2017false}
    and it is inherently connected with the so-called ``noise-sensitivity
    '' phase transition \cite{donoho2011noise}. In comparison, the optimal and BHq sequences can both reach power 1, after Type-I errors are above certain thresholds.
  \item
  Complementary to LASSO, the performance of  BHq sequences is closed to the theoretical upper bounds at low SNRs or large sparsity levels, while it deviates from the upper bounds in other scenarios.
\end{enumerate}

\section{Proof of Main Results\label{sec:Proof-of-Main-Results}}

\subsection{Asymptotic Separability\label{subsec:Proof-of-Asymp-sepa}}

In this section, we are going to prove Proposition \ref{prop:prox}.

From (\ref{eq:slope_prox}) we have the following scaling property:
$\mathcal{M}_{\vlambda}(\vy;\tau)=\frac{1}{\tau}\mathcal{M}_{\tau\vlambda}(\vy;1)$.
On the other hand, for any $\tau>0$, if $\vlambda$ is a converging
sequence with limiting measure $\mealim_{\Lambda}$, it is not hard
to show $\tau\vlambda$ is also a converging sequence, with limiting
measure $\mealim_{\tau\Lambda}$. Thus, to study the asymptotic limit
of (\ref{eq:slope_prox}) under $(\vy,\vlambda,\tau)$, it suffices
to consider $(\vy,\tau\vlambda,1)$. As a result, without loss of
generality, we will assume $\tau=1$ in the rest of our proof.

\subsubsection{Some preliminary facts about SLOPE}

The asymptotic separability stems from the following unique properties
of the SLOPE proximal minimization problem (\ref{eq:slope_prox}),
which are proved in \cite[Sec. 2]{bogdan2013statistical}.
\begin{fact}
\label{fact:SLOPE_prox_properties}For any $\vlambda,\vy\in\R^{p}$,
with $\lambda_{i}\geq0$, for all $i\in[p]$, it holds that \emph{}%
\begin{enumerate}
\item[(i)] \emph{(Sign consistency)} For any $i\in[p]$, $[\tprox_{\vlambda}(\vy)]_{i}$
has the same sign as $y_{i}$. Moreover, 
\[
[\tprox_{\vlambda}(\vy)]_{i}=\sgn(y_{i})[\tprox_{\vlambda}(|\vy|)]_{i}.
\]
\item[(ii)] \emph{(Permutation-invariance)} For any permutation matrix $\Pi$,
$\Pi\tprox_{\vlambda}(\vy)=\tprox_{\vlambda}(\Pi\vy).$
\item[(iii)] \emph{(Monotonicity and Lipschitz continuity)} For any $i,j\in[p]$,
if $y_{i}\leq y_{j}$, then $0\leq[\tprox_{\vlambda}(\vy)]_{j}-[\tprox_{\vlambda}(\vy)]_{i}\leq y_{j}-y_{i}$
and for any $y_{i}$, $[\tprox_{\vlambda}(\vy)]_{i}\leq|y_{i}|$.
\end{enumerate}
\end{fact}
An immediate yet important implication of Fact \ref{fact:SLOPE_prox_properties}
is the following lemma:
\begin{lem}
\label{lem:embedding}For any $\vlambda,\vy\in\R^{p}$, with $\lambda_{i}\geq0$,
there always exists an odd, non-decreasing and 1-Lipschitz function
$g_{p}$ such that for all $i\in[p]$, $g_{p}(y_{i})=[\tprox_{\vlambda}(\vy)]_{i}$.
\end{lem}
The proof of Lemma \ref{lem:embedding} is given in Appendix \ref{subsec:Proof-of-Lemma-embedding}.
By Lemma \ref{lem:embedding} we know $\tprox_{\vlambda}(\vy)$ is
actually the restriction of a function $g_{p}\in\Lipndset$ onto the
support of $\mealim_{\vy}$. Moreover, from the permutation invariance
property (Fact \ref{fact:SLOPE_prox_properties} (ii)), such $g_{p}$
is only determined by the empirical measure $\mealim_{\vlambda}$
and $\mealim_{\vy}$. We could expect if $\mealim_{\vlambda}$ and
$\mealim_{\vy}$ both converge to some limiting distributions, $g_{p}$
will also converge to certain limiting scalar function. This is exactly
the essential meaning of asymptotic separability.

\begin{figure}
\begin{centering}
\subfloat[$p=32$]{\begin{centering}
\includegraphics[scale=0.5]{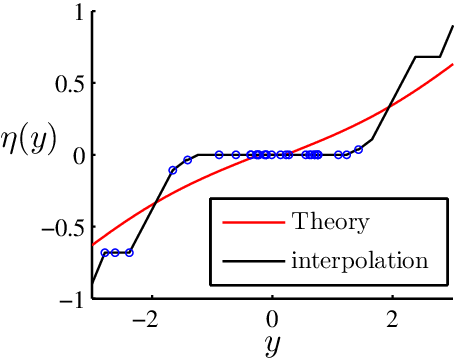}
\par\end{centering}
}\subfloat[$p=128$]{\begin{centering}
\includegraphics[scale=0.5]{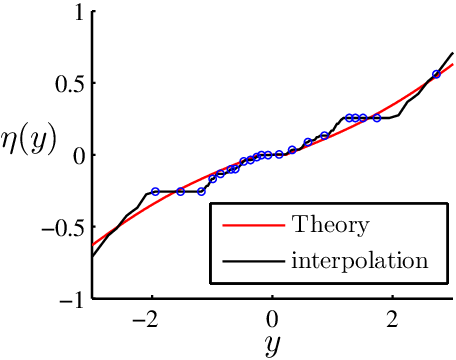}
\par\end{centering}
}\subfloat[$p=1024$]{\begin{centering}
\includegraphics[scale=0.5]{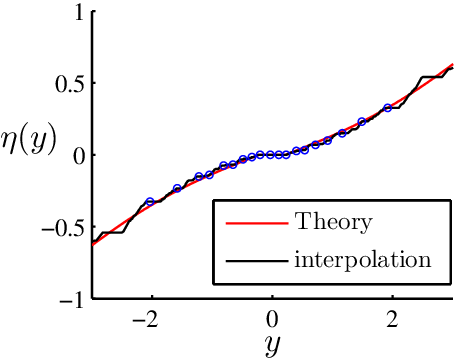}
\par\end{centering}
}
\par\end{centering}
\caption{\label{fig:Comparison-between-linearinter}Comparison between $\protect\prox(y)$
(red curve), linear interpolation (black curve) and $\{(y_{i},[\protect\tprox_{\protect\vlambda}(\protect\vy)]_{i})\}_{i\in[p]}$
(blue dots), under three different values of $p$. Here $y_{i}\protect\iid\mathcal{N}(0,2)$
and $\lambda_{i}$ are i.i.d. samples from BHq distribution \cite{bogdan2013statistical}.}
\end{figure}

Before proceeding, let us take a look at a numerical justification
shown in Fig. \ref{fig:Comparison-between-linearinter}. Here we choose
$g_{p}$ to be the linear interpolation of the following set of points:
$\{(|y_{i}|,|\hat{y}_{i}|),(-|y_{i}|,-|\hat{y}_{i}|)\}_{i=1}^{p}\cup(0,0),$
where $\hat{\vy}:=\tprox_{\vlambda}(\vy)$. It is easy to check (as
is shown in the proof of Lemma \ref{lem:embedding}) such linear interpolation
is a qualified candidate for $g_{p}$ in Lemma \ref{lem:embedding}.
We compare it with the limiting scalar function $\prox(y)$ predicted
by Proposition \ref{prop:prox}. It is clear that as $p$ becomes
larger, $g_{p}(y)$ gets increasingly close to $\prox(y)$.

\subsubsection{An equivalent form of (\ref{eq:slope_prox})}

Based on Lemma \ref{lem:embedding}, we then go on to show the equivalence
between (\ref{eq:slope_prox}) and the following problem:

\begin{equation}
\min_{g\in\Lipndset}\underbrace{\frac{1}{2}\E_{\mu_{\vy}}[Y-g(Y)]^{2}+\int_{0}^{1}F_{\vlambda}^{-1}(u)F_{|g(\vy)|}^{-1}(u)du}_{\bydef L_{p}(g)}.\label{eq:finite_SLOPE}
\end{equation}
This is formalized in the following lemma, whose proof is given in
Appendix \ref{subsec:Proof-of-Lemma_slope_embedding}.
\begin{lem}
\label{lem:slope_prox_embedding}Denote $\mathcal{M}_{\vlambda}^{*}(\vy)$
as the optimal value of (\ref{eq:finite_SLOPE}). Then it holds that
$\frac{\mathcal{M}_{\vlambda}(\vy;1)}{p}=\mathcal{M}_{\vlambda}^{*}(\vy)$.
Besides, any optimal solution $g_{p}^{*}(y)$ of (\ref{eq:finite_SLOPE})
satisfies: $g_{p}^{*}(\vy)=\tprox_{\vlambda}(\vy)$.
\end{lem}
Comparing (\ref{eq:infinite_SLOPE}) and (\ref{eq:finite_SLOPE}),
it could be now understood that the optimization in (\ref{eq:infinite_SLOPE})
is the limit of (\ref{eq:finite_SLOPE}), as $\mealim_{\vlambda}\to\mu_{\Lambda}$
and $\mealim_{\vy}\to\mu_{Y}$. Therefore, from Lemma \ref{lem:slope_prox_embedding},
we could expect $\frac{1}{p}\mathcal{M}_{\vlambda}(\vy;1)=\mathcal{M}_{\vlambda}^{*}(\vy)\to\mathcal{M}_{\mealim_{\Lambda}}(\mealim_{Y},1)$.
On the other hand, $g_{p}^{*}(\cdot)$, which is the optimal solution
of (\ref{eq:finite_SLOPE}) should also converge to the optimal solution
of (\ref{eq:infinite_SLOPE}): $\prox(\cdot;\mealim_{Y},\mealim_{\Lambda})$.
Thus for any $\vlambda$, $\vy$ satisfying $\mealim_{\vlambda}\approx\mu_{\Lambda}$
and $\mealim_{\vy}\approx\mu_{Y}$, we would have $\tprox_{\vlambda}(\vy)=g_{p}^{*}(\vy)\approx\prox(\vy;\mealim_{Y},\mealim_{\Lambda})$,\emph{
i.e.}, asymptotic separability holds. The final step of the proof
is to make the above intuition accurate and rigorous. %

\subsubsection{Taking the limit of (\ref{eq:finite_SLOPE})}

Recall that we have assumed $\tau=1$. For notational simplicity,
denote $\mathcal{M}_{\vlambda}(\vy):=\mathcal{M}_{\vlambda}(\vy;1)$
and $\mathcal{M}_{\mealim_{\Lambda}}(\mealim_{Y}):=\mathcal{M}_{\mealim_{\Lambda}}(\mealim_{Y},1)$.
Define $L(g)$ as the objective function of (\ref{eq:infinite_SLOPE}),
\emph{i.e.},
\begin{equation}
L(g)\bydef\frac{1}{2}\E_{\mu_{Y}}[Y-g(Y)]^{2}+\int_{0}^{1}F_{\Lambda}^{-1}(u)F_{|g(Y)|}^{-1}(u)du.\label{eq:Lg_def}
\end{equation}
and $g^{*}$ as the corresponding optimal solution. By Lemma \ref{lem:lg_unif_convergence}
in Appendix \ref{subsec:Other-Auxiliary-Results-prop-prox}, we have
$\sup_{g\in\Lipndset}|L(g)-L_{p}(g)|\to0$, where $L_{p}(g)$ is the
objective function of (\ref{eq:finite_SLOPE}). Therefore, (\ref{eq:Moreau_conv})
immediately follows, since 
\begin{align*}
|\tfrac{1}{p}\mathcal{M}_{\vlambda}(\vy;1)-\mathcal{M}_{\mealim_{\Lambda}}(\mealim_{Y},1)| & =|\sup_{g\in\Lipndset}L(g)-\sup_{g\in\Lipndset}L_{p}(g)|\\
 & \leq\sup_{g\in\Lipndset}|L(g)-L_{p}(g)|.
\end{align*}
On the other hand,
\begin{equation}
\begin{aligned}L_{p}(g^{*})-L_{p}(g_{p}^{*}) & =L_{p}(g^{*})-L(g^{*})+L(g_{p}^{*})-L_{p}(g_{p}^{*})\\
 & \;+L(g^{*})-L(g_{p}^{*})\\
 & \leq|L_{p}(g^{*})-L(g^{*})|+|L(g_{p}^{*})-L_{p}(g_{p}^{*})|,
\end{aligned}
\label{eq:Lp_optimality}
\end{equation}
where in the last step we use the optimality of $g^{*}$. By the strong
convexity of (\ref{eq:finite_SLOPE}), we have
\begin{equation}
L_{p}(g^{*})-L_{p}(g_{p}^{*})\geq\frac{1}{2}\E_{\mu_{\vy}}|g_{p}^{*}(Y)-g^{*}(Y)|^{2}.\label{eq:Lp_strong_convx}
\end{equation}
Combining (\ref{eq:Lp_optimality}) and (\ref{eq:Lp_strong_convx})
gives
\[
\E_{\mu_{\vy}}|g_{p}^{*}(Y)-g^{*}(Y)|^{2}\leq2\sup_{g\in\Lipndset}|L(g)-L_{p}(g)|.
\]
By Lemma \ref{lem:lg_unif_convergence} again, we have $\E_{\mu_{\vy}}|g_{p}^{*}(Y)-g^{*}(Y)|^{2}\to0$,
as $p\to\infty$. This is exactly (\ref{eq:prox_sep}), since $g_{p}^{*}(\vy)=\tprox_{\vlambda}(\vy)$
by Lemma \ref{lem:slope_prox_embedding} and $g^{*}(y)=\prox(y;\mealim_{Y},\mealim_{\tau\Lambda})$.
Finally, the uniqueness of $g^{*}(\cdot)$ (up to a set of measure
0 with respect to $\mealim_{Y}$) is proved in Lemma \ref{lem:gstar_uniqueness}.
This completes our proof.

\subsection{Asymptotic Estimation Performance\label{subsec:Asymptotic-Estimation-Performanc-proof}}

\subsubsection{Convex Gaussian Min-max Theorem}

Our proof hinges on the Convex Gaussian Min-max Theorem (CGMT). For
completeness, we briefly summarize the key idea here. The CGMT studies
a minimax optimization problem (PO) of the form:
\begin{equation}
\Phi(\mG)=\min_{\vv\in\comset_{\vv}}\max_{\vu\in\comset_{\vu}}\vu^{\T}\mG\vv+\psi(\vv,\vu),\label{eq:PO}
\end{equation}
where $\comset_{\vv}\subset\R^{p}$, $\comset_{\vu}\subset\R^{n}$
are two compact sets, $\psi:\comset_{\vv}\times\comset_{\vu}\to\R$
is a continuous convex-concave function w.r.t. $(\vv,\vu)$ and $G_{ij}\iid\calN(0,1)$.
Problem (\ref{eq:PO}) can be associated with the following auxiliary
optimization (AO) problem:
\begin{equation}
\phi(\vg,\vh)=\min_{\vv\in\comset_{\vv}}\max_{\vu\in\comset_{\vu}}\|\vv\|_{2}\vg^{\T}\vu+\|\vu\|_{2}\vh^{\T}\vv+\psi(\vv,\vu),\label{eq:AO}
\end{equation}
where $\vg\sim\calN(\mathbf{0},\mI_{n})$ and $\vh\sim\calN(\mathbf{0},\mI_{p})$.
Roughly speaking, CGMT shows that $\frac{1}{p}\Phi(\mG)\approx\frac{1}{p}\phi(\vg,\vh)$
and the optimal solutions of (\ref{eq:PO}) and (\ref{eq:AO}) have
approximately the same empirical distributions in the large $p$ limit.
Usually, (AO) is easier to analyze, so it provides a convenient handle
for analyzing (PO). For a detailed descriptions, readers can refer
to \cite[Theorem 3]{thrampoulidis2015regularized}.

\subsubsection{Proof of Theorem \ref{thm:asymp_char}\label{subsec:Proof-of-Theorem-asympchar}}

The first step is to recast (\ref{eq:slope_opt}) into the minimax
form as in (\ref{eq:PO}). Letting $\vv=\vx-\sgl$, (\ref{eq:slope_opt})
can be equivalently written as:
\begin{equation}
\begin{aligned} & \min_{\vv}\underbrace{\frac{1}{n}\Big[\frac{1}{2}\|\dmtx\vv-\noise\|^{2}+\regu(\vv+\vbeta)\Big]}_{:=C(\vv)}\\
= & \min_{\vv}\max_{\vu}\frac{1}{n}\left[\frac{\vu^{\T}}{\sqrt{n}}\left(\sqrt{n}\dmtx\right)\vv-\vu^{\T}\noise-\frac{\|\vu\|^{2}}{2}+\regu(\vv+\vbeta)\right].
\end{aligned}
\label{eq:slope_PO}
\end{equation}
Denote $\h{\vv}\bydef\argmin{\vv}C(\vv)$ and correspondingly, $\sol=\h{\vv}+\sgl$.
Now (\ref{eq:slope_PO}) has the same form as (\ref{eq:PO}) and the
corresponding (AO) is:
\begin{equation}
\begin{aligned} & \min_{\vv}\max_{\vu}\frac{1}{n}\left[-\frac{\|\vu\|}{\sqrt{n}}\vh^{\T}\vv-\frac{\|\vv\|}{\sqrt{n}}\vg^{\T}\vu-\vu^{\T}\noise-\frac{\|\vu\|^{2}}{2}+\regu(\vv+\vbeta)\right]\\
= & \min_{\vv}\max_{\theta\geq0}\theta\left(\left\Vert \frac{\|\vv\|}{n}\vg+\frac{\vw}{\sqrt{n}}\right\Vert -\frac{\vh^{\T}\vv}{n}\right)-\frac{1}{2}\theta^{2}+\frac{\regu(\vv+\vbeta)}{n}\\
= & \min_{\vv}\underbrace{\tfrac{1}{2}\Big(\sqrt{\tfrac{\|\vv\|^{2}}{n}\tfrac{\|\vg\|^{2}}{n}+\tfrac{\|\noise\|^{2}}{n}+2\tfrac{\|\vv\|}{\sqrt{n}}\tfrac{\vg^{\T}\noise}{n}}-\tfrac{\vh^{\T}\vv}{n}\Big)_{+}^{2}+\tfrac{\regu(\vv+\vbeta)}{n}}_{:=L(\vv)},
\end{aligned}
\label{eq:slope_AO}
\end{equation}
where $\vg\sim\calN(\mathbf{0},\mI_{n})$ and $\vh\sim\calN(\mathbf{0},\mI_{p})$.
Let $D\subset\R^{p}$ be any closed set. Then by CGMT we can show
for any $t\in\R$, 
\begin{equation}
\P(\min_{\vv\in D}C(\vv)\leq t)\leq2\P(\min_{\vv\in D}L(\vv)\leq t)\label{eq:CGMT_C_L1_leq}
\end{equation}
and if $D$ is also convex,
\begin{equation}
\P(\min_{\vv\in D}C(\vv)\geq t)\leq2\P(\min_{\vv\in D}L(\vv)\geq t).\label{eq:CGMT_C_L2_geq}
\end{equation}
The proof of (\ref{eq:CGMT_C_L1_leq}) and (\ref{eq:CGMT_C_L2_geq})
is the same as \cite[Corollary 5.1]{miolane2018distribution} and
is omitted here. We are going to apply (\ref{eq:CGMT_C_L1_leq}) and
(\ref{eq:CGMT_C_L2_geq}) to prove (\ref{eq:weak_convergence}). We
will follow the proof steps in \cite{miolane2018distribution}.

First define the following minimax problem:

\begin{equation}
\PsiOpt\bydef\min_{\sigma\geq\sigma_{\noisei}}\max_{\theta\geq0}\underbrace{\frac{\theta}{2}\Big(\frac{\sigma_{\noisei}^{2}}{\sigma}+\sigma\Big)-\frac{\theta^{2}}{2}+\frac{1}{\delta}\Big[\Moreau(\sigma,\theta)-\frac{\theta\sigma}{2}\Big]}_{:=\Psi(\sigma,\theta)},\label{eq:Psi_sigma_theta_def}
\end{equation}
where $\Moreau(\sigma,\theta)\bydef\frac{\theta}{2\sigma}\E[Y-\prox(Y)]^{2}+\int_{0}^{1}F_{\Lambda}^{-1}(u)F_{|\eta(Y)|}^{-1}(u)du,$
with $\prox(\cdot)=\eta(\cdot;\mealim_{Y},\mealim_{\sigma\Lambda/\theta})$.
To prove (\ref{eq:weak_convergence}), we adopt the same perturbation
argument as in \cite{thrampoulidis2015regularized,miolane2018distribution,thrampoulidis2018precise}.
In particular, for a pseudo-Lipschiz function $\psi(\cdot,\cdot)$,
define the following set of $\vv$:
\begin{equation}
D_{\nu}:=\big\{\vv\in\R^{p}:|\E_{\meajoint{\vv+\sgl}{\sgl}}\psi-\E_{\mu^{*}}\psi|\ge\nu\big\},\label{eq:Dnu_def_1}
\end{equation}
where $\nu>0$ and $\mu^{*}$ denotes the joint measure of $\big(\eta(\YOpt;\mealim_{\YOpt},\mealim_{\sigOpt\Lambda/\theOpt}),B\big)$.
Here $(\sigOpt,\theOpt)$ is the optimal solution of (\ref{eq:Psi_sigma_theta_def})
and $\YOpt=B+\sigOpt H$, with $H\sim\mathcal{N}(0,1)$ independent
of $B\sim\mealim_{B}$. Recall that $\sol=\h{\vv}+\sgl$, so for any
$\nu>0$ and $\veps>0$ 
\begin{equation}
\begin{aligned}\P\Big(\big|\tfrac{1}{p}\sum_{i=1}^{p}\psi(\soli_{i},\,\sgli_{i})-\E\big[\psi\big(\eta(\YOpt;\mealim_{\YOpt},\mealim_{\sigOpt\Lambda/\theOpt}),B\big)\big]\big|\ge\nu\Big)= & \P\big(\h{\vv}\in D_{\nu}\big)\\
\leq & \P\big(\min_{\vv\in D_{\nu}}C(\vv)\leq\min_{\vv}C(\vv)+\veps\big).
\end{aligned}
\label{eq:pseudo_p_conv_2}
\end{equation}
This indicates that if we can show for any $\nu>0$ and some $\veps>0$,
$\min_{\vv\in D_{\nu}}C(\vv)\leq\min_{\vv}C(\vv)+\veps$ occurs with
vanishing probability, then (\ref{eq:weak_convergence}) will immediately
follow (with $\tauOpt=\sigOpt/\theOpt$). In this way, proving (\ref{eq:weak_convergence})
is reformulated as the perturbation analysis of $C_{\vlambda}(\vv)$,
which can be done as follows. For any $\veps\geq0$ and $K>0$ , we
have 
\begin{align}
 & \P\big(\min_{\vv\in D_{\nu}}C(\vv)\leq\min_{\vv}C(\vv)+\veps\big)\nonumber \\
 & \leq\P\big(\min_{\vv\in D_{\nu}}C(\vv)\leq\PsiOpt+2\veps\big)+\P\big(\min_{\vv}C(\vv)>\PsiOpt+\veps\big)\nonumber \\
 & \tleq{\text{}}\P\big(\min_{\vv\in D_{\nu}\bigcap\bdball_{\sqrt{n}K}}C(\vv)\leq\PsiOpt+2\veps\big)+\P\big(\min_{\vv\in\bdball_{\sqrt{n}K}}C(\vv)>\PsiOpt+\veps\big)+2\P\big(\hat{\vv}\notin\bdball_{\sqrt{n}K}\big)\nonumber \\
 & \tleq{\text{(a)}}2\P\big(\min_{\vv\in D_{\nu}\bigcap\bdball_{\sqrt{n}K}}L(\vv)\leq\PsiOpt+2\veps\big)+2\P\big(\min_{\vv\in\bdball_{\sqrt{n}K}}L(\vv)>\PsiOpt+\veps\big)+2\P\big(\hat{\vv}\notin\bdball_{\sqrt{n}K}\big),\label{eq:Cv_perturbation_3terms}
\end{align}
where (a) is due to (\ref{eq:CGMT_C_L1_leq}) and (\ref{eq:CGMT_C_L2_geq}).
Here $\PsiOpt$ is the optimal value of (\ref{eq:Psi_sigma_theta_def}).
In Appendix \ref{subsec:Auxiliary-Results-for-main-theorem}, we show
all the three probabilities on the RHS of (\ref{eq:Cv_perturbation_3terms})
vanish for $K\geq\tfrac{\sigOpt}{\sqrt{\delta}}+\tfrac{\theta_{\min}}{4}$
(with $\theta_{\min}$ given in Lemma \ref{lem:scalar_minimax}) :
\begin{enumerate}
\item[(i)] From (\ref{eq:min_Lv_concentrate}) of Lemma \ref{lem:L_local_stability},
$\P\big(|\min_{\vv\in\bdball_{\sqrt{n}K}}L(\vv)-\PsiOpt|\geq\veps\big)\to0$
for any $\veps>0$.
\item[(ii)] From (\ref{eq:Cv_opt_bd_1}) of Lemma \ref{lem:C_optsol_bd}, $\P\big(\hat{\vv}\notin\bdball_{\sqrt{n}K}\big)\to0$.
\item[(iii)] From Lemma \ref{lem:pesudo_Lip_inclusion}, for any $\nu>0$ there
exists $\veps_{0}>0$ such that for any $\veps\leq\veps_{0}$, $\P\big(\min_{\vv\in D_{\nu}\bigcap\bdball_{\sqrt{n}K}}L(\vv)\leq\PsiOpt+2\veps\big)\to0$.
\end{enumerate}
After substituting (i)-(iii) back to (\ref{eq:Cv_perturbation_3terms}),
we deduce that for any $\nu>0$, there always exist $\veps_{0}>0$
such that for any $\veps\leq\veps_{0}$, the RHS of (\ref{eq:pseudo_p_conv_2})
converges to 0 as $p\to\infty$. Therefore, 
\[
\frac{1}{p}\sum_{i=1}^{p}\psi(\soli_{i},\,\sgli_{i})\pconv\E\big[\psi\big(\eta(\YOpt;\mealim_{\YOpt},\mealim_{\sigOpt\Lambda/\theOpt}),B\big)\big],
\]
On the other hand, by Lemma \ref{lem:scalar_minimax} in Appendix
\ref{subsec:Properties-of-Limiting-scalar-prob}, $(\sigOpt,\theOpt)$
is the unique solution of the following fixed point equation of $(\sigma,\theta)$:
\begin{align}
\sigma^{2} & =\sigma_{\noisei}^{2}+\frac{1}{\delta}\E[\eta(Y;\mealim_{Y},\mealim_{\sigma\Lambda/\theta})-B]^{2}\label{eq:sigmatau_1_0}\\
\theta & =\sigma\Big[1-\frac{1}{\delta}\E\eta'(Y;\mealim_{Y},\mealim_{\sigma\Lambda/\theta})\Big].\label{eq:sigmatau_2_0}
\end{align}
Therefore, letting $\tauOpt=\sigOpt/\theOpt$, we can see $(\sigOpt,\tauOpt)$
is also a solution of (\ref{eq:sigmatau_1})-(\ref{eq:sigmatau_2}).
Finally, we show such $(\sigOpt,\tauOpt)$ is the unique solution
of (\ref{eq:sigmatau_1})- (\ref{eq:sigmatau_2}). By Lemma \ref{lem:scalar_minimax},
$(\sigOpt,\theOpt)$ is the unique solution of (\ref{eq:sigmatau_1_0})-(\ref{eq:sigmatau_2_0})
and it satisfies $\sigOpt\geq\sigma_{\noisei}$, $\theOpt\geq\theta_{\min}>0$.
Suppose there exist two different solutions $(\sigma_{1},\tau_{1})$
and $(\sigma_{2},\tau_{2})$ to (\ref{eq:sigmatau_1})-(\ref{eq:sigmatau_2}),
then since $\sigma_{1},\tau_{1},\sigma_{2},\tau_{2}>0$, $(\sigma_{1},\sigma_{1}/\tau_{1})$
and $(\sigma_{2},\sigma_{2}/\tau_{2})$ are two different solutions
to (\ref{eq:sigmatau_1_0})-(\ref{eq:sigmatau_2_0}), leading to a
contradiction. This concludes our proof.

\subsection{Asymptotic Variable Selection Performance \label{subsec:Proof-of-Testing}}

In this section, our goal is to prove Proposition \ref{prop:valid_testing_2}.
We first prove the convergence of $R_{0}^{(p)}$.

\subsubsection{Probabilistic upper bound of $R_{0}^{(p)}$}

To prove the convergence of $R_{0}^{(p)}$, the first step is to establish
the following probabilistic upper bound.
\begin{lem}
\label{lem:R0p_prob_ubd}For any $\veps>0$, $R_{0}^{(p)}\leq\P\big(\prox(\YOpt)=0\big)+\veps$,
with probability approaching 1, as $p\to\infty$.
\end{lem}
The proof of Lemma \ref{lem:R0p_prob_ubd} is given in Appendix \ref{subsec:Proof-of-Lemma-prob-ubd},
which uses a standard approximation argument (see \emph{e.g.,} the
proof of Lemma 2.2 (iii) and (v) in \cite{van2000asymptotic}).

\subsubsection{Probabilistic lower bound of $R_{0}^{(p)}$}

The second step is to prove the following matching probabilistic lower
bound for $R_{0}^{(p)}$:
\begin{lem}
\label{lem:R0_matching_lbd}Under the same setting as Proposition
\ref{prop:valid_testing_2}, for any $\veps>0$, 
\begin{equation}
R_{0}^{(p)}\geq\P\big(\prox(\YOpt)=0\big)-\veps,\label{eq:R0p_lower_bd}
\end{equation}
with probability approaching 1, as $p\to\infty$.
\end{lem}
The proof of Lemma \ref{lem:R0_matching_lbd}, which can be found
in Appendix \ref{subsec:Proof-of-Lemma-mathcing_lbd}, is the mostly
technically involved part, so we provide more detailed explanations
here.

{} A key strategy we adopt is using the vector 
\begin{equation}
\hat{\sg}\bydef\dmtx^{\T}(\vy-\dmtx\sol)\label{eq:sg_def_2}
\end{equation}
as an indicator of zero coordinates of $\sol$ . To give the formal
statements, we need to first introduce the notion of \emph{majorization}.
\begin{defn}
\label{def:majorization}For two vectors $\va,\vb\in\R^{p}$, we say
$\va$ is majorized by $\vb$ (denoted as $\va\majorized\vb$), if
for any $j\in[p]$, $\sum_{i=j}^{p}|a|_{(j)}\leq\sum_{i=j}^{p}|b|_{(j)}.$
On the other hand, we say $\va$ is \emph{strictly} majorized by $\vb$
(denoted as $\va\stmajor\vb$), if for any $j\in[p]$, $\sum_{i=j}^{p}|a|_{(j)}<\sum_{i=j}^{p}|b|_{(j)}.$
\end{defn}
Denote $|\vx|_{(1:k)}$ as a vector formed by the largest $k$ components
of $|\vx|$. Let us call $|\vx|_{(1:k)}$ as the \emph{$k$-dominant
subvector} of $\vx$. The following is the key lemma for establishing
the probabilistic lower bound of $R_{0}^{(p)}$.

\begin{lem}
\label{lem:majorize_zero-sufficient}For the optimization problem
(\ref{eq:slope_opt}), suppose for some $k\in[p]$, $|\hat{\sg}|_{(1:k)}\stmajor\vlambda_{1:k}$
, where $\hat{\sg}=\dmtx^{\T}(\vy-\dmtx\sol).$ Then we have $|\sol|_{(1:k)}=\boldsymbol{0}_{k}$.
\end{lem}
The proof of Lemma \ref{lem:majorize_zero-sufficient} can be found
in Appendix \ref{subsec:Proof-of-Lemma-majorize-zero}. %
This characterization transform the original problem of searching
zero coordinates of $\sol$ into a new problem of discovering whether
there is a strict majorization relation between $k$-dominant subvectors
of $\hat{\sg}$ and $\vlambda$. %
{} The nice thing making this strategy work is that the majorization
relation between two vectors is fully captured by their empirical
distributions. Besides, in our setting, the empirical distributions
$\mealim_{\vlambda}$ and $\mealim_{\hat{\sg}}$ both have simple
limits: by our assumption, $\mealim_{\vlambda}\to\mealim_{\Lambda}$
and in Proposition \ref{prop:subgradient_conv} of Appendix \ref{subsec:Asymptotic-Properties-of-sg},
we show $\mealim_{\hat{\sg}}\to\mu_{\hat{S}}$, with $\mealim_{\hat{S}}$
being the law of $\frac{\YOpt-\prox(\YOpt)}{\tauOpt}$.

A major part of proof of Lemma \ref{lem:R0_matching_lbd} is to show
if condition \ref{enu:R_strict_majorization} is satisfied and $\P\big(\prox(\YOpt)=0\big)>0$,
then for $k=\lfloor p[\P\big(\prox(\YOpt)=0\big)-\veps]\rfloor$,
where $\veps>0$ can be arbitrarily small, we have $|\hat{\sg}|_{(1:k)}\stmajor\vlambda_{(1:k)}$
with probability approaching 1, as $p\to\infty$. Then an application
of Lemma \ref{lem:majorize_zero-sufficient} will give us the desired
probabilistic lower bound for $R_{0}^{(p)}$ shown in Lemma \ref{lem:R0_matching_lbd}.
\begin{rem}
\label{rem:R1_explanation}Let us briefly explain why $\sg$ in (\ref{eq:sg_def_2})
is related with the zero coordinates of $\sol$. By the first order
condition of (\ref{eq:slope_opt}), we can get $\hat{\sg}\in\partial J_{\vlambda}(\sol)$,\emph{
i.e.}, $\sg$ defined in (\ref{eq:sg_def_2}) is a subgradient of
$J_{\vlambda}(\vx)$ at $\vx=\sol$. For non-smooth regularizer like
$J_{\vlambda}$, the subgradient $\partial J_{\vlambda}$ at $\vx$
can reveal some information for detecting the zero coordinates of
$\vx$. A simple example is LASSO: $J_{\vlambda}(\vx)=\lambda\|\vx\|_{1}$.
In this case, we have $x_{i}=0$ as long as $[\partial J_{\vlambda}(\vx)]_{i}\in[0,\lambda)$.
This identity is used in \cite{miolane2018distribution} to obtain
the limiting sparsity level of LASSO estimator. Here, we extend this
idea to SLOPE estimator, while a key difference is that unlike the
LASSO case, the zero coordinates are not determined locally: whether
$x_{i}=0$ or not is not completely determined by $[\partial J_{\vlambda}(\vx)]_{i}$.
This is mainly a consequence of non-separability of $J_{\vlambda}(\vx)$.
\end{rem}
\begin{rem}
We can now explain where condition \ref{enu:R_strict_majorization}
originates from. By Lemma \ref{lem:majorize_zero-sufficient}, we
know a sufficient condition for $R_{0}^{(p)}/p\geq\P\big(\prox(\YOpt)=0\big)$
is that $|\hat{\sg}|_{(1:k)}\stmajor\vlambda_{(1:k)}$ holds for some
$k$ satisfying $k/p\approx\P\big(\prox(\YOpt)=0\big):=\probthresh$.
In the asymptotic limit, this translate into the following limiting
form: for any $t\in[0,\probthresh)$,
\begin{equation}
\int_{t}^{\probthresh}F_{|\YOpt-\prox(\YOpt)|/\tauOpt}^{-1}(u)du<\int_{t}^{\probthresh}F_{\Lambda}^{-1}(u)du.\label{eq:strict_majorized_condition_1}
\end{equation}
In fact, (\ref{eq:strict_majorized_condition_1}) is exactly \ref{enu:R_strict_majorization},
since for any $u\in[0,\probthresh]$ we have $F_{|\YOpt-\prox(\YOpt)|/\tauOpt}^{-1}(u)=F_{|\YOpt|/\tauOpt}^{-1}(u)$.
\end{rem}
After combining the probabilistic upper and lower bounds, we conclude
that $R_{0}^{(p)}\pconv\P\big(\prox(\YOpt)=0\big)$.

\subsubsection{Convergence of $V^{(p)}$}

Finally, we prove the convergence of $V^{(p)}$. We have the following
lemma, which shows $V^{(p)}\pconv\P\big(\prox(\YOpt)\neq0,B=0\big)$
can be implied by $R_{0}^{(p)}\pconv\P\big(\prox(\YOpt)=0\big)$.
\begin{lem}
\label{lem:Vp_convergence_implied}For any $\veps>0$, 
\[
\big|V^{(p)}-\P\big[\prox(\YOpt)\neq0,B=0\big]\big|\leq|\P\big(\prox(\YOpt)=0\big)-R_{0}^{(p)}|+\veps,
\]
with probability approaching 1 as $p\to\infty$.
\end{lem}
The proof of Lemma \ref{lem:Vp_convergence_implied} can be found
in Appendix \ref{subsec:Proof-of-Lemma-imply-Vp}. Since the convergence
of $R_{0}^{(p)}$ has been established, we finish our proof.

\subsection{Optimal Estimation\label{subsec:min_MSE_proof}}

In this section, we prove the fundamental estimation performance of
SLOPE, as stated in Proposition \ref{prop:min_MSE}. The proof of
part (a) and (b), which justifies the uniqueness of $\proxfree_{\sigma}$
and the existence of $\sigma_{0}$, can be found in Lemma \ref{lem:f_sigma_unique_Lsigma_conti}
and Lemma \ref{lem:min_sol_infA_lbd} in Appendix \ref{subsec:Auxiliary-Results-for-estimation}.
Here we focus on proving part (c), which is the core part of Proposition
\ref{prop:min_MSE}.

From discussions before, finding the minimum MSE is equivalent to
solving (\ref{eq:inf_sigmastar}). Indeed, we have
\begin{equation}
\inf_{\mealim_{\Lambda}\in\Lamspace}\lim_{p\to\infty}\frac{\|\sol-\sgl\|_{2}^{2}}{p}=\delta(\optsig^{2}-\sigma_{\noisei}^{2}),\label{eq:minMSE_sigopt}
\end{equation}
where $\optsig$ is the optimal value of (\ref{eq:inf_sigmastar}),
\emph{i.e.}, $\optsig\bydef\inf_{\mealim_{\Lambda}\in\Lamspace}\sigOpt$.

\subsubsection{A reformulation of $\inf_{\protect\mealim_{\Lambda}\in\protect\Lamspace}\protect\sigOpt$}

We start by noting that $\optsig$ can be equivalently expressed as:
\begin{equation}
\begin{aligned}\optsig= & \inf\{\sigma\mid(\sigma,\tau)\in\mathcal{D}_{L},\text{ for some }\text{\ensuremath{\tau>0}}\},\end{aligned}
\label{eq:sigopt_1}
\end{equation}
where 
\[
\mathcal{D}_{L}\bydef\big\{(\sigma,\tau)\in\R_{>0}^{2}:\,\exists\mealim_{\Lambda}\in\Lamspace\text{ s.t. }(\sigma,\tau)\text{ satisfies }\eqref{eq:sigmatau_1}\text{-}\eqref{eq:sigmatau_2}\big\}.
\]
Geometrically, computing $\optsig$ is equivalent to searching for
the leftmost point in $\mathcal{D}_{L}$, which is the set of all
realizable $(\sigma,\tau)$ pair. However, characterizing $\mathcal{D}_{L}$
is difficult, since it is determined in a convoluted way via (\ref{eq:sigmatau_1})-(\ref{eq:sigmatau_2}).
To simplify, consider instead the following equation of $(\proxfree,\sigma,\tau)$
\begin{align}
\sigma^{2} & =\sigma_{\noisei}^{2}+\frac{1}{\delta}\E[\proxfree(B+\sigma H)-B]^{2}\label{eq:fixed_point_equation_f_sigmatau_1}\\
1 & =\tau\Big[1-\frac{1}{\delta}\E\proxfree'(B+\sigma H)\Big]\label{eq:fixed_point_equation_f_sigmatau_2}
\end{align}
where $(\proxfree,\sigma,\tau)\in\Lipndset\times\R_{>0}\times\R_{>0}$
and $H\sim\mathcal{N}(0,1)$ is independent of $B\sim\mealim_{B}$.
Let us emphasize that although (\ref{eq:fixed_point_equation_f_sigmatau_1})-(\ref{eq:fixed_point_equation_f_sigmatau_2})
has a similar form as (\ref{eq:sigmatau_1})-(\ref{eq:sigmatau_2}),
a key difference is that unlike $\eta$ in (\ref{eq:sigmatau_1})-(\ref{eq:sigmatau_2}),
$\proxfree$ in (\ref{eq:fixed_point_equation_f_sigmatau_1})-(\ref{eq:fixed_point_equation_f_sigmatau_2})
is not dependent on other parameters such as $(B,H,\sigma,\tau)$.

Now define the following set of $(\sigma,\tau)$:
\[
\mathcal{D}_{F}\bydef\big\{(\sigma,\tau)\in\R_{>0}^{2}:\,\exists f\in\Lipndset\text{ s.t. }(f,\sigma,\tau)\text{ satisfies }\eqref{eq:fixed_point_equation_f_sigmatau_1}\text{-}\eqref{eq:fixed_point_equation_f_sigmatau_2}\big\}.
\]
A key step of our proof is to show $\mathcal{D}_{L}=\mathcal{D}_{F}$.
This can be done as follows. Clearly, we have $\mathcal{D}_{L}\subseteq\mathcal{D}_{F}$,
since $\prox\in\Lipndset$. To prove $\mathcal{D}_{F}\subseteq\mathcal{D}_{L}$,
we need to utilize Proposition \ref{prop:function_space}. Suppose
$(\sigma,\tau)\in\mathcal{D}_{F}$ and let $(f,\sigma,\tau)$ be the
corresponding solution of (\ref{eq:fixed_point_equation_f_sigmatau_1})-(\ref{eq:fixed_point_equation_f_sigmatau_2}).
If $\delta>1$, we have $\Lamspace=\Wspace(\R)$ and by Proposition
\ref{prop:function_space} we can take $\Lambda\sim\frac{|Y|-f(|Y|)}{\tau}\in\Wspace(\R)$
so that $\eta(\cdot;\mealim_{Y},\mealim_{\tau\Lambda})=f$; if $\delta\leq1$,
then from (\ref{eq:fixed_point_equation_f_sigmatau_2}) we know $f(y)\neq y$
and $\P\big(\frac{|Y|-f(|Y|)}{\tau}\neq0\big)>0$, so $\frac{|Y|-f(|Y|)}{\tau}\in\Lamspace$
and we can still take $\Lambda\sim\frac{|Y|-f(|Y|)}{\tau}$ which
gives us $\eta(\cdot;\mealim_{Y},\mealim_{\tau\Lambda})=f$. This
means $(\sigma,\tau)\in\mathcal{D}_{L}$. As a result, we conclude
that $\mathcal{D}_{F}\subseteq\mathcal{D}_{L}$ and thus $\mathcal{D}_{L}=\mathcal{D}_{F}$.
Then substituting $\mathcal{D}_{L}=\mathcal{D}_{F}$ into (\ref{eq:sigopt_1}),
we get the following reformulation of $\optsig$:
\begin{equation}
\begin{aligned}\optsig= & \inf\{\sigma\mid(\sigma,\tau)\in\mathcal{D}_{F},\text{ for some }\text{\ensuremath{\tau>0}}\}.\end{aligned}
\label{eq:sigopt_2}
\end{equation}

\subsubsection{Lower Bound of MSE}

Note that any $(\proxfree,\sigma)\in\Lipndset\times\R_{>0}$ satisfying
(\ref{eq:fixed_point_equation_f_sigmatau_1})-(\ref{eq:fixed_point_equation_f_sigmatau_2})
for some $\tau>0$, should also satisfy 
\begin{equation}
\begin{aligned}\E[\proxfree(B+\sigma H)-B]^{2}= & \delta(\sigma^{2}-\sigma_{\noisei}^{2})\\
\delta^{-1}\E\proxfree'(B+\sigma H)\leq & 1.
\end{aligned}
\label{eq:sigma2_cond1}
\end{equation}
Therefore, if we consider the following set of $\sigma$:
\begin{equation}
\mathcal{A}\bydef\big\{\sigma>0:\,\exists\proxfree\in\Lipndset,\text{s.t. \ensuremath{(f,\sigma)}\text{ satisfies }\eqref{eq:sigma2_cond1}}\big\},\label{eq:sigmaset_def_1}
\end{equation}
then from (\ref{eq:sigopt_2}) we have 
\begin{equation}
\optsig\geq\inf\mathcal{A}.\label{eq:sig_opt_lbd_1}
\end{equation}
Compared with $\optsig$, the lower bound $\inf\mathcal{A}$ in (\ref{eq:sig_opt_lbd_1})
is easier to obtain, since the variable $\tau$ is dropped. In Lemma
\ref{lem:min_sol_infA_lbd} in Appendix \ref{subsec:Auxiliary-Results-for-estimation},
we show that $\inf\mathcal{A}=\sigma_{0}$. Therefore, $\optsig\geq\sigma_{0}$.
Together with (\ref{eq:minMSE_sigopt}), we prove (\ref{eq:MSElbd}).

\subsubsection{Reaching the Lower Bound}

We now show lower bound $\optsig\geq\sigma_{0}$ is tight, if $\delta^{-1}\E\big[\proxfree_{\sigma_{0}}'(B+\sigma_{0}H)\big]<1$.
Recall that $\proxfree_{\sigma_{0}}$ is the unique optimal solution
of (\ref{eq:opt_esti_prox}) when $\sigma=\sigma_{0}$ and $(\proxfree_{\sigma_{0}},\sigma_{0})$
satisfies (\ref{eq:sigma2_cond1}). Let $\tau_{0}=\big[1-\delta^{-1}\E\proxfree_{\sigma_{0}}'(B+\sigma_{0}H)\big]^{-1}$.
It is not hard to see when $\delta^{-1}\E\big[\proxfree_{\sigma_{0}}'(B+\sigma_{0}H)\big]<1$,
$\tau_{0}\in(0,\infty)$ and $(\proxfree_{\sigma_{0}},\sigma_{0},\tau_{0})$
is a solution of (\ref{eq:fixed_point_equation_f_sigmatau_1})-(\ref{eq:fixed_point_equation_f_sigmatau_2}).
This indicates $(\sigma_{0},\tau_{0})\in\mathcal{D}_{F}$ and thus
from (\ref{eq:sigopt_2}) we have $\optsig\leq\sigma_{0}$. Together
with the lower bound $\optsig\geq\sigma_{0}$, we get $\optsig=\sigma_{0}$.

On the other hand, by Proposition \ref{prop:function_space} we know
if $\mealim_{\Lambda}$ is taken as the law of $\frac{1}{\tau_{0}}\big(|\optY|-\proxfree_{\sigma_{0}}(|\optY|)\big),$
then $\proxfree_{\sigma_{0}}(y)=\eta(y;\mealim_{\optY},\mealim_{\tau_{0}\Lambda})$.
Since $(\proxfree_{\sigma_{0}},\sigma_{0},\tau_{0})$ is a solution
of equation (\ref{eq:fixed_point_equation_f_sigmatau_1})-(\ref{eq:fixed_point_equation_f_sigmatau_2}),
we know $(\sigOpt,\tauOpt)=(\sigma_{0},\tau_{0})$, where $(\sigOpt,\tauOpt)$
is the solution of fixed-point equation (\ref{eq:sigmatau_1})-(\ref{eq:sigmatau_2})
under this choice of $\mealim_{\Lambda}$. According to (\ref{eq:limit_MSE}),
we get $\lim_{p\to\infty}\frac{\|\sol-\sgl\|_{2}^{2}}{p}=\delta(\sigma_{0}^{2}-\sigma_{\noisei}^{2})$.
This completes our proof.

\subsection{Optimal Variable Selection\label{subsec:max_power_proof}}

In this section, we are going to prove Proposition \ref{prop:max_power}.
First, part (a) and (b) can be proved in an analogous way as in Proposition
\ref{prop:max_power}, which is summarized in Lemma \ref{lem:f_sigma_unique_L_alpha_sigma_conti}.
Here we focus on part (c).

\subsubsection{Upper bound of $\protect\limpower(\alpha)$}

Directly solving the original optimization (\ref{eq:slope_test_opt})
is not easy. Instead, replacing by a new objective function in (\ref{eq:slope_test_opt}),
we will first consider the following problem:
\begin{equation}
\begin{aligned}\overline{\mathcal{\limpower}}(\alpha)\bydef & \sup_{\mealim_{\Lambda}\in\tLamspace}\P\Big(|B+\sigOpt H|\geq\Phi^{-1}(1-\tfrac{\alpha}{2})\sigOpt\mid B\neq0\Big)\\
\st & \P(|\sigOpt H|\geq\ythresh)\leq\alpha
\end{aligned}
\label{eq:slope_test_opt_ubd}
\end{equation}
where $\ythresh=\sup_{y\geq0}\{y\mid\prox(y;\mealim_{\YOpt},\mealim_{\tauOpt\Lambda})=0\}$.
It is not hard to show for any $\alpha\in[0,1]$, we have $\overline{\mathcal{\limpower}}(\alpha)\geq\limpower(\alpha)$.
This is because the constraint $\P(|\sigOpt H|\geq\ythresh)\leq\alpha$
in (\ref{eq:slope_test_opt}) implies $\ythresh\geq\Phi^{-1}(1-\frac{\alpha}{2})\sigOpt$
and hence the objective function of (\ref{eq:slope_test_opt}) is
upper bounded by that of (\ref{eq:slope_test_opt_ubd}) for any $\mealim_{\Lambda}\in\tLamspace$.

Problem (\ref{eq:slope_test_opt_ubd}) can be further simplified.
By direct differentiation, one can check for any fixed $b\neq0$ and
$c\geq0$, the function $\sigma\mapsto\P(|b+\sigma H|\geq c\sigma)$
is non-increasing on $\R_{\geq0}$, where $H\sim\mathcal{N}(0,1)$.
This then implies, by conditioning on $B$, that $\sigma\mapsto\P(|B+\sigma H|\geq c\sigma\mid B\neq0)$
is non-increasing on $\R_{\geq0}$ for any distribution of $B$ satisfying
$\P(B\neq0)>0$.%
{} Therefore, solving maximization problem in (\ref{eq:slope_test_opt_ubd})
is equivalent to solving the minimization problem of $\sigOpt$. Meanwhile,
by the definition of $\ythresh$ and the fact that $\prox(y;\mealim_{\YOpt},\mealim_{\tauOpt\Lambda})\in\Lipndset$,
we know: $\P(|\sigOpt H|\geq\ythresh)\leq\alpha$ if and only if $\prox(y;\mealim_{\YOpt},\mealim_{\tauOpt\Lambda})=0$,
for all $|y|\leq\Phi^{-1}(1-\frac{\alpha}{2})\sigOpt$. As a result,
solving (\ref{eq:slope_test_opt_ubd}) is equivalent to solving:
\begin{equation}
\begin{aligned}\toptsig\bydef & \inf_{\mealim_{\Lambda}\in\tLamspace}\sigOpt\\
\st & \prox(y;\mealim_{\YOpt},\mealim_{\tauOpt\Lambda})=0,\forall|y|\leq\Phi^{-1}(1-\tfrac{\alpha}{2})\sigOpt
\end{aligned}
\label{eq:slope_test_opt_ubd1}
\end{equation}
and $\overline{\mathcal{\limpower}}(\alpha)$ in (\ref{eq:slope_test_opt_ubd})
can be expressed in terms of $\toptsig$ in (\ref{eq:slope_test_opt_ubd1})
as:
\begin{equation}
\overline{\mathcal{\limpower}}(\alpha)=\P\Big(|B+\toptsig H|\geq\Phi^{-1}(1-\tfrac{\alpha}{2})\toptsig\mid B\neq0\Big).\label{eq:power_ubd_express}
\end{equation}

So far, we arrive at the optimization problem (\ref{eq:slope_test_opt_ubd1}),
which is similar to the one that we have analyzed in the estimation
setting {[}c.f. (\ref{eq:inf_sigmastar}){]}. Yet there are two differences:
(i) a constraint on $\prox$ is added to ensure type-I error is bounded
by $\alpha$, (ii) a constraint on $\mealim_{\Lambda}$ is added to
guarantee valid limits of Type-I error and power exist (see Proposition
\ref{prop:valid_testing_2}). It turns out that the strategy we used
can still be applied. The results are parallel to Proposition \ref{prop:min_MSE}
part (c) and are summarized in Lemma \ref{lem:testing_tightness_cond_1}
in Appendix \ref{subsec:Auxiliary-Results-for-testing}, where it
is shown that 
\begin{equation}
\toptsig\geq\tminsigsol\label{eq:sigopt_lbd}
\end{equation}
and the lower bound can be achieved when $\mealim_{\Lambda}=\mealim_{\text{opt},\alpha}$,
if $\delta^{-1}\E\big[\proxfree_{\alpha}'(\toptY)\big]<1$. After
combining (\ref{eq:sigopt_lbd}) with

\begin{equation}
\limpower(\alpha)\leq\overline{\mathcal{\limpower}}(\alpha)=\P\Big(|B+\toptsig H|\geq\Phi^{-1}(1-\tfrac{\alpha}{2})\toptsig\mid B\neq0\Big),\label{eq:power_ubd2}
\end{equation}
and (\ref{eq:slope_test_opt_0}), we get (\ref{eq:powerubd}).

\subsubsection{Reaching the upper bound}

Now we show for any $\alpha\in[0,1]$, the upper bound (\ref{eq:powerubd})
is tight, if $\delta^{-1}\E\big[\proxfree_{\alpha}'(\toptY)\big]<1$
and $\tythreshzero=\Phi^{-1}(1-\frac{\alpha}{2})\tminsigsol$. Also
it is attained by $\mealim_{\Lambda}=\mealim_{\text{opt},\alpha}$.
The case of $\alpha=0$ is easy. Indeed, in this case, both sides
of (\ref{eq:powerubd}) equal to 0. We just need to verify the case
of $\alpha\in(0,1]$. By Lemma \ref{lem:testing_tightness_cond_1}
and (\ref{eq:power_ubd2}), we know it suffices to show $\limpower(\alpha)=\overline{\mathcal{\limpower}}(\alpha)$
and also $\lim_{p\to\infty}\text{Power}=\limpower(\alpha)$, when
$\mealim_{\Lambda}=\mealim_{\text{opt},\alpha}$. We verify this in
Lemma \ref{lem:testing_tightness_cond_2} in Appendix \ref{subsec:Auxiliary-Results-for-testing},
which completes our proof.

\section{Concluding Remarks \label{sec:Conclusions}}
We have established the asymptotic characterization of SLOPE in the high-dimensional
regime. Although SLOPE is a high-dimensional regularized regression method,
asymptotically its statistical performance can be fully characterized by a few scalar
random variables.
The precise characterization enabled us to derive the fundamental performance limits of SLOPE for both estimation and variable selection settings. Also we showed how to design the optimal regularizing sequences that achieve these limits.

Finally, let us point out some generalizations of current results that worth exploring in the future.
\begin{enumerate}
  \item
  One major technical assumption in the current paper is that the sensing matrix is generated from i.i.d. Gaussian. There are two possible ways to relax this assumption. The first one is to consider the Gaussian design with correlated columns, which is the setting analyzed in \cite{figueiredo2016ordered}. Under this scenario, SLOPE enjoys the nice properties of selecting all the variables associated with highly correlated columns. It would be interesting to derive a precise explanation for this phenomenon. The second direction is staying in the i.i.d. setting, while generalizing to other ensembles, \emph{e.g.}, sub-Gaussian distribution. This is to verify the so-called \emph{universality} phenomenon and some works have been done in the setting where the regularizer is separable \cite{montanari2017universality,panahi2017universal}. It would be interesting to generalize these results to non-separable regularizers such as SLOPE.
  \item
  The optimal designs of $\vlambda$ sequences considered in this paper are based on the assumption that the true distribution of unknown signal is known. The natural question is: can we design $\vlambda$ sequences without (or just with partial) such prior knowledge? One related problem is designing a regularizing sequence such that the false discovery rate is always controlled under a given level. In this setting, the realistic assumption is that we do not know the sparsity of underlying signal. For this purpose, a design of $\vlambda$ is proposed in \cite{bogdan2013statistical} based on some qualitative insights. It would be nice to have quantitative results utilizing the exact characterizations derived here.
  \item
  From numerical simulations, we can find that in several cases, the performance of practical $\vlambda$ sequences such as LASSO and BHq is comparable to the optimal performance. Is it possible that the optimal performance of SLOPE can actually be approximately achieved, when we are restricted to certain sub-classes of regularizing sequences? A key step is establishing some easy-to-evaluate bounds for the performance gap between practical and optimal sequences. One benefit of using practical sequences is that we can apply some purely data-dependent methods such as cross-validation to search for the optimal tuning parameter. Note that since general $\vlambda$ sequence includes order $\mathcal{O}(p)$ parameters, the grid search approach that is usually used in data-dependent method is not plausible here.
\end{enumerate}


\appendix

\subsection{Proof of Lemma \ref{lem:embedding}\label{subsec:Proof-of-Lemma-embedding}}

First assume $0\leq y_{1}\leq\cdots\leq y_{p}$. Denote $\hat{\vy}:=\tprox_{\vlambda}(\vy)$.
Then consider the linear interpolation of the points $\{(y_{i},\hat{y}_{i})\}_{i=0}^{p}$,
where $(y_{0},\hat{y}_{0})=(0,0)$:
\begin{equation}
g_{p}^{+}(y)=\begin{cases}
y_{i-1}+\frac{\hat{y}_{i}-\hat{y}_{i-1}}{y_{i}-y_{i-1}}(y-y_{i-1}) & y\in(y_{i-1},y_{i}),\\
\hat{y}_{i} & y=y_{i},\\
\hat{y}_{p}+(y-y_{p}) & y>y_{p}.
\end{cases}\label{eq:gpplus_y}
\end{equation}
By Fact \ref{fact:SLOPE_prox_properties} (iii), we know $g_{p}^{+}(y)$
is non-decreasing and 1-Lipschitz continuous on $\R_{\geq0}$.

For general $\vy$, we first obtain the linear interpolation $g_{p}^{+}(y)$
of the points $\{(|y|_{(i)},|\hat{y}|_{(i)})\}_{i=0}^{p}$ as above.
Then $g_{p}(y)$ can be constructed as follows:
\[
g_{p}(y)=\begin{cases}
g_{p}^{+}(y) & y\geq0,\\
-g_{p}^{+}(-y) & y<0.
\end{cases}
\]
Clearly, such $g_{p}(y)$ is an odd, non-decreasing and 1-Lipschitz
function. Also by Fact \ref{fact:SLOPE_prox_properties} (i) and (ii),
one can easily check $g_{p}(y_{i})=\hat{y}_{i}$, for all $i\in[p]$.
This finishes the proof.

\subsection{Proof of Lemma \ref{lem:slope_prox_embedding}\label{subsec:Proof-of-Lemma_slope_embedding}}

For notational simplicity, denote $\mathcal{M}_{\vlambda}(\vy):=\mathcal{M}_{\vlambda}(\vy;1)$.
Let $g_{p}^{*}(\vy)$ be any minimizer of (\ref{eq:finite_SLOPE}).
Since $\mathcal{M}_{\vlambda}(\vy)$ is the minimum value of (\ref{eq:slope_prox}),
\begin{align}
\mathcal{M}_{\vlambda}(\vy) & \leq\frac{1}{2}\|\vy-g_{p}^{*}(\vy)\|_{2}^{2}+\sum_{i=1}^{p}\lambda_{i}|g_{p}^{*}(\vy)|_{(i)}\nonumber \\
 & =p\mathcal{M}_{\vlambda}^{*}(\vy).\label{eq:Moreau_ineq_1}
\end{align}
Next, we show $p\mathcal{M}_{\vlambda}^{*}(\vy)\leq\mathcal{M}_{\vlambda}(\vy)$.
Given Lemma \ref{lem:embedding}, this is immediate. Indeed,
\begin{align}
\mathcal{M}_{\vlambda}^{*}(\vy) & \leq L_{p}(g_{p})\nonumber \\
 & =\frac{1}{2}\|\vy-\tprox_{\vlambda}(\vy)\|_{2}^{2}+\sum_{i=1}^{p}\lambda_{i}|\tprox_{\vlambda}(\vy)|_{(i)}\nonumber \\
 & =\frac{\mathcal{M}_{\vlambda}(\vy)}{p},\label{eq:Moreau_ineq_2}
\end{align}
where $g_{p}$ is the function we construct in Lemma \ref{lem:embedding},
which satisfies $g_{p}\in\Lipndset$ and $g_{p}(\vy)=\tprox_{\vlambda}(\vy)$.
Combining (\ref{eq:Moreau_ineq_1}) and (\ref{eq:Moreau_ineq_2}),
we get $\mathcal{M}_{\vlambda}^{*}(\vy)=\frac{\mathcal{M}_{\vlambda}(\vy)}{p}$.

Substituting $\mathcal{M}_{\vlambda}^{*}(\vy)=\frac{\mathcal{M}_{\vlambda}(\vy)}{p}$
into (\ref{eq:Moreau_ineq_1}), we have for any minimizer $g_{p}^{*}$
of (\ref{eq:finite_SLOPE}), $g_{p}^{*}(\vy)$ is also a minimizer
of (\ref{eq:slope_prox}). Since $\tprox_{\vlambda}(\vy)$ is the
unique minimizer of (\ref{eq:slope_prox}), we know any minimizer
of (\ref{eq:finite_SLOPE}) should satisfy: $g_{p}^{*}(\vy)=\tprox_{\vlambda}(\vy)$.

\subsection{Auxiliary Results for Proving Proposition \ref{prop:prox}\label{subsec:Other-Auxiliary-Results-prop-prox}}

\begin{lem}
\label{lem:lg_unif_convergence}Suppose $\{\vy^{(p)}\}_{p\in\mathbb{Z}^{+}}$
and $\{\vlambda^{(p)}\}_{p\in\mathbb{Z}^{+}}$ are converging sequences
with limiting measure $\mu_{Y}$ and $\mu_{\Lambda}$. Then
\begin{equation}
\sup_{g\in\Lipndset}|L(g)-L_{p}(g)|\to0,\label{eq:Lg_unif_convergence}
\end{equation}
where $L(g)$ and $L_{p}(g)$ are defined in (\ref{eq:Lg_def}) and
(\ref{eq:finite_SLOPE}).
\end{lem}
\begin{IEEEproof}
The first step is to establish the following uniform convergence of
a class of Pseudo-Lipschitz functions. Let $\Psi$ be the set of all
functions $\psi:\R\to\R$ satisfying: $\psi(0)=0$ and $|\psi(x)-\psi(y)|\leq(1+|x|+|y|)|x-y|$
for any $x,y\in\R$. Then
\begin{equation}
\sup_{\psi\in\Psi}|\E_{\mu_{\vy}}\psi(Y)-\E_{\mu_{Y}}\psi(Y)|\to0.\label{eq:pseudo_testfn_unif_conv}
\end{equation}
To prove (\ref{eq:pseudo_testfn_unif_conv}), first for any $\psi\in\Psi$
consider the truncation:
\[
\hat{\psi}_{A}(y)=\begin{cases}
\psi(-A) & y<-A,\\
\psi(y) & |y|\leq A,\\
\psi(A) & y>A,
\end{cases}
\]
where $A>0$ is a constant. It is easy to check $\hat{\psi}_{A}(y)$
is $(1+2A)$-Lipschitz continuous, so
\[
\big|\E_{\mu_{\vy}}\hat{\psi}_{A}(Y)-\E_{\mu_{Y}}\hat{\psi}_{A}(Y)\big|\tleq{\text{(a)}}(1+2A)W_{1}(\mu_{\vy},\mu_{Y})\tleq{\text{(b)}}(1+2A)W_{2}(\mu_{\vy},\mu_{Y}),
\]
where (a) follows from Kantonovich duality theorem (\cite[Theorem 1.3]{villani2003topics})
and (b) follows from Holder's inequality. Therefore,
\begin{equation}
\begin{aligned}\big|\E_{\mu_{\vy}}\psi(Y)-\E_{\mu_{Y}}\psi(Y)\big| & \leq\big|\E_{\mu_{\vy}}\hat{\psi}_{A}(Y)-\E_{\mu_{Y}}\hat{\psi}_{A}(Y)\big|\\
 & \;+\big|\E_{\mu_{\vy}}\hat{\psi}_{A}(Y)-\E_{\mu_{\vy}}\psi(Y)\big|+\big|\E_{\mu_{Y}}\hat{\psi}_{A}(Y)-\E_{\mu_{Y}}\psi(Y)\big|\\
 & \tleq{}(1+2A)W_{2}(\mu_{\vy},\mu_{Y})\\
 & \;+2\E_{\mu_{\vy}}[\indicatorfn_{|Y|\geq A}(Y^{2}+|Y|)]+2\E_{\mu_{Y}}[\indicatorfn_{|Y|\geq A}(Y^{2}+|Y|)].
\end{aligned}
\label{eq:test_fn_bd_1}
\end{equation}
For any $\veps>0$, each term on the RHS of (\ref{eq:test_fn_bd_1})
can be bounded as follows. Since $(\E_{\mu_{Y}}|Y|)^{2}\leq\E_{\mu_{Y}}Y^{2}<\infty$,
by DCT there always exists $A>0$ such that $\E_{\mu_{Y}}[\indicatorfn_{|Y|\geq A}(Y^{2}+|Y|)]\leq\frac{\veps}{4}$.
On the other hand, for any given $A>0$ and $\veps>0$, there always
exists $p_{0}\in\mathbb{N}$ such that for any $p\geq p_{0}$, (i)
$W_{2}(\mu_{\vy},\mu_{Y})\leq\frac{\veps}{4(1+2A)}$, since $W_{2}(\mu_{\vy},\mu_{Y})\to0$
and (ii) $\E_{\mu_{\vy}}[\indicatorfn_{|Y|\geq A}(Y^{2}+|Y|)]\leq\E_{\mu_{Y}}[\indicatorfn_{|Y|\geq A}(Y^{2}+|Y|)]+\frac{\veps}{4}$
by Theorem 7.12 (iv) in \cite{villani2003topics}. Note that the RHS
of (\ref{eq:test_fn_bd_1}) does not depend on $\psi$, so for any
$\veps>0$, there exists $p_{0}\in\mathbb{N}$ such that for any $p\geq p_{0}$,
$\sup_{\psi\in\Psi}\big|\E_{\mu_{\vy}}\psi(Y)-\E_{\mu_{Y}}\psi(Y)\big|\leq\veps$.
Therefore, (\ref{eq:pseudo_testfn_unif_conv}) is proved.

We are now ready to show (\ref{eq:Lg_unif_convergence}). Recalling
the definitions of $L(g)$ and $L_{p}(g)$ in (\ref{eq:Lg_def}) and
(\ref{eq:finite_SLOPE}), we have
\begin{equation}
\begin{aligned}\sup_{g\in\Lipndset}|L(g)-L_{p}(g)| & \leq\frac{1}{2}\sup_{g\in\Lipndset}\big|\E_{\mu_{\vy}}[Y-g(Y)]^{2}-\E_{\mu_{Y}}[Y-g(Y)]^{2}\big|\\
 & \;+\sup_{g\in\Lipndset}\int_{0}^{1}|F_{\vlambda}^{-1}(u)-F_{\Lambda}^{-1}(u)|F_{|g(Y)|}^{-1}(u)du\\
 & \;+\sup_{g\in\Lipndset}\int_{0}^{1}F_{\vlambda}^{-1}(u)\big|F_{|g(Y)|}^{-1}(u)-F_{|g(\vy)|}^{-1}(u)\big|du.
\end{aligned}
\label{eq:Lg_unif_convergence_1}
\end{equation}
Therefore, it remains to control each term on the RHS of (\ref{eq:Lg_unif_convergence_1}).
The first term can be handled by using (\ref{eq:pseudo_testfn_unif_conv}),
since $y\mapsto[y-g(y)]^{2}$ belongs to $\Psi$ for $g\in\Lipndset$;
the second term can be controlled as : $\text{Term II}\leq W_{2}(\mu_{\vlambda},\mu_{\Lambda})\sqrt{\E_{\mealim_{Y}}Y^{2}}$
using (\ref{eq:Wasserstein_R}) and Cauchy-Swartz inequality; similarly
for the third term, we have
\[
\begin{aligned}\text{Term III}\leq & \sup_{g\in\Lipndset}\sqrt{\E_{\mu_{\vlambda}}\Lambda^{2}}W_{2}\big(\mu_{|g(\vy)|},\mu_{|g(Y)|}\big)\\
\leq & \sqrt{\E_{\mu_{\vlambda}}\Lambda^{2}}W_{2}\big(\mu_{\vy},\mu_{Y}\big),
\end{aligned}
\]
where the last inequality follows from the definition of Wasserstein
distance and the Lipschitz continuity of $g$:
\[
\begin{aligned}W_{2}\big(\mu_{|g(\vy)|},\mu_{|g(Y)|}\big)^{2}= & \inf_{\pi\in\Pi(\mu_{|g(\vy)|},\mu_{|g(Y)|})}\int(g-h)^{2}d\pi(g,h)\\
= & \inf_{\pi\in\Pi(\mu_{\vy},\mu_{Y})}\int[|g(x)|-|g(y)|]^{2}d\pi(x,y)\\
\leq & \inf_{\pi\in\Pi(\mu_{\vy},\mu_{Y})}\int(x-y)^{2}d\pi(x,y)\\
= & W_{2}\big(\mu_{\vy},\mu_{Y}\big)^{2}.
\end{aligned}
\]
Substituting the above bounds back to (\ref{eq:Lg_unif_convergence_1})
and using the assumption that $W_{2}(\mu_{\vlambda},\mu_{\Lambda}),W_{2}\big(\mu_{\vy},\mu_{Y}\big)\to0$,
we obtain the desired results.
\end{IEEEproof}
\begin{lem}
\label{lem:gstar_uniqueness}The optimization problem (\ref{eq:infinite_SLOPE})
has an optimal solution and it is unique (up to a set of measure 0
with respect to $\mealim_{Y}$).
\end{lem}
\begin{IEEEproof}
Without loss of generality, we assume $\tau=1$. The objective function
$L(g)$ of (\ref{eq:infinite_SLOPE}) is defined on the following
$L^{2}$ space:
\begin{equation}
\Lyspace_{\mu_{Y}}\bydef\{g(y)\mid g(y)\text{ is measurable and }\|g\|_{\mealim_{Y}}<\infty\}\label{eq:L2space_mealim_Y}
\end{equation}
where $\|g\|_{\mealim_{Y}}\bydef[\E_{\mu_{Y}}g^{2}(Y)]^{1/2}$. It
is known that in $L^{2}$ space (and more generally in all normed
linear spaces), the convention is to work with \emph{equivalence class}
of functions \cite[p.135-136]{royden2010real}. The equivalence class
of a function $f\in\Lyspace_{\mu_{Y}}$, denoted as $[f]$, is the
collection of all functions $g\in\Lyspace_{\mu_{Y}}$ satisfying $\|g-f\|_{\mealim_{Y}}=0$.
As a notational convention, we will write $[f]$ as $f$, and the
set $\{[f]:f\in\Lipndset\}$ as $\Lipndset$. Also $\|g-f\|_{\mealim_{Y}}=0$
will be denoted as $g=f$.

We first show $L(g)$ is 1-strongly convex on $\Lyspace_{\mu_{Y}}$,
\emph{i.e.}, for any $g_{1},g_{2}\in\Lyspace_{\mu_{Y}}$,
\begin{equation}
L(\theta g_{1}+(1-\theta)g_{2})\leq\theta L(g_{1})+(1-\theta)L(g_{2})-\frac{\theta(1-\theta)}{2}\|g_{2}(Y)-g_{1}(Y)\|^{2}.\label{eq:Lg_strong_convexity}
\end{equation}
First, for any $\theta\in[0,1]$,
\begin{align*}
\int_{0}^{1}F_{\Lambda}^{-1}(u)F_{|\theta g_{1}(Y)+(1-\theta)g_{2}(Y)|}^{-1}(u)du & \leq\int_{0}^{1}F_{\Lambda}^{-1}(u)F_{\theta|g_{1}(Y)|+(1-\theta)|g_{2}(Y)|}^{-1}(u)du\\
 & =\theta\int_{0}^{1}F_{\Lambda}^{-1}(u)F_{|g_{1}(Y)|}^{-1}(u)du+(1-\theta)\int_{0}^{1}F_{\Lambda}^{-1}(u)F_{|g_{2}(Y)|}^{-1}(u)du,
\end{align*}
which implies that $L_{1}(g):=\int_{0}^{1}F_{\Lambda}^{-1}(u)F_{|g(Y)|}^{-1}(u)du$
is convex. Also, it is not hard to check $L_{2}(g):=\frac{1}{2}\E_{\mu}[Y-g(Y)]^{2}$
is 1-strongly convex by definition (\ref{eq:Lg_strong_convexity}).
Then the strong convexity of $L(g)$ follows, since $L(g)=L_{1}(g)+L_{2}(g)$.
On the other hand, we can show $L(g)$ is continuous on $\Lyspace_{\mu_{Y}}$.
Indeed,
\begin{align*}
|L(g_{2})-L(g_{1})| & \leq\frac{1}{2}\|g_{2}-g_{1}\|_{\mealim_{Y}}\cdot\|2y-g_{1}-g_{2}\|_{\mealim_{Y}}\\
 & \;+\sqrt{\E\Lambda^{2}}\left\Vert |g_{2}|-|g_{1}|\right\Vert _{\mealim_{Y}}\\
 & \leq\|g_{2}-g_{1}\|_{\mealim_{Y}}\big(2\|y\|_{\mealim_{Y}}+2\|g_{1}\|_{\mealim_{Y}}+\|g_{2}-g_{1}\|_{\mealim_{Y}}+\sqrt{\E\Lambda^{2}}\big).
\end{align*}
Since $\|y\|_{\mealim_{Y}},\|g_{1}\|_{\mealim_{Y}},\|g_{2}\|_{\mealim_{Y}}<\infty$,
we conclude that $L(g)$ is continuous.

Next we are going to show the set $\Lipndset$ is convex, bounded
and closed in $\Lyspace_{\mu_{Y}}$. The convexity can be directly
checked by definition. Choose any $g_{1},g_{2}\in\Lipndset$. Then
there exists $S\subseteq\R$ with $\mealim_{Y}(S)=1$ such that for
any $y_{1},y_{2}\in S$, $y_{1}\leq y_{2}$ and any $y\in S$, we
have $0\leq g_{i}(y_{2})-g_{i}(y_{1})\leq y_{2}-y_{1}$ and $g_{i}(y)=-g_{i}(-y)$,
where $i=1,2$. Then for any $\theta\in[0,1]$, function $\theta g_{1}+(1-\theta)g_{2}$
also satisfy (i) and (ii) on $S$, so $\theta g_{1}+(1-\theta)g_{2}\in\Lipndset$.
The boundedness directly follows from the fact that for any $g\in\Lipndset$,
$g(y)\leq|y|$ on some $S\subseteq\R$ with $\mealim_{Y}(S)=1$ .
To show closedness, suppose $g_{k}(y)\in\Lipndset,k=1,2,\ldots$ is
a sequence of functions that converge to some $g(y)\in\Lyspace_{\mealim_{Y}}$.
Then by Riesz-Fischer Theorem, there exists a sub-sequence of $\{g_{k}(y)\}_{k\in\mathbb{Z}^{+}}$
that converges point-wise to $g(y)$ on some $S\subseteq\R$ with
$\mealim_{Y}(S)=1$. By this $\mealim_{Y}$-almost everywhere convergence
of $g_{k}(y)$ to $g(y)$, we know there exists some $S'\subseteq\R$
with $\mealim_{Y}(S')=1$, such that for any $y_{1},y_{2}\in S'$,$y_{1}\leq y_{2}$
and any $y\in S'$, it holds that $0\leq g(y_{2})-g(y_{1})\leq y_{2}-y_{1}$
and $g(y)=-g(-y)$. Therefore, $g(y)\in\Lipndset$ and thus $\Lipndset$
is closed.

The final step is to apply Theorem 17 in \cite[Chap. 8]{royden2010real}
to conclude that (\ref{eq:infinite_SLOPE}) has an optimal solution
$g^{*}\in\Lipndset$. Also the uniqueness of $g^{*}$ can be easily
checked by the strong convexity of $L(g)$. Suppose there exists two
different optimal solutions, $g_{1}^{*},g_{2}^{*}$ with $L(g_{1}^{*})=L(g_{2}^{*})$
and $g_{1}^{*}\neq g_{2}^{*}$. Then by (\ref{eq:Lg_strong_convexity}),
for $g=\frac{g_{1}+g_{2}}{2}\in\Lipndset$ we have $L(g)<L(g_{1}^{*})=L(g_{2}^{*})$,
which leads to a contradiction.
\end{IEEEproof}

The following result provides the explicit formula for calculating
Wasserstein-2 distance between probability measure on $\R$. Readers
can find a proof in Theorem 2.18 in \cite{villani2003topics}.%

\begin{lem}
\label{lem:Wasserstein_R}Suppose $\mu_{1},\mu_{2}\in\Wspace(\R)$
and the corresponding quantile functions are $F_{1}^{-1}$ and $F_{2}^{-1}$.
Then
\begin{equation}
W_{2}(\mu_{1},\mu_{2})^{2}=\int_{0}^{1}(F_{1}^{-1}(t)-F_{2}^{-1}(t))^{2}dt.\label{eq:Wasserstein_R}
\end{equation}
\end{lem}

\subsection{Auxiliary Results for Proving Theorem \ref{thm:asymp_char}\label{subsec:Auxiliary-Results-for-main-theorem}}

In this section, we prove three auxiliary lemmas used in the proof
of Theorem \ref{thm:asymp_char}.

The first two results are on the asymptotic properties of auxiliary
problem (\ref{eq:slope_AO}). To state these asymptotic results, similar
as Theorem \ref{thm:asymp_char}, we will consider a sequence of auxiliary
problems described by the instances $\{\vg^{(p)},\vh^{(p)},\sgl^{(p)},\noise^{(p)},\vlambda^{(p)}\}_{p\in\mathbb{Z}^{+}}$.
They satisfy the following: (i) $\vg^{(p)}\sim\mathcal{\mathcal{N}}(\boldsymbol{0},\mI_{n})$,
$\vh^{(p)}\sim\mathcal{\mathcal{N}}(\boldsymbol{0},\mI_{p})$, $p\in\mathbb{Z}^{+}$
are all independent, (ii)$\{\sgl^{(p)}\}_{p\in\mathbb{Z}^{+}}$, $\{\noise^{(p)}\}_{p\in\mathbb{Z}^{+}}$,
$\{\vlambda^{(p)}\}_{p\in\mathbb{Z}^{+}}$ are the same converging
sequences as in Theorem \ref{thm:asymp_char}. Here the requirement
that $\{\vg^{(p)}\}_{p\in\mathbb{Z}^{+}}$ and $\{\vh^{(p)}\}_{p\in\mathbb{Z}^{+}}$
are independent is not completely necessary, since we are only aiming
for results regarding convergence in probability. The independence
assumption simply allows us to directly apply some results obtained
in Appendix \ref{subsec:Moreau-Envelope-lim}. %

The first lemma is about the minimum value and the minimizer of $L(\vv)$
over a bounded Euclidean ball. Recall that 
\begin{equation}
L(\vv)\bydef\frac{1}{2}\left(\sqrt{\frac{\|\vv\|^{2}}{n}\frac{\|\vg\|^{2}}{n}+\frac{\|\noise\|^{2}}{n}+2\frac{\|\vv\|}{\sqrt{n}}\frac{\vg^{\T}\noise}{n}}-\frac{\vh^{\T}\vv}{n}\right)_{+}^{2}+\frac{\regu(\vv+\vbeta)}{n}.\label{eq:Lv}
\end{equation}

\begin{lem}
\label{lem:L_local_stability}Let $\PsiOpt$ and $(\sigOpt,\theOpt)$
be the optimal value and solution of the minimax problem in (\ref{eq:Psi_sigma_theta_def})
and $\theta_{\min}>0$ be the lower bound of $\theOpt$ obtained in
Lemma \ref{lem:scalar_minimax}. For any $\veps>0$ and $K\geq\tfrac{\sigOpt}{\sqrt{\delta}}+\tfrac{\theta_{\min}}{4}$,
we have
\begin{equation}
\P\Big(\Big|\min_{\vv\in\bdball_{\sqrt{n}K}}L(\vv)-\PsiOpt\Big|\leq\veps\Big)\to1\label{eq:min_Lv_concentrate}
\end{equation}
and
\begin{equation}
\P\Big(\min_{\vv\in\outbdball_{\sqrt{32n\veps/\gamma}}(\approxdif)\cap\bdball_{\sqrt{n}K}}L(\vv)\geq\PsiOpt+\veps\Big)\to1,\label{eq:Lv_perturbation_1}
\end{equation}
where $\gamma=\frac{\theta_{\min}\sigma_{\noisei}^{2}}{4(K^{2}+\sigma_{\noisei}^{2})^{3/2}}$
and 
\begin{equation}
\approxdif:=\eta(\sgl+\sigOpt\vh)-\sgl.\label{eq:v_star_p}
\end{equation}
Here in (\ref{eq:v_star_p}), $\prox(\cdot):=\eta(\sgl+\sigOpt\vh;\mealim_{\YOpt},\mealim_{\sigOpt\Lambda/\theOpt})$
and $\YOpt=B+\sigOpt H$, with $H\sim\mathcal{N}(0,1)$ independent
of $B\sim\mealim_{B}$.
\end{lem}
\begin{IEEEproof}
We follow the proof of Proposition B.2 in \cite{miolane2018distribution}.
First introduce the event $\mathcal{A}=\bigcap_{i=1}^{5}\mathcal{A}_{i}$,
where
\begin{equation}
\begin{aligned}\mathcal{A}_{1} & :=\Big\{\tfrac{\|\vg\|^{2}}{n},\tfrac{\|\vh\|^{2}}{p},\tfrac{\|\noise\|^{2}}{\sigma_{\noisei}^{2}n}\in[1-\varsigma,1+\varsigma],\left|\tfrac{\vg^{\T}\noise}{n}\right|\leq\varsigma\Big\},\\
\mathcal{A}_{2} & :=\{\|\approxdif\|/\sqrt{p}\leq\sigOpt\},\\
\mathcal{A}_{3} & :=\Big\{\sqrt{\tfrac{\|\approxdif\|^{2}}{n}+\sigma_{\noisei}^{2}}-\tfrac{\vh^{\T}\approxdif}{n}\geq\tfrac{\theta_{\min}}{2}\Big\},\\
\mathcal{A}_{4} & :=\{|\widetilde{L}(\approxdif)-\PsiOpt|\leq\veps\},\\
\mathcal{A}_{5} & :=\Big\{\sup_{\sigma\in[\sigma_{\noisei},\sqrt{\sigma_{\noisei}^{2}+K^{2}}]}\big|\big(\Moreau_{p}(\sigma,\theOpt)-\tfrac{\sigma\theOpt\|\vh\|^{2}}{2p}\big)-\big(\Moreau(\sigma,\theOpt)-\tfrac{\sigma\theOpt}{2}\big)\big|\leq\delta\veps\Big\},
\end{aligned}
\label{eq:highprob_event_5}
\end{equation}
with $\veps>0$ and $\varsigma\in(0,\tfrac{1}{2})$. In (\ref{eq:highprob_event_5}),
$\Moreau_{p}$ and $\Moreau$ are the same as in (\ref{eq:Fp_ctau})
and (\ref{eq:F_lim_form}) and $\widetilde{L}(\vv)$ is defined as:
\begin{equation}
\widetilde{L}(\vv)\bydef\frac{1}{2}\left(\sqrt{\frac{\|\vv\|^{2}}{n}+\sigma_{\noisei}^{2}}-\frac{\vh^{\T}\vv}{n}\right)_{+}^{2}+\frac{\regu(\vv+\vbeta)}{n}.\label{eq:tilde_L_v}
\end{equation}
Based on the event $\mathcal{A}$, our subsequent analysis will become
fully deterministic: we will condition on fixed $\vh$ and $\vg$
in $\mathcal{A}$. Before doing so, let us first show each of the
events $\mathcal{A}_{1}\sim\mathcal{A}_{5}$ occurs with probability
approaching 1 as $p\to\infty$, so $\P(\mathcal{A})\to1$. This will
ensure all the results obtained by conditioning on $\mathcal{A}$
 hold with probability approaching 1.

$\mathcal{A}_{1}$: By the law of large number and the fact that $\{\noise\}_{p\in\mathbb{N}}$
is a converging sequence with limiting variance $\sigma_{\noisei}^{2}>0$,
it is not hard to show $\P(\mathcal{A}_{1})\to1$.

$\mathcal{A}_{2}$: From (\ref{eq:fixedequa_lim1_as}), we have
\begin{equation}
\frac{\|\approxdif\|^{2}}{p}\asconv\E[B-\eta(B+\sigOpt H)]^{2}.\label{eq:v_norm2_asconv}
\end{equation}
Then together with (\ref{eq:fixed_point_equation_sigmatheta}), we
get $\frac{\|\approxdif\|^{2}}{p}\asconv(\sigOpt)^{2}-\sigma_{\noisei}^{2}$.
Therefore, $\P(\mathcal{A}_{2})\to1$, since $\sigma_{\noisei}^{2}>0$
by assumption.

$\mathcal{A}_{3}$: From (\ref{eq:fixedequa_lim1_as}) and (\ref{eq:fixedequa_lim2_as}),
we can get 
\begin{equation}
\begin{aligned}\sqrt{\frac{\|\approxdif\|^{2}}{n}+\sigma_{\noisei}^{2}}-\frac{\vh^{\T}\approxdif}{n} & \asconv\sqrt{\frac{1}{\delta}\E[B-\eta(B+\sigOpt H)]^{2}+\sigma_{\noisei}^{2}}-\frac{\sigOpt\E\eta'(B+\sigOpt H)}{\delta}=\theOpt,\end{aligned}
\label{eq:thetastar_asconv}
\end{equation}
where the last step follows from (\ref{eq:fixed_point_equation_sigmatheta}).
From Lemma \ref{lem:scalar_minimax}, there exists $\theta_{\min}>0$
such that $\theOpt\geq\theta_{\min}$. Therefore, $\P(\mathcal{A}_{3})\to1$.

$\mathcal{A}_{4}$: From the definition of $\Moreau(\sigma,\theta)$
in (\ref{eq:F_lim_form}),
\begin{align}
\Moreau(\sigOpt,\theOpt) & =\frac{\theOpt}{2\sigOpt}\E[(\YOpt)-\eta(\YOpt)]^{2}+\int_{0}^{1}F_{\Lambda}^{-1}(u)F_{|\eta(\YOpt)|}^{-1}(u)du.\nonumber \\
 & =\frac{\theOpt}{2\sigOpt}\Big[\E\big(\eta(\YOpt)-B\big)^{2}-2(\sigOpt)^{2}\E\big(\eta'(\YOpt)\big)\Big]+\frac{\theOpt\sigOpt}{2}\nonumber \\
 & \quad+\int_{0}^{1}F_{\Lambda}^{-1}(u)F_{|\eta(\YOpt)|}^{-1}(u)du\nonumber \\
 & =\frac{\theOpt\delta}{2\sigOpt}\big(-(\sigOpt)^{2}-\sigma_{\noisei}^{2}+2\sigOpt\sigma_{\noisei}\big)+\int_{0}^{1}F_{\Lambda}^{-1}(u)F_{|\eta(\YOpt)|}^{-1}(u)du+\frac{\theOpt\sigOpt}{2},\label{eq:F_sigmaopt_thetaopt}
\end{align}
where in the last step we use (\ref{eq:fixed_point_equation_sigmatheta}).
Then substituting (\ref{eq:F_sigmaopt_thetaopt}) into (\ref{eq:Psi_sigma_theta_def}),
we get
\begin{equation}
\PsiOpt=\Psi(\sigOpt,\theOpt)=\frac{(\theOpt)^{2}}{2}+\int_{0}^{1}F_{\Lambda}^{-1}(u)F_{|\eta(\YOpt)|}^{-1}(u)du.\label{eq:Psi_sigmastar_thetastar}
\end{equation}
On the other hand,
\begin{align}
\widetilde{L}(\approxdif) & =\frac{1}{2}\left(\sqrt{\frac{\|\approxdif\|^{2}}{n}+\sigma_{\noisei}^{2}}-\frac{\vh^{\T}\approxdif}{n}\right)_{+}^{2}+\frac{\regu(\approxdif+\vbeta)}{n}\nonumber \\
 & \asconv\frac{(\theOpt)^{2}}{2}+\int_{0}^{1}F_{\Lambda}^{-1}(u)F_{|\eta(\YOpt)|}^{-1}(u)du.\label{eq:Ltilde_vstar_asconv}
\end{align}
where we use (\ref{eq:thetastar_asconv}). From (\ref{eq:Psi_sigmastar_thetastar})
and (\ref{eq:Ltilde_vstar_asconv}), we have $\widetilde{L}(\approxdif)\asconv\PsiOpt$.
Therefore, $\P(\mathcal{A}_{4})\to1$ for any $\veps>0$.

$\mathcal{A}_{5}$: From (\ref{eq:F_asconv}) and strong law of large
number for triangular array \cite[Theorem 2.1]{hu1997strong}, we
have $\Moreau_{p}(\sigma,\theOpt)-\tfrac{\sigma\theOpt\|\vh\|^{2}}{2p}\asconv\Moreau(\sigma,\theOpt)-\tfrac{\sigma\theOpt}{2}$
for any $\sigma\in[\sigma_{\noisei},\sqrt{\sigma_{\noisei}^{2}+K^{2}}]$.
From (\ref{eq:F_deri1}) in Lemma \ref{lem:F_sigma_theta_contdiff},
we know $\Moreau(\cdot,\theOpt)$ is Lipschitz continuous on $[\sigma_{\noisei},\sqrt{\sigma_{\noisei}^{2}+K^{2}}]$.
This indicates $\sigma\mapsto\Moreau(\sigma,\theOpt)-\tfrac{\sigma\theOpt}{2}$
is also Lipschitz continuous on $[\sigma_{\noisei},\sqrt{\sigma_{\noisei}^{2}+K^{2}}]$.
On the other hand, in the proof of Lemma \ref{lem:F_sigma_theta_contdiff}
{[}line below (\ref{eq:par_Fp_sigma}){]}, we show $\Moreau_{p}(\cdot,\theOpt)$
is continuously differentiable on $[\sigma_{\noisei},\infty)$ with
derivative satisfying $\left|\frac{\partial\Moreau_{p}(\sigma,\theOpt)}{\partial\sigma}\right|\leq\frac{3\theOpt\|\sgl\|^{2}+\sigma^{2}(2+\theOpt)\|\vh\|^{2}}{\sigma^{2}p}$.
Then by \cite[Theorem 2.1]{hu1997strong} again and the fact that
$\{\sgl\}_{p\in\mathbb{Z}^{+}}$ is a converging sequence, we have
almost surely $\limsup_{p\to\infty}\left|\frac{\partial\Moreau_{p}(\sigma,\theOpt)}{\partial\sigma}-\tfrac{\theOpt\|\vh\|^{2}}{2p}\right|\leq C$
for any $\sigma\in[\sigma_{\noisei},\sqrt{\sigma_{\noisei}^{2}+K^{2}}]$,
where $C>0$ is some constant. This indicates that almost surely,
$\sigma\mapsto\frac{\partial\Moreau_{p}(\sigma,\theOpt)}{\partial\sigma}-\tfrac{\theOpt\|\vh\|^{2}}{2p}$
is $C$-Lipschitz continuous on $[\sigma_{\noisei},\sqrt{\sigma_{\noisei}^{2}+K^{2}}]$
for any large enough $p$. Then by the same epsilon net argument as
in the proof of Lemma \ref{lem:F_sigma_theta_contdiff}, we can show
\[
\sup_{\sigma\in[\sigma_{\noisei},\sqrt{\sigma_{\noisei}^{2}+K^{2}}]}\Big|\big(\Moreau_{p}(\sigma,\theOpt)-\tfrac{\sigma\theOpt\|\vh\|^{2}}{2p}\big)-\big(\Moreau(\sigma,\theOpt)-\tfrac{\sigma\theOpt}{2}\big)\Big|\asconv0.
\]
Therefore, $\P(\mathcal{A}_{5})\to1$ for any $\veps>0$.

Now we are ready to start the deterministic analysis conditioned on
the event $\mathcal{A}$. It is more convenient to work with $\widetilde{L}(\vv)$
than $L(\vv)$, since it is locally strongly convex (the precise meaning
will be given below). In the sequel, we will start by studying the
limiting properties of $\widetilde{L}(\vv)$ and then associate them
to $L(\vv)$, by showing $L(\vv)$ can be well-approximated by $\widetilde{L}(\vv)$
as $p\to\infty$. For $\vv\in\bdball_{\sqrt{n}K}$, $\widetilde{L}(\vv)$
can be equivalently written as:
\begin{align}
\widetilde{L}(\vv) & =\max_{\theta\geq0}\theta\Big(\sqrt{\frac{\|\vv\|^{2}}{n}+\sigma_{\noisei}^{2}}-\frac{\vh^{\T}\vv}{n}\Big)-\frac{\theta^{2}}{2}+\frac{\regu(\vv+\vbeta)}{n}\nonumber \\
 & =\max_{\theta\geq0}\theta\Big(\min_{\sigma\in[\sigma_{\noisei},\sqrt{\sigma_{\noisei}^{2}+K^{2}}]}\big\{\frac{\sigma}{2}+\frac{\|\vv\|^{2}/n+\sigma_{\noisei}^{2}}{2\sigma}\big\}-\frac{\vh^{\T}\vv}{n}\Big)-\frac{\theta^{2}}{2}+\frac{\regu(\vv+\vbeta)}{n}\nonumber \\
 & =\max_{\theta\geq0}\min_{\sigma\in[\sigma_{\noisei},\sqrt{\sigma_{\noisei}^{2}+K^{2}}]}\frac{\theta}{2}\Big(\frac{\sigma_{\noisei}^{2}}{\sigma}+\sigma\Big)-\frac{\theta^{2}}{2}+\frac{1}{n}\Big[\frac{\theta\|\vv\|^{2}}{2\sigma}-\theta\vh^{\T}\vv+\regu(\vv+\vbeta)\Big].\label{eq:L_tilde_v_equi_1}
\end{align}
Therefore,
\begin{align}
\min_{\vv\in\bdball_{\sqrt{n}K}}\widetilde{L}(\vv) & \tgeq{\text{(a)}}\min_{\sigma\in[\sigma_{\noisei},\sqrt{\sigma_{\noisei}^{2}+K^{2}}]}\frac{\theOpt}{2}\Big(\frac{\sigma_{\noisei}^{2}}{\sigma}+\sigma\Big)-\frac{(\theOpt)^{2}}{2}+\frac{1}{\delta}\Big[\Moreau_{p}(\sigma,\theOpt)-\frac{\sigma\theOpt\|\vh\|^{2}}{2p}\Big],\nonumber \\
 & \tgeq{\text{(b)}}\min_{\sigma\in[\sigma_{\noisei},\sqrt{\sigma_{\noisei}^{2}+K^{2}}]}\Psi(\sigma,\theOpt)-\veps\nonumber \\
 & \geq\PsiOpt-\veps\nonumber \\
 & \tgeq{\text{(c)}}\widetilde{L}(\approxdif)-2\veps,\label{eq:Ltilde_v_min_lbd}
\end{align}
where (a) follows from (\ref{eq:L_tilde_v_equi_1}) and (\ref{eq:Fp_ctau}),
(b) is due to $\mathcal{A}_{5}$ and (c) is due to $\mathcal{A}_{4}$.
Besides, since $\tfrac{\|\approxdif\|_{2}}{\sqrt{n}}\leq\tfrac{\sigOpt}{\sqrt{\delta}}$
under $\mathcal{A}_{2}$ and $\tfrac{\sigOpt}{\sqrt{\delta}}\leq K-\frac{\theta_{\min}}{4}$
by assumption, we have $\tfrac{\|\approxdif\|_{2}}{\sqrt{n}}\leq K$
and thus $\min_{\vv\in\bdball_{\sqrt{n}K}}\widetilde{L}(\vv)\leq\widetilde{L}(\approxdif)\leq\PsiOpt+\veps.$
Therefore, combining it with (\ref{eq:Ltilde_v_min_lbd}) yields
\begin{equation}
\Big|\min_{\vv\in\bdball_{\sqrt{n}K}}\widetilde{L}(\vv)-\PsiOpt\Big|\leq2\veps.\label{eq:Ltilde_min_concentrate}
\end{equation}
On the other hand, $\vv\mapsto\sqrt{\tfrac{\|\vv\|^{2}}{n}+\sigma_{\noisei}^{2}}-\tfrac{\vh^{\T}\vv}{n}$
is $\tfrac{1}{\sqrt{n}}(\sqrt{\tfrac{3}{2\delta}}+1)$-Lipschitz continuous
under $\mathcal{A}_{1}$ and $\sqrt{\tfrac{\|\approxdif\|^{2}}{n}+\sigma_{\noisei}^{2}}-\tfrac{\vh^{\T}\approxdif}{n}\geq\tfrac{\theta_{\min}}{2}$
under $\mathcal{A}_{3}$. Therefore, for $r=(\sqrt{\tfrac{3}{2\delta}}+1)^{-1}\tfrac{\theta_{\min}}{4}$,
$\sqrt{\tfrac{\|\vv\|^{2}}{n}+\sigma_{\noisei}^{2}}-\tfrac{\vh^{\T}\vv}{n}\geq\tfrac{\theta_{\min}}{4}$
for all $\vv\in\bdball_{\sqrt{n}r}(\approxdif)$. Then using Lemma
F.14 in \cite{miolane2018distribution} and $\tfrac{\|\approxdif\|_{2}}{\sqrt{n}}+r\leq\tfrac{\sigOpt}{\sqrt{n}}+r<K$
(hence $\bdball_{\sqrt{n}r}(\approxdif)\subset\bdball_{\sqrt{n}K}$)
under $\mathcal{A}_{2}$, we can show $\widetilde{L}(\vv)$ is $\frac{\gamma}{n}$-strongly
convex on $\bdball_{\sqrt{n}r}(\approxdif)$, where $\gamma=\frac{\theta_{\min}\sigma_{\noisei}^{2}}{4(K^{2}+\sigma_{\noisei}^{2})^{3/2}}$.
In other words, $\widetilde{L}(\vv)$ is locally strongly convex in
$\bdball_{\sqrt{n}r}(\approxdif)$. Also since $\bdball_{\sqrt{n}r}(\approxdif)\subset\bdball_{\sqrt{n}K}$,
together with (\ref{eq:Ltilde_v_min_lbd}) we have $\min_{\vv\in\bdball_{\sqrt{n}r}(\approxdif)}\widetilde{L}(\vv)\geq\min_{\vv\in\bdball_{\sqrt{n}K}}\widetilde{L}(\vv)\geq\widetilde{L}(\approxdif)-2\veps$.
By Lemma B.1 in \cite{miolane2018distribution} we know if $0<2\veps\leq\tfrac{(\sqrt{n}r)^{2}\gamma}{8n}$
, then $\|\tilde{\vv}-\approxdif\|^{2}\leq\frac{4n\veps}{\gamma}\leq\frac{nr^{2}}{4}$,
where $\tilde{\vv}=\argmin{\vv\in\bdball_{\sqrt{n}K}}\widetilde{L}(\vv)$.
Moreover, for any $\vv\in\outbdball_{4\sqrt{n\veps/\gamma}}(\approxdif)$,
we have $\widetilde{L}(\vv)\geq\min_{\vv\in\bdball_{\sqrt{n}K}}\widetilde{L}(\vv)+2\veps$.
This implies
\begin{equation}
\min_{\vv\in\outbdball_{4\sqrt{n\veps/\gamma}}(\approxdif)\cap\bdball_{\sqrt{n}K}}\widetilde{L}(\vv)\geq\min_{\vv\in\bdball_{\sqrt{n}K}}\widetilde{L}(\vv)+2\veps.\label{eq:Ltilde_perturbation}
\end{equation}

Finally, we show $L(\vv)$ is well-approximated by $\widetilde{L}(\vv)$
under event $\mathcal{A}_{1}$. Note that $L(\vv)-\widetilde{L}(\vv)=g(\Delta)$,
where $g(t)=\tfrac{1}{2}[(\sqrt{x+t}-y)_{+}^{2}-(\sqrt{x}-y)_{+}^{2}]$,
with $x=\frac{\|\vv\|^{2}}{n}+\sigma_{\noisei}^{2}$, $y=\frac{\vh^{\T}\vv}{n}$
and $\Delta=\frac{\|\vv\|^{2}}{n}\left(\frac{\|\vg\|^{2}}{n}-1\right)+\left(\frac{\|\noise\|^{2}}{n}-\sigma_{\noisei}^{2}\right)+2\frac{\|\vv\|}{\sqrt{n}}\frac{\vg^{\T}\noise}{n}$.
Also it is not hard to show under event $\mathcal{A}_{1}$, $|g(t)|\leq\frac{|t|}{2}\left(1+\frac{|y|}{\sigma_{\noisei}}\right)$,
$|y|\leq\sqrt{\tfrac{3}{2\delta}}K$ and $|\Delta|\leq(K^{2}+\sigma_{\noisei}^{2}+2K)\varsigma$
for any $\vv\in\comset_{\vv}(K)$. Therefore,
\begin{equation}
\sup_{\vv\in\bdball_{\sqrt{n}K}}|L(\vv)-\widetilde{L}(\vv)|\leq\underbrace{\tfrac{K^{2}+\sigma_{\noisei}^{2}+2K}{2}\left(1+\sqrt{\tfrac{3}{2\delta}}\tfrac{K}{\sigma_{\noisei}}\right)}_{:=C_{K}}\varsigma\label{eq:diff_Lv}
\end{equation}
and thus
\begin{equation}
\Big|\min_{\vv\in\bdball_{\sqrt{n}K}}L(\vv)-\min_{\vv\in\bdball_{\sqrt{n}K}}\widetilde{L}(\vv)\Big|\leq C_{K}\varsigma.\label{eq:diff_opt_LV}
\end{equation}
Now we are ready to turn back to $L(\vv)$ to show (\ref{eq:min_Lv_concentrate})
and (\ref{eq:Lv_perturbation_1}). Substituting (\ref{eq:diff_Lv})
and (\ref{eq:diff_opt_LV}) into (\ref{eq:Ltilde_min_concentrate})
and (\ref{eq:Ltilde_perturbation}) gives
\begin{equation}
\Big|\min_{\vv\in\bdball_{\sqrt{n}K}}L(\vv)-\PsiOpt\Big|\leq C_{K}\varsigma+\veps\label{eq:L_min_concentrate}
\end{equation}
and 
\begin{align}
\min_{\vv\in\outbdball_{4\sqrt{n\veps/\gamma}}(\approxdif)\cap\bdball_{\sqrt{n}K}}L(\vv) & \geq\min_{\vv\in\bdball_{\sqrt{n}K}}L(\vv)+2(\veps-C_{K}\varsigma).\nonumber \\
 & \tgeq{\text{(a)}}\PsiOpt+\veps-3C_{K}\varsigma,\label{eq:L_perturbation}
\end{align}
where in (a) we use (\ref{eq:L_min_concentrate}). For any $\veps>0$,
choose $\varsigma\leq\min\big\{\tfrac{1}{2},\tfrac{\veps}{6C_{K}}\big\}$
in (\ref{eq:L_min_concentrate}) and (\ref{eq:L_perturbation}). Then
(\ref{eq:min_Lv_concentrate}) and (\ref{eq:Lv_perturbation_1}) immediately
follows, since $\P(\mathcal{A})\to1$.
\end{IEEEproof}
The second lemma is on the asymptotic empirical distribution of the
optimal solution of auxiliary problem.
\begin{lem}
\label{lem:pesudo_Lip_inclusion}Let $\psi(\cdot,\cdot)$ be a pseudo-Lipschitz
function with constant $L$ and $D_{\nu}:=\big\{\vv\in\R^{p}:|\E_{\meajoint{\vv+\sgl}{\sgl}}\psi-\E_{\mu^{*}}\psi|\ge\nu\big\}$
as defined in (\ref{eq:Dnu_def_1}). Then for any $K\geq\tfrac{\sigOpt}{\sqrt{\delta}}+\tfrac{\theta_{\min}}{4}$,
$\nu>0$ and $\veps\leq\frac{\gamma\nu^{2}}{192L^{2}(1+2\delta K^{2}+32(\E B^{2}+\sigma_{\noisei}^{2}))\delta}$,
\begin{equation}
\P\big(\min_{\vv\in D_{\nu}\bigcap\bdball_{\sqrt{n}K}}L(\vv)\leq\PsiOpt+2\veps\big)\to0,\label{eq:Lv_perturbation_Dnu_2}
\end{equation}
where $\PsiOpt$ is defined in (\ref{eq:Psi_sigma_theta_def}).
\end{lem}
\begin{IEEEproof}
For any $\nu>0$, we will consider the following event:
\[
\mathcal{E}=\big\{|\E_{\meajoint{\approxdif+\sgl}{\sgl}}\psi-\E_{\mu^{*}}\psi|\leq\tfrac{\nu}{2}\big\}\bigcap\big\{\tfrac{1}{p}\|\approxdif\|^{2},\tfrac{1}{p}\|\sgl\|^{2}\leq4(\E B^{2}+\sigma_{\noisei}^{2})\big\},
\]
where $\approxdif:=\eta(\sgl+\sigOpt\vh)-\sgl$ is the same as in
(\ref{eq:v_star_p}) and $\mu^{*}$ is the joint measure of $\big(\prox(B+\sigOpt H),B\big)$,
with $\prox(\cdot):=\eta(\cdot;\mealim_{\YOpt},\mealim_{\sigOpt\Lambda/\theOpt})$
and $H\sim\mathcal{N}(0,1)$ independent of $B\sim\mealim_{B}$.

We first show $\P(\mathcal{E})\to1$, as $p\to\infty$. From (\ref{eq:mu_B_H_W2_convergence})
in the proof of Lemma \ref{lem:conv_F}, we have $W_{2}\big(\meajoint{\vh}{\sgl},H\otimes B\big)\asconv0$,
with $H\sim\mathcal{N}(0,1)$ and $B\sim\mealim_{B}$. Meanwhile,
$(h,b)\mapsto\big(\eta(b+\sigOpt h),b\big)$ is a $\sqrt{3+2(\sigOpt)^{2}}$-Lipschitz
continuous mapping. Hence similar as (\ref{eq:mu_y_W2_convergence}),
$W_{2}\big(\meajoint{\approxdif+\sgl}{\sgl},\mu^{*}\big)\asconv0$
and by Theorem 7.12 (iv) in \cite{villani2003topics}, $\E_{\meajoint{\approxdif+\sgl}{\sgl}}\psi\asconv\E_{\mu^{*}}\psi$.
Similarly, we can show
\[
\tfrac{1}{p}\|\approxdif\|^{2}\asconv\E\big[\eta(B+\sigOpt H)-B\big]^{2}<4(\E B^{2}+\sigma_{\noisei}^{2}).
\]
Also since $\{\sgl\}_{p\in\mathbb{N}}$ is a converging sequence,
$\tfrac{1}{p}\|\sgl\|^{2}\to\E B^{2}$. As a result, $\P(\mathcal{E})\to1$
for any $\nu>0$.

Next we show conditioned on $\mathcal{E}$, it holds that
\begin{equation}
D_{\nu}\bigcap\bdball_{\sqrt{n}K}\subseteq\outbdball_{\sqrt{n}\veps_{0}}(\approxdif)\bigcap\bdball_{\sqrt{n}K},\label{eq:setD_eps_subset_1}
\end{equation}
where $K\geq\tfrac{\sigOpt}{\sqrt{\delta}}+\tfrac{\theta_{\min}}{4}$,
$D_{\nu}:=\big\{\vv\in\R^{p}:|\E_{\meajoint{\vv+\sgl}{\sgl}}\psi-\E_{\mu^{*}}\psi|\ge\nu\big\}$
and $\veps_{0}^{2}=\tfrac{\nu^{2}}{3L^{2}(1+2\delta K^{2}+32(\E B^{2}+\sigma_{\noisei}^{2}))\delta}$
. Since $\psi$ is $L$ pseudo-Lipschitz, we can get
\begin{align}
|\E_{\meajoint{\vv+\sgl}{\sgl}}\psi-\E_{\meajoint{\approxdif+\sgl}{\sgl}}\psi| & =\big|\tfrac{1}{p}\sum_{i=1}^{p}\psi(v_{i}+\sgli_{i},\sgli_{i})-\tfrac{1}{p}\sum_{i=1}^{p}\psi(\approxdifi_{i}+\sgli_{i},\sgli_{i})\big|\nonumber \\
 & \leq\tfrac{L}{p}\sum_{i=1}^{p}\big(1+\sqrt{(v_{i}+\sgli_{i})^{2}+\sgli_{i}^{2}}+\sqrt{(\approxdifi_{i}+\sgli_{i})^{2}+\sgli_{i}^{2}}\big)\big|v_{i}-\approxdifi_{i}\big|\nonumber \\
 & \tleq{\text{(a)}}\tfrac{L}{p}\big[3(p+2\|\vv\|^{2}+2\|\approxdif\|^{2}+6\|\sgl\|^{2})\big]^{1/2}\|\vv-\approxdif\|,\label{eq:pseu_lipschitz_1}
\end{align}
where in (a) we use Cauchy-Swartz inequality and $1+\sqrt{x}+\sqrt{y}\leq\sqrt{3(1+x+y)}$.
Meanwhile, conditioned on event $\mathcal{E}$, if $\vv\in D_{\nu}$,
then
\begin{equation}
|\E_{\meajoint{\vv+\sgl}{\sgl}}\psi-\E_{\meajoint{\approxdif+\sgl}{\sgl}}\psi|\geq\big||\E_{\meajoint{\vv+\sgl}{\sgl}}\psi-\E_{\mu^{*}}\psi|-|\E_{\meajoint{\approxdif+\sgl}{\sgl}}\psi-\E_{\mu^{*}}\psi|\big|\geq\tfrac{\nu}{2}.\label{eq:traingle_1}
\end{equation}
Combining (\ref{eq:pseu_lipschitz_1}) and (\ref{eq:traingle_1}),
we know conditioned on event $\mathcal{E}$, $\tfrac{1}{n}\|\vv-\approxdif\|^{2}\geq\tfrac{\nu^{2}}{3L^{2}(1+2\delta K^{2}+32(\E B^{2}+\sigma_{\noisei}^{2}))\delta}=\veps_{0}^{2}$
for any $\vv\in D_{\nu}\bigcap\bdball_{\sqrt{n}K}$.

From (\ref{eq:setD_eps_subset_1}), we have
\begin{equation}
\P\big(\min_{\vv\in D_{\nu}\bigcap\bdball_{\sqrt{n}K}}L(\vv)\leq\PsiOpt+2\veps\big)\leq\P\big(\min_{\vv\in\outbdball_{\sqrt{n}\veps_{0}}(\approxdif)\bigcap\bdball_{\sqrt{n}K}}L(\vv)\leq\PsiOpt+2\veps\big)+\P(\mathcal{E}^{C}).\label{eq:Lv_Dnu_perturbation}
\end{equation}
On the other hand, from (\ref{eq:Lv_perturbation_1}) in Lemma \ref{lem:L_local_stability}
we have
\begin{equation}
\P\big(\min_{\vv\in\outbdball_{\sqrt{n}\veps_{0}}(\approxdif)\bigcap\bdball_{\sqrt{n}K}}L(\vv)\leq\PsiOpt+2\veps\big)\to0,\label{eq:Lv_perturbation_sphere2}
\end{equation}
if $\sqrt{n}\veps_{0}\geq8\sqrt{\frac{n\veps}{\gamma}}$. Therefore,
for any $\nu>0$ we can choose $\veps\leq\frac{\gamma\nu^{2}}{192L^{2}(1+2\delta K^{2}+32(\E B^{2}+\sigma_{\noisei}^{2}))\delta}$
and from (\ref{eq:Lv_Dnu_perturbation}) and (\ref{eq:Lv_perturbation_sphere2})
we can get (\ref{eq:Lv_perturbation_Dnu_2}).
\end{IEEEproof}
The last lemma in this section shows that the optimal solution of
the original problem is bounded with probability converging to 1.
\begin{lem}
\label{lem:C_optsol_bd}For $K\geq\tfrac{\sigOpt}{\sqrt{\delta}}+\tfrac{\theta_{\min}}{4}$,
we have as $p\to\infty$,
\begin{equation}
\P(\hat{\vv}\notin\bdball_{\sqrt{n}K})\to0.\label{eq:Cv_opt_bd_1}
\end{equation}
\end{lem}
\begin{IEEEproof}
To show $\hat{\vv}=\argmin{\vv}C(\vv)$ is bounded with probability
approaching 1, we use the following property: for any $a\leq K$,
\begin{equation}
\min_{\vv\in\outbdball_{\sqrt{n}a}\bigcap\bdball_{\sqrt{n}K}}C(\vv)>\min_{\vv\in\bdball_{\sqrt{n}K}}C(\vv)+\veps\Rightarrow\min_{\vv\in\outbdball_{\sqrt{n}a}}C(\vv)>\min_{\vv\in\bdball_{\sqrt{n}K}}C(\vv)+\veps.\label{eq:convexity_inclusion}
\end{equation}
One can prove (\ref{eq:convexity_inclusion}) by contradiction. If
$\min_{\vv\in\outbdball_{\sqrt{n}a}}C(\vv)\leq\min_{\vv\in\bdball_{\sqrt{n}K}}C(\vv)+\veps$,
it must hold that $\min_{\vv\in\bdball_{\sqrt{n}K}^{C}}C(\vv)\leq\min_{\vv\in\bdball_{\sqrt{n}K}}C(\vv)+\veps$.
Then by convexity of $C(\vv)$, we can always find $\vv_{0}\in\outbdball_{\sqrt{n}a}\bigcap\bdball_{\sqrt{n}K}$
such that $C(\vv_{0})\leq\min_{\vv\in\bdball_{\sqrt{n}K}}C(\vv)+\veps$,
which leads to a contradiction with $\min_{\vv\in\outbdball_{\sqrt{n}a}\bigcap\bdball_{\sqrt{n}K}}C(\vv)>\min_{\vv\in\bdball_{\sqrt{n}K}}C(\vv)+\veps$.
As a result, for any $a\leq K$,
\begin{equation}
\begin{aligned}\P\Big(\min_{\vv\in\bdball_{\sqrt{n}K}^{C}}C(\vv)\leq\min_{\vv\in\bdball_{\sqrt{n}K}}C(\vv)+\veps\Big) & \leq\P\Big(\min_{\vv\in\outbdball_{\sqrt{n}a}}C(\vv)\leq\min_{\vv\in\bdball_{\sqrt{n}K}}C(\vv)+\veps\Big)\\
 & \tleq{\text{(a)}}\P\Big(\min_{\vv\in\outbdball_{\sqrt{n}a}\bigcap\bdball_{\sqrt{n}K}}C(\vv)\leq\min_{\vv\in\bdball_{\sqrt{n}K}}C(\vv)+\veps\Big),
\end{aligned}
\label{eq:convexity_inclusion_2}
\end{equation}
where (a) follows from (\ref{eq:convexity_inclusion}). Now we choose
$a\in\big(\tfrac{\sigOpt}{\sqrt{\delta}},K\big)$. Then in the probability
space of auxiliary problem (\ref{eq:slope_AO}), under event $\mathcal{A}_{2}$
{[}c.f. (\ref{eq:highprob_event_5}){]} we can get $\outbdball_{\sqrt{n}a}\subset\outbdball_{\sqrt{n}\Delta_{a}}(\approxdif)$,
with $\Delta_{a}=a-\tfrac{\sigOpt}{\sqrt{\delta}}$. Therefore, for
any $\veps>0$ we have
\begin{align}
 & \P\Big(\min_{\vv\in\outbdball_{\sqrt{n}a}\bigcap\bdball_{\sqrt{n}K}}C(\vv)\leq\min_{\vv\in\bdball_{\sqrt{n}K}}C(\vv)+\veps\Big)\nonumber \\
\leq & \P\Big(\min_{\vv\in\outbdball_{\sqrt{n}a}\bigcap\bdball_{\sqrt{n}K}}C(\vv)\leq\PsiOpt+2\veps\Big)+\P\Big(\min_{\vv\in\bdball_{\sqrt{n}K}}C(\vv)\geq\PsiOpt+\veps\Big)\nonumber \\
\tleq{\text{(a)}} & \P\Big(\min_{\vv\in\outbdball_{\sqrt{n}a}\bigcap\bdball_{\sqrt{n}K}}L(\vv)\leq\PsiOpt+2\veps\Big)+\P\Big(\min_{\vv\in\bdball_{\sqrt{n}K}}L(\vv)\geq\PsiOpt+\veps\Big)\nonumber \\
\leq & \P\Big(\min_{\vv\in\outbdball_{\sqrt{n}\Delta_{a}}(\approxdif)\bigcap\bdball_{\sqrt{n}K}}L(\vv)\leq\PsiOpt+2\veps\Big)+\P\Big(\min_{\vv\in\bdball_{\sqrt{n}K}}L(\vv)\geq\PsiOpt+\veps\Big)+\P(\mathcal{A}_{2}^{C}),\label{eq:Cv_bd_sol_perturbation}
\end{align}
where $\PsiOpt$ is defined in (\ref{eq:Psi_sigma_theta_def}), in
(a) we use (\ref{eq:CGMT_C_L1_leq}) and (\ref{eq:CGMT_C_L2_geq}).
Combining (\ref{eq:convexity_inclusion_2}) and (\ref{eq:Cv_bd_sol_perturbation}),
we have for any $\veps>0$,
\begin{equation}
\begin{aligned}\P(\hat{\vv}\notin\bdball_{\sqrt{n}K})\leq & \P\Big(\min_{\vv\in\bdball_{\sqrt{n}K}^{C}}C(\vv)\leq\min_{\vv\in\bdball_{\sqrt{n}K}}C(\vv)+\veps\Big)\\
\leq & \P\Big(\min_{\vv\in\outbdball_{\sqrt{n}\Delta_{a}}(\approxdif)\bigcap\bdball_{\sqrt{n}K}}L(\vv)\leq\PsiOpt+2\veps\Big)+\P\Big(\min_{\vv\in\bdball_{\sqrt{n}K}}L(\vv)>\PsiOpt+\veps\Big)+\P(\mathcal{A}_{2}^{C}).
\end{aligned}
\label{eq:prob_v_notin_Bk}
\end{equation}
It remains to show all the three terms on the RHS of (\ref{eq:prob_v_notin_Bk})
converge to 0 for some $\veps>0$. From (\ref{eq:Lv_perturbation_1}),
we know for $\veps>0$ if we choose an $a$ such that $\sqrt{n}\Delta_{a}\geq8\sqrt{\frac{n\veps}{\gamma}}$,
where $\gamma=\frac{\theta_{\min}\sigma_{\noisei}^{2}}{4(K^{2}+\sigma_{\noisei}^{2})^{3/2}}$,
then 
\begin{equation}
\P\Big(\min_{\vv\in\outbdball_{\sqrt{n}\Delta_{a}}(\approxdif)\bigcap\bdball_{\sqrt{n}K}}L(\vv)\leq\PsiOpt+2\veps\Big)\to0.\label{eq:Lv_perturbation_sphere_1}
\end{equation}
Clearly, such $a$ always exists if $\veps\in\big(0,\tfrac{\gamma^{2}\theta_{\min}^{2}}{1024}\big)$
since $K\geq\tfrac{\sigOpt}{\sqrt{\delta}}+\tfrac{\theta_{\min}}{4}$.
On the other hand, in the proof of Lemma \ref{lem:L_local_stability}
we show $\P(\mathcal{A}_{2}^{C})\to0$ and from (\ref{eq:min_Lv_concentrate})
we have for any $\veps>0$, $\P\Big(\min_{\vv\in\bdball_{\sqrt{n}K}}L(\vv)>\PsiOpt+\veps\Big)\to0$.
Substituting these results back to (\ref{eq:prob_v_notin_Bk}), we
reach (\ref{eq:Cv_opt_bd_1}).
\end{IEEEproof}

\subsection{Proof of Lemma \ref{lem:R0p_prob_ubd}\label{subsec:Proof-of-Lemma-prob-ubd}}

To obtain the probabilistic upper bound for $R_{0}^{(p)}$, we can
approximate indicator function $\indicatorfn_{x=0}$ by a series of
envelope functions:
\begin{equation}
\psi_{h}(x)=\begin{cases}
1-h^{-1}x & 0\leq x<h,\\
1+h^{-1}x & -h\leq x<0,\\
0 & |x|\geq h,
\end{cases}\label{eq:indicator_approximation}
\end{equation}
where $h>0$. We can see $\psi_{h}(x)$ is an upper bound of $\indicatorfn_{x=0}$
and satisfies $0\leq\psi_{h}(x)-\indicatorfn_{x=0}\leq\indicatorfn_{0<|x|<h}$,
so
\begin{align}
R_{0}^{(p)}-\P\big[\prox(\YOpt)=0\big] & =\E_{\mealim_{\sol}}\indicatorfn_{x=0}-\E_{\mu_{\hat{B}}}\indicatorfn_{x=0}\nonumber \\
 & \leq\E_{\mealim_{\sol}}\psi_{h}(x)-\E_{\mu_{\hat{B}}}\psi_{h}(x)+\E_{\mu_{\hat{B}}}\psi_{h}(x)-\E_{\mu_{\hat{B}}}\indicatorfn_{x=0}\nonumber \\
 & \leq\big|\E_{\mealim_{\sol}}\psi_{h}(x)-\E_{\mu_{\hat{B}}}\psi_{h}(x)\big|+\P\big[|\prox(\YOpt)|\in(0,h)\big],\label{eq:R0p_upper_bd0}
\end{align}
where $\mu_{\hat{B}}$ denotes the distribution of $\prox(\YOpt)$.
Moreover, $\psi_{h}(x)$ is $h^{-1}$-Lipschitz (and hence pseudo-Lipschitz
by definition), so (\ref{eq:weak_convergence}) can now be applied,
which gives us $\big|\E_{\mealim_{\sol}}\psi_{h}(x)-\E_{\mu_{\hat{B}}}\psi_{h}(x)\big|\pconv0$
for any fixed $h>0$. Meanwhile, by continuity of probability, we
have $\lim_{h\to0}\P\big(|\prox(\YOpt)|\in(0,h)\big)=0$. As a result,
on both sides of (\ref{eq:R0p_upper_bd0}) taking $p\to\infty$ and
then $h\to0$, we get for any $\veps>0$,
\begin{equation}
R_{0}^{(p)}\leq\P\big(\prox(\YOpt)=0\big)+\veps\label{eq:R0p_upper_bd}
\end{equation}
with probability approaching 1 as $p\to\infty$.

\subsection{Proof of Lemma \ref{lem:R0_matching_lbd}\label{subsec:Proof-of-Lemma-mathcing_lbd}}

If $\P\big(\prox(\YOpt)=0\big)=0$, then since $R_{0}^{(p)}\geq0$,
(\ref{eq:R0p_lower_bd}) trivially holds. Thus, it only remains to
address the case when $\P\big(\prox(\YOpt)=0\big)>0$. Towards this
end, we will utilize Lemma \ref{lem:majorize_zero-sufficient}. To
apply Lemma \ref{lem:majorize_zero-sufficient}, we need to verify
for sufficiently large $k\in[p]$, $|\hat{\sg}|_{(1:k)}\stmajor\vlambda_{1:k}$
with probability approaching 1. For any $p,k,\ell\in\mathbb{Z}^{+}$,
with $0\leq k<\ell\leq p$,
\begin{align}
\sum_{i=k+1}^{\ell}|\hat{s}|_{(i)}-\sum_{i=k+1}^{\ell}\lambda_{i}= & \int_{k/p}^{\ell/p}\empiquant{|\hat{\vs}|}(u)du-\int_{k/p}^{\ell/p}\empiquant{|\vlambda|}(u)(u)du\nonumber \\
\leq & \int_{k/p}^{\ell/p}F_{|\hat{S}|}^{-1}(u)du-\int_{k/p}^{\ell/p}F_{\Lambda}^{-1}(u)du+\big|\int_{k/p}^{\ell/p}\empiquant{|\hat{\vs}|}(u)du-\int_{k/p}^{\ell/p}F_{|\hat{S}|}^{-1}(u)du\big|\nonumber \\
 & +\big|\int_{k/p}^{\ell/p}F_{\Lambda}^{-1}(u)du-\int_{k/p}^{\ell/p}\empiquant{|\vlambda|}(u)du\big|\nonumber \\
\leq & \int_{k/p}^{\ell/p}F_{|\hat{S}|}^{-1}(u)du-\int_{k/p}^{\ell/p}F_{\Lambda}^{-1}(u)du+W_{2}(\mealim_{\hat{\vs}},\mu_{\hat{S}})+W_{2}(\mealim_{\vlambda},\mu_{\Lambda}),\label{eq:discrete_majorization_bound}
\end{align}
where in the last step, we use (\ref{eq:Wasserstein_R}). Here, $\mu_{\hat{S}}$
is the law of $\tauOpt^{-1}[\YOpt-\prox(\YOpt)]$. By condition \ref{enu:R_strict_majorization},
we know for any $\veps\in(0,\probthresh]$ (recall that $\probthresh=\P\big(\prox(\YOpt)=0\big)$),
there exists $\varsigma>0$ such that 
\begin{align}
 & \max_{t\in[0,\probthresh-\veps]}\int_{t}^{\probthresh}F_{|\hat{S}|}^{-1}(u)du-\int_{t}^{\probthresh}F_{\Lambda}^{-1}(u)du\nonumber \\
\teq{\text{(a)}} & \tauOpt^{-1}\max_{t\in[0,\probthresh-\veps]}\int_{t}^{\probthresh}F_{|\YOpt|}^{-1}(u)du-\int_{t}^{\probthresh}F_{\tauOpt\Lambda}^{-1}(u)du\nonumber \\
\tleq{\text{(b)}} & -\tauOpt^{-1}\varsigma,\label{eq:continuous_majorization_upperbd_1}
\end{align}
where to reach (a), we use $\hat{S}\teq{\text{law}}\tauOpt^{-1}[\YOpt-\prox(\YOpt)]$
and the fact that $\eta(y)=0$ for $|y|\leq F_{|\YOpt|}^{-1}(\probthresh)$
and (b) is due to \ref{enu:R_strict_majorization} and the fact that
$t\mapsto\int_{t}^{\probthresh}F_{|\YOpt|}^{-1}(u)du$ and $t\mapsto\int_{t}^{\probthresh}F_{\tauOpt\Lambda}^{-1}(u)du$
are both continuous. For any fixed $\veps\in(0,\probthresh]$, $\lfloor p(\probthresh-\veps)\rfloor\leq\lfloor p\probthresh\rfloor-1$
for large enough $p$. Then substituting (\ref{eq:continuous_majorization_upperbd_1})
into (\ref{eq:discrete_majorization_bound}) we can get for large
enough $p$ and any $0\leq k\leq\lfloor p(\probthresh-\veps)\rfloor$,
\begin{align}
 & \sum_{i=k+1}^{\lfloor p\probthresh\rfloor}|\hat{s}|_{(i)}-\sum_{i=k+1}^{\lfloor p\probthresh\rfloor}\lambda_{i}\nonumber \\
\tleq{\text{(a)}} & \int_{k/p}^{\probthresh}F_{|\hat{S}|}^{-1}(u)du-\int_{k/p}^{\probthresh}F_{\Lambda}^{-1}(u)du-\big[\int_{\lfloor p\probthresh\rfloor/p}^{\probthresh}F_{|\hat{S}|}^{-1}(u)du-\int_{\lfloor p\probthresh\rfloor/p}^{\probthresh}F_{\Lambda}^{-1}(u)du\big]\nonumber \\
 & +W_{2}(\mealim_{\hat{\vs}},\mu_{\hat{S}})+W_{2}(\mealim_{\vlambda},\mu_{\Lambda})\nonumber \\
\tleq{\text{(b)}} & -\tauOpt^{-1}\varsigma+\tauOpt^{-1}\big(\int_{\lfloor p\probthresh\rfloor/p}^{\probthresh}F_{|\YOpt|}^{-1}(u)du+\int_{\lfloor p\probthresh\rfloor/p}^{\probthresh}F_{\tauOpt\Lambda}^{-1}(u)du\big)+W_{2}(\mealim_{\hat{\vs}},\mu_{\hat{S}})+W_{2}(\mealim_{\vlambda},\mu_{\Lambda}),\label{eq:discrete_majorization_bound_2}
\end{align}
where (a) follows from (\ref{eq:discrete_majorization_bound}) and
(b) follows from (\ref{eq:continuous_majorization_upperbd_1}). We
now show the last four terms in (\ref{eq:discrete_majorization_bound_2})
vanish as $p\to\infty$: $\int_{\lfloor p\probthresh\rfloor/p}^{\probthresh}F_{|\YOpt|}^{-1}(u)du$
and $\int_{\lfloor p\probthresh\rfloor/p}^{\probthresh}F_{\tauOpt\Lambda}^{-1}(u)du$
converges to 0 by DCT; $W_{2}(\mealim_{\vs},\mu_{S})\pconv0$ is proved
in Proposition \ref{prop:subgradient_conv}; $W_{2}(\mealim_{\vlambda},\mu_{\Lambda})\to0$
since $\{\vlambda\}_{p\in\mathbb{Z}^{+}}$ is a converging sequence.
As a result, for any fixed $\veps\in(0,\probthresh]$, there exists
$\veps>0$ such that 
\begin{equation}
\max_{0\leq k\leq\lfloor p(\probthresh-\veps)\rfloor}\sum_{i=k+1}^{\lfloor p\probthresh\rfloor}|\hat{s}|_{(i)}-\sum_{i=k+1}^{\lfloor p\probthresh\rfloor}\lambda_{i}\leq-\varsigma,\label{eq:discrete_majorization_bound_3}
\end{equation}
with probability approaching 1, as $p\to\infty$. Now we show conditioned
on (\ref{eq:discrete_majorization_bound_3}), there exists $k_{0}\in[\lfloor p(\probthresh-\veps)\rfloor,\lfloor p\probthresh\rfloor-1]$
such that 
\begin{equation}
\max_{0\leq k\leq k_{0}}\sum_{i=k+1}^{k_{0}+1}|\hat{s}|_{(i)}-\sum_{i=k+1}^{k_{0}+1}\lambda_{i}<0.\label{eq:discrete_majorization_bound_4}
\end{equation}
In other words, $|\hat{\vs}|_{(1:k_{0}+1)}\stmajor\vlambda_{1:k_{0}+1}$.
Such $k_{0}$ can be retrieved as follows:
\begin{enumerate}
\item[Step 0.] Let $k_{s}$ denote the candidate for $k_{0}$ and initialize $k_{s}=\lfloor p(\probthresh-\veps)\rfloor$.
From (\ref{eq:discrete_majorization_bound_3}) we know at the initial
step,
\begin{equation}
\max_{0\leq k\leq k_{s}}\sum_{i=k+1}^{\lfloor p\probthresh\rfloor}|\hat{s}|_{(i)}-\sum_{i=k+1}^{\lfloor p\probthresh\rfloor}\lambda_{i}\leq-\varsigma.\label{eq:discrete_majorization_bound_5}
\end{equation}
\item[Step 1.] If $k_{s}+1=\lfloor p\probthresh\rfloor$, then we output $k_{0}=k_{s}$;
otherwise we go to step 2.
\item[Step 2.] If $\sum_{i=k_{s}+2}^{\lfloor p\probthresh\rfloor}|\hat{s}|_{(i)}-\sum_{i=k_{s}+2}^{\lfloor p\probthresh\rfloor}\lambda_{i}>-\frac{\varsigma}{2},$
then together with (\ref{eq:discrete_majorization_bound_5}) we get
$\max_{0\leq k\leq k_{s}}\sum_{i=k+1}^{k_{s}+1}|\hat{s}|_{(i)}-\sum_{i=k+1}^{k_{s}+1}\lambda_{i}\leq-\frac{\varsigma}{2}$.
Hence, we output $k_{0}=k_{s}$; otherwise, we update $k_{s}$ and
$\varsigma$ as: $k_{s}\leftarrow k_{s}+1$, $\varsigma\leftarrow\frac{\varsigma}{2}$
and return back to step 1. Clearly, (\ref{eq:discrete_majorization_bound_5})
still holds under the updated $k_{s}$ and $\varsigma$.
\end{enumerate}
Then from Lemma \ref{lem:majorize_zero-sufficient}, (\ref{eq:discrete_majorization_bound_4})
implies that $|\sol|_{(1:k_{0}+1)}=0$ and thus $R_{0}^{(p)}\geq\probthresh-\veps$
since $k_{0}\geq p(\probthresh-\veps)-1$ by construction. Summing
up, if $\probthresh>0$ then for any $\veps\in(0,\probthresh]$, $R_{0}^{(p)}\geq\probthresh-\veps=\P(\prox(\YOpt)=0)-\veps$
with probability approaching 1 as $p\to\infty$. Therefore (\ref{eq:R0p_lower_bd})
is verified.

\subsection{Proof of Lemma \ref{lem:majorize_zero-sufficient}\label{subsec:Proof-of-Lemma-majorize-zero}}

The key is to establish the following result: for any $p$-dimensional
vectors $\va$ and $\vlambda$ with $0\leq\lambda_{1}\leq\cdots\leq\lambda_{p}$,
if $|\va-\tprox_{\vlambda}(\va)|_{(1:k)}\stmajor\vlambda_{1:k}$ for
some $k\in[p]$, then $|\tprox_{\vlambda}(\va)|_{(1:k)}=\boldsymbol{0}$.
To prove this, it suffices to show $|\tprox_{\vlambda}(\va)|_{(k)}=0$.
Assume $|\tprox_{\vlambda}(\va)|_{(k)}\neq0$. Define the index set
$I_{k}\bydef\{i\mid|\tprox_{\vlambda}(\va)|_{(i)}=|\tprox_{\vlambda}(\va)|_{(k)}\}$
and denote $\ell:=\min I_{k}$ and $m:=\max I_{k}$. According to
the formula for $\partial J_{\vlambda}$ \cite[Fact V.3]{bu2020algorithmic},
we have: if $|\tprox_{\vlambda}(\va)|_{(k)}\neq0$, then for any $\vg\in\partial J_{\vlambda}\big(\tprox_{\vlambda}(\va)\big)$
it holds that $|\vg|_{(\ell:m)}$$\majorized\vlambda_{\ell:m}$ and
$\sum_{i=\ell}^{m}|\vg|_{(i)}=\sum_{i=\ell}^{m}\lambda_{i}$. On the
other hand, by the first order optimality condition, $\va-\tprox_{\vlambda}(\va)\in\partial J_{\vlambda}\big(\tprox_{\vlambda}(\va)\big)$.
Hence, $|\va-\tprox_{\vlambda}(\va)|_{(\ell:m)}\majorized\vlambda_{\ell:m}$,
\emph{i.e.}, for any $\ell\leq q\leq m$,
\begin{equation}
\sum_{i=q}^{m}|\va-\tprox_{\vlambda}(\va)|_{(i)}\leq\sum_{i=q}^{m}\lambda_{i}\label{eq:equivalence_class_subgradient_2}
\end{equation}
and also
\begin{equation}
\sum_{i=\ell}^{m}|\va-\tprox_{\vlambda}(\va)|_{(i)}=\sum_{i=\ell}^{m}\lambda_{i}.\label{eq:equivalence_class_subgradient_1}
\end{equation}
Therefore, 
\begin{equation}
\begin{aligned}\sum_{i=\ell}^{m}|\va-\tprox_{\vlambda}(\va)|_{(i)}= & \sum_{i=\ell}^{k}|\va-\tprox_{\vlambda}(\va)|_{(i)}+\sum_{i=k+1}^{m}|\va-\tprox_{\vlambda}(\va)|_{(i)}\\
< & \sum_{i=\ell}^{k}\lambda_{i}+\sum_{i=k+1}^{m}\lambda_{i}\\
= & \sum_{i=\ell}^{m}\lambda_{i},
\end{aligned}
\label{eq:equivalence_class_subgradient_3}
\end{equation}
where the inequality follows from the condition that $|\va-\tprox_{\vlambda}(\va)|_{(1:k)}\stmajor\vlambda_{1:k}$
and (\ref{eq:equivalence_class_subgradient_2}). Clearly, (\ref{eq:equivalence_class_subgradient_3})
contradicts (\ref{eq:equivalence_class_subgradient_1}). Therefore,
$|\tprox_{\vlambda}(\va)|_{(k)}=0$ and thus $|\tprox_{\vlambda}(\va)|_{(1:k)}=\boldsymbol{0}_{k}$.

Now we are ready to prove Lemma \ref{lem:majorize_zero-sufficient}.
To apply the above result, we just need to express $\sol$ as 
\begin{equation}
\sol=\tprox_{\vlambda}(\sol+\hat{\sg}),\label{eq:proximal_characterization}
\end{equation}
using the first order optimality condition of (\ref{eq:slope_opt}).
Therefore, we can let $\va=\sol+\hat{\sg}$ and thus $\hat{\sg}=\va-\tprox_{\vlambda}(\va)$.
The desired result then immediately follows.

\subsection{Proof of Lemma \ref{lem:Vp_convergence_implied}\label{subsec:Proof-of-Lemma-imply-Vp}}

Similar as proof of Lemma \ref{lem:R0p_prob_ubd}, the idea is again
approximating the indicator function $\indicatorfn_{x\neq0,y=0}$
by some Lipschitz continuous functions. Here we use:
\[
\phi_{h}(x,y)=[1-\psi_{h}(x)]\psi_{h}(y),
\]
where $\psi_{h}(x)$ is defined in (\ref{eq:indicator_approximation}).
We have 
\begin{equation}
|\phi_{h}(x,y)-\indicatorfn_{x\neq0,y=0}|\leq\indicatorfn_{0<|x|<h}+\indicatorfn_{0<|y|<h}.\label{eq:indicator_2Dapprox_bd}
\end{equation}
Therefore, for any $h>0$,
\begin{align}
 & \big|V^{(p)}-\P\big[\prox(\YOpt)\neq0,B=0\big]\big|\nonumber \\
= & \big|\E_{\meajoint{\sol}{\sgl}}\indicatorfn_{x\neq0,y=0}-\E_{\meajoint{\hat{B}}B}\indicatorfn_{x\neq0,y=0}\big|\nonumber \\
\leq & \E_{\meajoint{\sol}{\sgl}}\big|\indicatorfn_{x\neq0,y=0}-\phi_{h}(x,y)\big|+\big|\E_{\meajoint{\sol}{\sgl}}\phi_{h}(x,y)-\E_{\meajoint{\hat{B}}B}\phi_{h}(x,y)\big|\nonumber \\
 & +\E_{\meajoint{\hat{B}}B}\big|\indicatorfn_{x\neq0,y=0}-\phi_{h}(x,y)\big|\nonumber \\
\leq & \frac{1}{p}\sum_{i=1}^{p}(\indicatorfn_{|\soli_{i}|<h}-\indicatorfn_{\soli_{i}=0})+\frac{1}{p}\sum_{i=1}^{p}(\indicatorfn_{|\sgli_{i}|<h}-\indicatorfn_{\sgli_{i}=0})+\P\big(0<|\prox(\YOpt)|<h\big)\nonumber \\
 & +\P(0<|B|<h)+\big|\E_{\meajoint{\sol}{\sgl}}\phi_{h}(x,y)-\E_{\meajoint{\hat{B}}B}\phi_{h}(x,y)\big|,\label{eq:Vp_upper_bd}
\end{align}
where $\meajoint{\hat{B}}B$ denotes the joint distribution of $(\prox(\YOpt),B)$
and in the last step we use (\ref{eq:indicator_2Dapprox_bd}) and
$\indicatorfn_{0<|x|<h}=\indicatorfn_{|x|<h}-\indicatorfn_{x=0}$.
Let us compute the limit of each term on the RHS of (\ref{eq:Vp_upper_bd}).
The last term converges in probability to zero due to (\ref{eq:weak_convergence}).
By continuity of probability, $\P\big(0<|\prox(\YOpt)|<h\big)$ and
$\P(0<|B|<h)$ converge to 0 as $h\to0$. Following similar steps
leading to (\ref{eq:R0p_upper_bd}), we can get $\frac{1}{p}\sum_{i=1}^{p}\indicatorfn_{|\soli_{i}|<h}\pconv\P\big(|\prox(\YOpt)|<h\big)$
and $\frac{1}{p}\sum_{i=1}^{p}\indicatorfn_{|\sgli_{i}|<h}\pconv\P(|B|<h)$
if $h$ satisfies $\P\big(|\prox(\YOpt)|=h\big)=\P(B=h)=0$. Therefore,
for such $h>0$, (\ref{eq:Vp_upper_bd}) yields the following: for
any $\veps>0$ there exists $p_{0}$ such that when $p>p_{0}$,
\begin{align}
\big|V^{(p)}-\P\big[\prox(\YOpt)\neq0,B=0\big]\big| & \leq\big[\P(|B|=0)+\P\big(|\prox(\YOpt)|=0\big)\big]-(r_{0}^{(p)}+R_{0}^{(p)})\nonumber \\
 & \:+2\P(0<|B|<h)+2\P\big(0<|\prox(\YOpt)|<h\big)+\frac{\veps}{2}.\nonumber \\
 & \leq2\P(0<|B|<h)+2\P\big(0<|\prox(\YOpt)|<h\big)+|\P\big(|\prox(\YOpt)|=0\big)-R_{0}^{(p)}|+\veps,\label{eq:Vp_pconv_0}
\end{align}
where in the last step we use Assumption \ref{a:sparsity_ratio}.
Then taking $h\to0$ along a sequence $\{h_{i}\}_{i\in\mathbb{Z}^{+}}$
with $\P\big(|\prox(\YOpt)|=h_{i}\big)=\P(B=h_{i})=0$ in (\ref{eq:Vp_pconv_0}),
we get for any $\veps>0$,
\begin{equation}
\begin{aligned}\big|V^{(p)}-\P\big[\prox(\YOpt)\neq0,B=0\big]\big|\leq & |\P\big(\prox(\YOpt)=0\big)-R_{0}^{(p)}|+\veps,\end{aligned}
\label{eq:Vp_pconv}
\end{equation}
with probability approaching 1 as $p\to\infty$. %

\subsection{Asymptotic Properties of $\hat{\protect\sg}$\label{subsec:Asymptotic-Properties-of-sg}}

In this section, we study the limiting properties of the following
vector: $\hat{\sg}=\dmtx^{\T}(\noise-\dmtx\hat{\vv}),$ where $\h{\vv}=\targmin{\vv}C(\vv)$.
Recall that $C(\vv)$ is the objective function of primary problem
defined in (\ref{eq:slope_PO}):
\[
C(\vv)=\frac{1}{n}\Big[\frac{1}{2}\|\dmtx\vv-\noise\|^{2}+\regu(\vv+\vbeta)\Big]
\]
and $\h{\vv}$ is the optimal solution. The main goal is to prove
Proposition \ref{prop:subgradient_conv}, which characterizes the
limiting empirical distribution of $(\hat{\sg},\sgl)$. We will follow
the proof strategy in \cite[Appendix E]{miolane2018distribution}.
\begin{prop}
\label{prop:subgradient_conv}Under the same setting as Theorem \ref{thm:asymp_char},
define $\mu_{\hat{S},B}$ as the joint measure of $(\tauOpt^{-1}[Y-\eta(Y;\mealim_{\YOpt},\mealim_{\tauOpt\Lambda})],B)$.
It holds that $W_{2}(\meajoint{\hat{\sg}}{\sgl},\mu_{\hat{S},B})\pconv0$.
\end{prop}
\begin{IEEEproof}
The first step is to obtain an alternative representation of $\hat{\sg}$.
Consider the event $E=\{\hat{\vv}\in\bdball_{\sqrt{n}K}\}$, for some
$K\geq\tfrac{\sigOpt}{\sqrt{\delta}}+\tfrac{\theta_{\min}}{4}$, where
$\theta_{\min}$ is given in Lemma \ref{lem:scalar_minimax}. It is
shown in Lemma \ref{lem:C_optsol_bd} that $\P(E)\to1$ as $p\to\infty$.
Under event $E$, we have
\begin{align}
\min_{\vv\in\R^{p}}C(\vv) & =\min_{\vv\in\bdball_{\sqrt{n}K}}\frac{1}{n}\Big[\frac{1}{2}\|\dmtx\vv-\noise\|^{2}+\regu(\vv+\vbeta)\Big]\nonumber \\
 & =\min_{\vv\in\bdball_{\sqrt{n}K}}\max_{\vs\in\dualball}\frac{1}{n}\Big[\frac{1}{2}\|\dmtx\vv-\noise\|^{2}+\sg^{\T}(\vv+\vbeta)\Big]\nonumber \\
 & \teq{\text{}}\max_{\vs\in\dualball}\underbrace{\min_{\vv\in\bdball_{\sqrt{n}K}}\frac{1}{n}\Big[\frac{1}{2}\|\dmtx\vv-\noise\|^{2}+\sg^{\T}(\vv+\vbeta)\Big],}_{\bydef\Sgobj(\sg)}\label{eq:alternative_sg}
\end{align}
where $\dualball$ is defined in (\ref{eq:prox_as_projection}) and
the last step follows from Sion's minimax theorem \cite{sion1958general}.
By the first order optimality condition of , we know $\hat{\sg}\in\partial\regu(\hat{\vv}+\vbeta)$.
On the other hand, it is not hard to show for any $\vx\in\R^{p}$
and $\vs\in\partial\regu(\vx)$, it holds that $\vs\in\dualball$
and $J_{\vlambda}(\vx)=\vs^{\T}\vx$. Therefore, $\hat{\sg}\in\dualball$
and $\hat{\sg}^{\T}(\hat{\vv}+\vbeta)=\regu(\hat{\vv}+\vbeta)$. Then
\[
\begin{aligned}\Sgobj(\hat{\sg})= & \min_{\vv\in\bdball_{\sqrt{n}K}}\frac{1}{n}\Big[\frac{1}{2}\|\dmtx\vv-\noise\|^{2}+\hat{\sg}^{\T}(\vv+\vbeta)\Big]\\
\teq{\text{(a)}} & \frac{1}{n}\Big[\frac{1}{2}\|\dmtx\hat{\vv}-\noise\|^{2}+\hat{\sg}^{\T}(\hat{\vv}+\vbeta)\Big]\\
= & \frac{1}{n}\Big[\frac{1}{2}\|\dmtx\hat{\vv}-\noise\|^{2}+\regu(\hat{\vv}+\vbeta)\Big]\\
\teq{\text{(b)}} & \max_{\vs\in\dualball}\Sgobj(\sg),
\end{aligned}
\]
where (a) follows from first order optimality condition and the fact
that $\hat{\vv}\in\bdball_{\sqrt{n}K}$ under event $E$ and (b) follows
from (\ref{eq:alternative_sg}) and $\hat{\vv}\in\min_{\vv\in\R^{p}}C(\vv)$.
This implies that under event $E$, which happens with probability
approaching 1, $\hat{\sg}\in\targmax{\vs\in\dualball}\Sgobj(\sg)$.
Therefore, in order to study the limiting behavior of $\hat{\sg}$,
we can instead study $\targmax{\vs\in\dualball}\Sgobj(\sg)$.

The analysis of $\targmax{\vs\in\dualball}\Sgobj(\sg)$ can be carried
out based on CGMT framework. First, similar as Proposition E.1 in
\cite{miolane2018distribution}, we can get for any closed set $D$
and $t\in\R$,
\begin{equation}
\P\big(\max_{\sg\in D}\Sgobj(\sg)\geq t\big)\leq2\P\big(\max_{\sg\in D}\SgAOobj(\sg)\geq t\big)\label{eq:CGMT_sg_1}
\end{equation}
and if $D$ is also convex,
\begin{equation}
\P\big(\max_{\sg\in D}\Sgobj(\sg)\leq t\big)\leq2\P\big(\max_{\sg\in D}\SgAOobj(\sg)\leq t\big).\label{eq:CGMT_sg_2}
\end{equation}
Here
\begin{equation}
\SgAOobj(\sg)=\min_{\vv\in\bdball_{\sqrt{n}K}}\tfrac{1}{2}\Big(\sqrt{\tfrac{\|\vv\|^{2}}{n}\tfrac{\|\vg\|^{2}}{n}+\tfrac{\|\noise\|^{2}}{n}+2\tfrac{\|\vv\|}{\sqrt{n}}\tfrac{\vg^{\T}\noise}{n}}-\tfrac{\vh^{\T}\vv}{n}\Big)_{+}^{2}+\tfrac{\sg^{\T}(\vv+\sgl)}{n}.\label{eq:S_AOfn_def}
\end{equation}
Then for any $\veps,\nu>0$ and $D_{\nu}\bydef\left\{ \sg:W_{2}(\meajoint{\sg}{\sgl},\mu_{\hat{S},B})\geq\nu\right\} \bigcap\dualball$,
we have 
\begin{equation}
\begin{aligned}\P\big(\max_{\sg\in D_{\nu}}\Sgobj(\sg)\geq\max_{\vs\in\dualball}\Sgobj(\sg)-\veps\big)\leq & \P\big(\max_{\sg\in D_{\nu}}\Sgobj(\sg)\geq\Psi^{*}-2\veps\big)+\P\big(\max_{\vs\in\dualball}\Sgobj(\sg)<\Psi^{*}-\veps\big)\\
\tleq{\text{(a)}} & 2\P\big(\max_{\sg\in D_{\nu}}\SgAOobj(\sg)\geq\Psi^{*}-2\veps\big)+2\P\big(\max_{\vs\in\dualball}\SgAOobj(\sg)<\Psi^{*}-\veps\big)
\end{aligned}
\label{eq:S_PO_perturbation_1}
\end{equation}
where $\Psi^{*}$ is defined in (\ref{eq:Psi_sigma_theta_def}) and
step (a) follows from (\ref{eq:CGMT_sg_1}) and (\ref{eq:CGMT_sg_2}).
Then combining (\ref{eq:S_PO_perturbation_1}) with Lemma \ref{lem:SAO_perturbation_and_concentration},
we get for any $\nu>0$, there exists $\veps>0$ such that $\P\big(\max_{\sg\in D_{\nu}}\Sgobj(\sg)\geq\max_{\vs\in\dualball}\Sgobj(\sg)-\veps\big)\to0.$
Therefore, for any $\nu>0$ and $\veps>0$,
\begin{align}
\P\big(W_{2}(\meajoint{\hat{\sg}}{\sgl},\mu_{\hat{S},B})\geq\nu\big) & =\P(\hat{\sg}\in D_{\nu})\nonumber \\
 & \leq\P(\hat{\sg}\in D_{\nu}\bigcap E)+\P(E^{C})\nonumber \\
 & \leq\P\Big[\hat{\sg}\in D_{\nu}\text{ and }\Sgobj(\hat{\sg})=\max_{\vs\in\dualball}\Sgobj(\sg)\Big]+\P(E^{C})\nonumber \\
 & \leq\P\Big[\max_{\sg\in D_{\nu}}\Sgobj(\sg)=\max_{\vs\in\dualball}\Sgobj(\sg)\Big]+\P(E^{C})\nonumber \\
 & \leq\P\big(\max_{\sg\in D_{\nu}}\Sgobj(\sg)\geq\max_{\vs\in\dualball}\Sgobj(\sg)-\veps\big)+\P(E^{C}).\label{eq:s_hat_prob_bound}
\end{align}
By the discussion above, the RHS of (\ref{eq:s_hat_prob_bound}) converges
to 0 for some $\veps>0$. Since, the LHS of (\ref{eq:s_hat_prob_bound})
does not depend on $\veps$, this concludes the proof.
\end{IEEEproof}
\begin{lem}
\label{lem:SAO_perturbation_and_concentration}Under the same setting
as Proposition \ref{prop:subgradient_conv}, as $p\to\infty$,
\begin{equation}
\max_{\vs\in\dualball}\SgAOobj(\sg)\pconv\Psi^{*},\label{eq:SAO_optvalue_concentrate}
\end{equation}
where $\SgAOobj(\sg)$ is defined in (\ref{eq:S_AOfn_def}) and $\Psi^{*}$
is defined in (\ref{eq:Psi_sigma_theta_def}). Denote $D_{\nu}:=\left\{ \sg:W_{2}(\meajoint{\sg}{\sgl},\mu_{\hat{S},B})\geq\nu\right\} \bigcap\dualball$.
For any $\nu>0$, there exists $\veps>0$ such that
\begin{equation}
\P\big(\max_{\sg\in D_{\nu}}\SgAOobj(\sg)\geq\Psi^{*}-\veps\big)\to0.\label{eq:SAO_perturb_bound}
\end{equation}
\end{lem}
\begin{IEEEproof}
For the similar reason as introducing $\widetilde{L}(\vv)$ when analyzing
$L(\vv)$ {[}c.f. (\ref{eq:Lv}) and (\ref{eq:tilde_L_v}) in the
proof of Lemma \ref{lem:L_local_stability}{]}, we consider the following
approximation of $\SgAOobj(\sg)$:
\begin{equation}
\widetilde{\SgAOobj}(\sg)=\min_{\vv\in\bdball_{\sqrt{n}K}}\tfrac{1}{2}\Big(\sqrt{\tfrac{\|\vv\|^{2}}{n}+\sigma_{\noisei}^{2}}-\tfrac{\vh^{\T}\vv}{n}\Big)_{+}^{2}+\tfrac{\sg^{\T}(\vv+\sgl)}{n}.\label{eq:tilde_S_AO_def}
\end{equation}
Note that $\SgAOobj(\sg)=\min_{\vv\in\bdball_{\sqrt{n}K}}L(\vv)-\Delta(\vv)$
and $\widetilde{\SgAOobj}(\sg)=\min_{\vv\in\bdball_{\sqrt{n}K}}\widetilde{L}(\vv)-\Delta(\vv)$,
where $\Delta(\vv)=\tfrac{\regu(\vv+\vbeta)-\sg^{\T}(\vv+\sgl)}{n}$.
Therefore, by (\ref{eq:diff_Lv})
\begin{equation}
\sup_{\vs\in\R^{p}}|\SgAOobj(\sg)-\widetilde{\SgAOobj}(\sg)|\leq\sup_{\vv\in\bdball_{\sqrt{n}K}}|L(\vv)-\widetilde{L}(\vv)|\pconv0.\label{eq:unif_conv_S_AOfn}
\end{equation}
On the other hand, similar as (\ref{eq:alternative_sg}) we have
\begin{align}
\min_{\vv\in\bdball_{\sqrt{n}K}}\widetilde{L}(\vv)= & \min_{\vv\in\bdball_{\sqrt{n}K}}\tfrac{1}{2}\Big(\sqrt{\tfrac{\|\vv\|^{2}}{n}+\sigma_{\noisei}^{2}}-\tfrac{\vh^{\T}\vv}{n}\Big)_{+}^{2}+\tfrac{\regu(\vv+\vbeta)}{n}\nonumber \\
= & \min_{\vv\in\bdball_{\sqrt{n}K}}\max_{\sg\in\dualball}\tfrac{1}{2}\Big(\sqrt{\tfrac{\|\vv\|^{2}}{n}+\sigma_{\noisei}^{2}}-\tfrac{\vh^{\T}\vv}{n}\Big)_{+}^{2}+\tfrac{\sg^{\T}(\vv+\sgl)}{n}\nonumber \\
= & \max_{\sg\in\dualball}\min_{\vv\in\bdball_{\sqrt{n}K}}\tfrac{1}{2}\Big(\sqrt{\tfrac{\|\vv\|^{2}}{n}+\sigma_{\noisei}^{2}}-\tfrac{\vh^{\T}\vv}{n}\Big)_{+}^{2}+\tfrac{\sg^{\T}(\vv+\sgl)}{n}\nonumber \\
= & \max_{\sg\in\dualball}\widetilde{\SgAOobj}(\sg),\label{eq:alternative_sg_1}
\end{align}
where $\dualball$ is defined in (\ref{eq:prox_as_projection}). Using
successively (\ref{eq:unif_conv_S_AOfn}) and (\ref{eq:alternative_sg_1}),
we have for any $\veps>0$,
\begin{equation}
\begin{aligned}\P\big(\max_{\vs\in\dualball}\SgAOobj(\sg)<\Psi^{*}-\veps\big) & \leq\P\big(\max_{\vs\in\dualball}\widetilde{\SgAOobj}(\sg)<\Psi^{*}-\tfrac{\veps}{2}\big)+\delta_{p}\\
 & =\P\big(\min_{\vv\in\bdball_{\sqrt{n}K}}\widetilde{L}(\vv)<\Psi^{*}-\tfrac{\veps}{2}\big)+\delta_{p},
\end{aligned}
\label{eq:SAO_concentrate_lbd}
\end{equation}
where $\delta_{p}\to0$ as $p\to\infty$. Similarly on the other direction,
we can also get for any $\veps>0$, there is some $\delta_{p}'\to0$
such that
\begin{equation}
\P\big(\max_{\vs\in\dualball}\SgAOobj(\sg)\geq\Psi^{*}+\veps\big)\leq\P\big(\min_{\vv\in\bdball_{\sqrt{n}K}}\widetilde{L}(\vv)\geq\Psi^{*}+\tfrac{\veps}{2}\big)+\delta_{p}'\label{eq:SAO_concentrate_ubd}
\end{equation}
Then combining (\ref{eq:SAO_concentrate_lbd}) and (\ref{eq:SAO_concentrate_ubd})
with (\ref{eq:Ltilde_min_concentrate}), we get $\max_{\vs\in\dualball}\SgAOobj(\sg)\pconv\Psi^{*}$
and from (\ref{eq:unif_conv_S_AOfn}) we get $\max_{\vs\in\dualball}\widetilde{\SgAOobj}(\sg)\pconv\Psi^{*}$.

Next, we show (\ref{eq:SAO_perturb_bound}). First, the following
bound holds
\begin{align}
\P\big(\max_{\sg\in D_{\nu}}\SgAOobj(\sg)\geq\Psi^{*}-\veps\big)\leq & \P\big(\max_{\sg\in D_{\nu}}\widetilde{\SgAOobj}(\sg)\geq\Psi^{*}-2\veps\big)+\P\big(\sup_{\sg\in D_{\nu}}|\SgAOobj(\sg)-\widetilde{\SgAOobj}(\sg)|\geq\veps\big)\nonumber \\
\leq & \P\big(\max_{\sg\in D_{\nu}}\widetilde{\SgAOobj}(\sg)\geq\max_{\vs\in\dualball}\widetilde{\SgAOobj}(\sg)-3\veps\big)+\P\big(\max_{\vs\in\dualball}\widetilde{\SgAOobj}(\sg)\geq\Psi^{*}+\veps\big)\nonumber \\
 & +\P\big(\sup_{\sg\in D_{\nu}}|\SgAOobj(\sg)-\widetilde{\SgAOobj}(\sg)|\geq\veps\big).\label{eq:S_AO_perturbation_1}
\end{align}
Recall that we have already shown that the last two terms on the RHS
of (\ref{eq:S_AO_perturbation_1}) vanish as $p\to\infty$. Therefore,
it remains to show the first term also converges to 0. The main step
is to establish there exist $\varsigma_{\max},\gamma>0$ such that
for any $\varsigma\in(0,\varsigma_{\max})$,
\begin{equation}
\P\big(\max_{\sg\in\dualball\cap\outbdball_{\sqrt{p}\varsigma}(\approxsg)}\widetilde{\SgAOobj}(\sg)\geq\max_{\vs\in\dualball}\widetilde{\SgAOobj}(\sg)-\gamma\varsigma\big)\to0,\label{eq:Stilde_AO_perturbation_1}
\end{equation}
where 
\[
\begin{aligned}\approxsg\bydef & \frac{1}{\tauOpt}\big[(\sgl+\sigOpt\vh)-(\sgl+\approxdif)\big]\\
= & \frac{1}{\tauOpt}\big[(\sgl+\sigOpt\vh)-\eta(\sgl+\sigOpt\vh;\mealim_{\YOpt},\mealim_{\tauOpt\Lambda})\big]
\end{aligned}
\]
and $\approxdif$ is defined in (\ref{eq:v_star_p}). The convergence
in (\ref{eq:Stilde_AO_perturbation_1}) can be proved in exactly the
same way as Theorem E.7 in \cite{miolane2018distribution}, which
deals with the case of LASSO. For simplicity, we do not re-present
the proof details here. Here $\dualball$ (the unit ball of the dual
norm of $\regu$) plays the role of the set $\{\vx\mid\|\vx\|_{\infty}\leq\lambda\}$
in \cite{miolane2018distribution}, which is the unit ball of the
dual norm of $\lambda\|\cdot\|_{1}$. Now consider the event $E=\{W_{2}\big(\meajoint{\approxsg}{\sgl},\mu_{\hat{S},B}\big)\leq\tfrac{\nu}{2}\}$.
Conditioned on $E$, it holds that for $\sg\in D_{\nu}$, 
\[
\begin{aligned}\frac{1}{p}\|\sg-\approxsg\|^{2}\tgeq{\text{}} & W_{2}\big(\meajoint{\sg}{\sgl},\meajoint{\approxsg}{\sgl}\big)^{2}\\
\tgeq{\text{}} & \big[W_{2}\big(\meajoint{\sg}{\sgl},\mu_{\hat{S},B}\big)-W_{2}\big(\meajoint{\approxsg}{\sgl},\mu_{\hat{S},B}\big)\big]^{2}\\
\geq & \frac{\nu^{2}}{4},
\end{aligned}
\]
which indicates that $D_{\nu}\subseteq\dualball\cap\outbdball_{\frac{\sqrt{p}\nu}{2}}(\approxsg)$.
Therefore, 
\begin{equation}
\begin{aligned}\P\big(\max_{\sg\in D_{\nu}}\widetilde{\SgAOobj}(\sg)\geq\max_{\vs\in\dualball}\widetilde{\SgAOobj}(\sg)-\veps\big)\leq & \P\big(\max_{\sg\in D_{\nu}}\widetilde{\SgAOobj}(\sg)\geq\max_{\vs\in\dualball}\widetilde{\SgAOobj}(\sg)-\veps\bigcap E\big)+\P(E^{C})\\
\leq & \P\Big(\max_{\sg\in\dualball\cap\outbdball_{\frac{\sqrt{p}\nu}{2}}(\approxsg)}\widetilde{\SgAOobj}(\sg)\geq\max_{\vs\in\dualball}\widetilde{\SgAOobj}(\sg)-\veps\Big)+\P(E^{C}).
\end{aligned}
\label{eq:Stilde_AO_perturbation_2}
\end{equation}
According to (\ref{eq:Stilde_AO_perturbation_1}), the first term
in (\ref{eq:Stilde_AO_perturbation_2}) vanishes if $\veps\leq\frac{\gamma\nu}{2}$.
For the second term, it is not hard to show $(H,B)\mapsto\big(\tauOpt^{-1}[(B+\sigOpt H)-\eta(B+\sigOpt H;\mealim_{\YOpt},\mealim_{\tauOpt\Lambda})],B\big)$
is a Lipschitz continuous mapping and from (\ref{eq:mu_B_H_W2_convergence}),
$W_{2}\big(\meajoint{\vh}{\sgl},H\otimes B\big)\pconv0$, so similar
as (\ref{eq:mu_y_W2_convergence}) we can get $W_{2}\big(\meajoint{\approxsg}{\sgl},\mu_{\hat{S},B}\big)\pconv0$
and thus $\P(E^{C})\to0$. Therefore, from (\ref{eq:Stilde_AO_perturbation_2}),
for any $\nu>0$, there exists $\veps_{0}>0$ such that any $\veps\leq\veps_{0}$
satisfies $\P\big(\max_{\sg\in D_{\nu}}\widetilde{\SgAOobj}(\sg)\geq\max_{\vs\in\dualball}\widetilde{\SgAOobj}(\sg)-\veps\big)$.
Substituting this back to (\ref{eq:S_AO_perturbation_1}), we finish
the proof.
\end{IEEEproof}

\subsection{Properties of Limiting Scalar Problem\label{subsec:Properties-of-Limiting-scalar-prob}}

{} It turns out that the limiting behavior of (\ref{eq:slope_opt})
is fully captured by (\ref{eq:Psi_sigma_theta_def}). In this section,
we study the key properties of (\ref{eq:Psi_sigma_theta_def}).
\begin{lem}
\label{lem:scalar_minimax}The minimax problem (\ref{eq:Psi_sigma_theta_def})
has a unique optimal solution $(\sigOpt,\theOpt)$, which is also
the unique solution to the equation:
\begin{equation}
\begin{aligned}\sigma^{2} & =\sigma_{\noisei}^{2}+\frac{1}{\delta}\E[\eta(Y;\mealim_{Y},\mealim_{\sigma\Lambda/\theta})-B]^{2},\\
\theta & =\sigma\Big[1-\frac{1}{\delta}\E\eta'(Y;\mealim_{Y},\mealim_{\sigma\Lambda/\theta})\Big],
\end{aligned}
\label{eq:fixed_point_equation_sigmatheta}
\end{equation}
where $Y=B+\sigma H$, with $H\sim\mathcal{N}(0,1)$ independent of
$B\sim\mealim_{B}$. Besides, there exists $\theta_{\min}>0$ such
that $\theta^{*}\geq\theta_{\min}$.
\end{lem}
\begin{IEEEproof}
The proof includes two steps: (I) show the saddle point of $\Psi(\sigma,\theta)$
exists and is unique and it is also the unique optimal solution of
the minimax problem (\ref{eq:Psi_sigma_theta_def}), (II) show $(\sigOpt,\theOpt)$
is the saddle point of $\Psi(\sigma,\theta)$ if and only if it is
the solution to (\ref{eq:fixed_point_equation_sigmatheta}).

We first show the set of saddle points of $\Psi(\sigma,\theta)$ is
nonempty and compact, using Proposition 5.5.7 of \cite{bertsekas2009convex}.
To apply this result, it suffices to check: (i) $\Psi(\cdot,\theta)$
and $-\Psi(\sigma,\cdot)$ are convex and closed for any fixed $\theta\geq0$
and $\sigma\geq\sigma_{\noisei}$, (ii) there exists some $\overline{\theta}\geq0,\,\overline{\sigma}\geq\sigma_{\noisei}$
and $\gamma_{1},\gamma_{2}\in\R$ such that the level sets $\{\sigma\geq\sigma_{\noisei}\mid\Psi(\sigma,\overline{\theta})\leq\gamma_{1}\}$
and $\{\theta\geq0\mid-\Psi(\overline{\sigma},\theta)\leq\gamma_{2}\}$
are both non-empty and compact. From Lemma \ref{lem:F_sigma_theta_contdiff},
$\Moreau(\sigma,\theta)$ is convex-concave and continuously differentiable
with respect to $\sigma$ and $\theta$. Therefore, condition (i)
is satisfied. Also partial derivatives of $\Moreau(\sigma,\theta)$
can be computed as
\begin{align}
\frac{\partial\Psi}{\partial\sigma} & =\frac{\theta}{2\sigma^{2}}\Big(\sigma^{2}-\sigma_{\noisei}^{2}-\frac{1}{\delta}\lim_{p\to\infty}\frac{1}{p}\E\|\sgl-\tprox_{\sigma\vlambda/\theta}(\sgl+\sigma\vh)\|^{2}\Big),\label{eq:partial_Phi_partial_sigma}\\
\frac{\partial\Psi}{\partial\theta} & =\frac{1}{2}\Big(\frac{\sigma_{\noisei}^{2}}{\sigma}+\sigma\Big)-\theta+\frac{1}{\delta}\Big[\lim_{p\to\infty}\E\frac{\|\sgl+\sigma\vh-\tprox_{\sigma\vlambda/\theta}(\sgl+\sigma\vh)\|^{2}}{2\sigma p}-\frac{\sigma}{2}\Big],\label{eq:partial_Phi_partial_theta}
\end{align}
using (\ref{eq:Psi_sigma_theta_def}), (\ref{eq:lim_par_EFp_sigma})
and (\ref{eq:lim_par_EFp_theta}). Next we show $\{\sigma\geq\sigma_{\noisei}\mid\Psi(\sigma,\overline{\theta})\leq\gamma_{1}\}$
is non-empty and compact for some $\overline{\theta}\geq0$ and $\gamma_{1}\in\R$.
First, we have
\begin{align}
\E\|\sgl-\tprox_{\sigma\vlambda/\theta}(\sgl+\sigma\vh)\|^{2} & \teq{\text{(a)}}\sigma^{2}\E\big\Vert\tfrac{\sgl}{\sigma}-\tprox_{\vlambda/\theta}(\tfrac{\sgl}{\sigma}+\vh)\|^{2}\nonumber \\
 & \tleq{\text{}}2\sigma^{2}\E\big(\tfrac{\|\sgl\|^{2}}{\sigma^{2}}+2\|\tprox_{\vlambda/\theta}(\vh)\|^{2}+2\|\tprox_{\vlambda/\theta}(\tfrac{\sgl}{\sigma}+\vh)-\tprox_{\vlambda/\theta}(\vh)\|^{2}\big)\nonumber \\
 & \tleq{\text{(b)}}4\sigma^{2}\E\big\Vert\tprox_{\vlambda/\theta}(\vh)\|^{2}+6\|\sgl\|^{2}\nonumber \\
 & \tleq{\text{(c)}}4\sigma^{2}\sum_{i=1}^{p}\E\max\big\{|h_{i}|-\tfrac{\bar{\lambda}}{\theta},0\big\}^{2}+6\|\sgl\|^{2},\label{eq:v_normsq_bd_finite_p}
\end{align}
where (a) follows from the identity $\tprox_{\tau\vlambda}(\vx)=\tau\tprox_{\vlambda}(\vx/\tau)$,
(b) follows from the non-expansiveness of proximal operator and (c)
is a consequence of (\ref{eq:prox_norm_bd_by_LASSO}), where $\bar{\lambda}=\tfrac{1}{p}\sum_{i=1}^{p}\lambda_{i}$.
Plugging the bound (\ref{eq:v_normsq_bd_finite_p}) into (\ref{eq:partial_Phi_partial_sigma})
gives
\begin{equation}
\tfrac{\partial\Psi}{\partial\sigma}\geq\tfrac{\theta}{2\sigma^{2}}\Big[\sigma^{2}\big(1-\tfrac{8}{\delta}\int_{\frac{\E\Lambda}{\theta}}^{\infty}(z-\tfrac{\E\Lambda}{\theta})d\Phi(z)\big)-\sigma_{\noisei}^{2}-\tfrac{6}{\delta}\E B^{2}\Big],\label{eq:partial_Phi_partial_sigma_lbd}
\end{equation}
where $\Phi(z)$ is CDF of standard Gaussian. When $\E\Lambda>0$
, from (\ref{eq:partial_Phi_partial_sigma_lbd}) we know there exists
$\theta_{1}>0$ and $\sigma_{1}\geq\sigma_{\noisei}$ such that $\frac{\partial\Psi(\sigma,\theta_{1})}{\partial\sigma}\geq\frac{\theta_{1}}{8}$
for all $\sigma\geq\sigma_{1}$; when $\E\Lambda=0$, by our assumption
we must have $\delta>1$, so from (\ref{eq:lim_partial_Phi_partial_sigma})
we have $\frac{\partial\Psi(\sigma,\theta)}{\partial\sigma}=\frac{\theta}{2\sigma^{2}}[(1-\tfrac{1}{\delta})\sigma^{2}-\sigma_{\noisei}^{2}]$
for any $\theta>0$, $\sigma\geq\sigma_{\noisei}$ implying $\frac{\partial\Psi(\sigma,\theta)}{\partial\sigma}\geq\tfrac{\theta}{4}(1-\tfrac{1}{\delta})$
for all $\sigma\geq\sqrt{\frac{2\delta}{\delta-1}}\sigma_{\noisei}$.
Therefore, there exists $\overline{\theta}>0$, $c>0$ and $K\geq\sigma_{\noisei}$
such that $\frac{\partial\Psi(\sigma,\overline{\theta})}{\partial\sigma}\geq c$
for any $\sigma\geq K$. This means that $\Psi(\sigma,\overline{\theta})>\Psi(K,\overline{\theta})$
for all $\sigma>K$, so the set $\{\sigma\geq\sigma_{\noisei}\mid\Psi(\sigma,\overline{\theta})\leq\Psi(K,\overline{\theta})\}\subset[\sigma_{\noisei},K]$
and it is non-empty (include at least one point $\sigma=K$) and closed
since $\Psi(\cdot,\overline{\theta})$ is a closed function. As a
result, we can take $\gamma_{1}=\Psi(K,\overline{\theta})$ and the
level set $\{\sigma\geq\sigma_{\noisei}\mid\Psi(\sigma,\overline{\theta})\leq\gamma_{1}\}$
is non-empty and compact. On the other hand, we can show $\{\theta\geq0\mid-\Psi(\sigma_{\noisei},\theta)\leq0\}$
is non-empty and compact. First since $\Psi(\sigma_{\noisei},\cdot)$
is 1-strongly concave and continuously differentiable, we have for
any $\theta\geq0$,
\begin{align*}
\Psi(\sigma_{\noisei},\theta) & \leq\Psi(\sigma_{\noisei},0)+\tfrac{\partial\Psi(\sigma_{\noisei},0)}{\partial\theta}\theta-\tfrac{1}{2}\theta^{2}\\
 & \leq\big(\sigma_{\noisei}+\tfrac{\E B^{2}}{2\sigma_{\noisei}\delta}\big)|\theta|-\tfrac{1}{2}\theta^{2},
\end{align*}
where in the last step we use $\Psi(\sigma_{\noisei},0)=0$ and $\frac{\partial\Psi(\sigma,0)}{\partial\theta}=\sigma_{\noisei}+\frac{\E B^{2}}{2\sigma_{\noisei}\delta}$,
which can be deduced from (\ref{eq:Psi_sigma_theta_def}) and (\ref{eq:partial_Phi_partial_theta}).
Then the level set $\{\theta\geq0\mid-\Psi(\sigma_{\noisei},\theta)\leq0\}\subset\Big[0,2\big(\sigma_{\noisei}+\tfrac{\E B^{2}}{2\sigma_{\noisei}\delta}\big)\Big]$
and it is non-empty (include at least one point $\theta=0$) and closed
since $\Psi(\sigma_{\noisei},\cdot)$ is a closed function. Letting
$\gamma_{2}=0$, we verify condition (ii).

Up to now, we have proved the existence and boundedness of saddle
points of $\Psi(\sigma,\theta)$. Next we prove the uniqueness. To
do this, it suffices to show the optimal solution of $\min_{\sigma\geq\sigma_{\noisei}}\max_{\theta\geq0}\Psi(\sigma,\theta)$
is bounded and unique, then the uniqueness of saddle points follows
due to the fact that each saddle point of $\Psi(\sigma,\theta)$ is
also an optimal solution of $\min_{\sigma\geq\sigma_{\noisei}}\max_{\theta\geq0}\Psi(\sigma,\theta)$
\cite[Proposition 3.4.1]{bertsekas2009convex}. First, we show that
any $\sigOpt$ should be bounded. Indeed, from the verification of
condition (ii) above, we know there exists $c>0$ and $K\geq\sigma_{\noisei}$
such that for any $\sigma\geq K$,
\begin{align*}
\max_{\theta\geq0}\Psi(\sigma,\theta) & \geq\Psi(\sigma,\bar{\theta})\geq\Psi(K,\bar{\theta})+c(\sigma-K),
\end{align*}
so we must have 
\[
\sigOpt\leq\underbrace{K+1+\tfrac{\max_{\sigma\in[\sigma_{\noisei},K]}\max_{\theta\geq0}\Psi(\sigma,\theta)-\Psi(K,\bar{\theta})}{c}}_{:=C_{1}},
\]
otherwise, $\min_{\sigma\geq\sigma_{\noisei}}\max_{\theta\geq0}\Psi(\sigma,\theta)>\min_{\sigma\in[\sigma_{\noisei},K]}\max_{\theta\geq0}\Psi(\sigma,\theta)$
leading to a contradiction. On the other hand, we can also show $\theOpt(\sigma):=\argmax{\theta\geq0}\Psi(\sigma,\theta)$
is uniformly bounded for $\sigma\in[\sigma_{\noisei},C_{1}]$. To
see this, note from (\ref{eq:partial_Phi_partial_theta}) 
\begin{align*}
\tfrac{\partial\Psi}{\partial\theta} & \leq-\theta+2\Big[\max_{\sigma\in[\sigma_{\noisei},C_{1}]}\sigma\big(1+\tfrac{1}{\delta}\big)+\tfrac{\E B^{2}+\sigma_{\noisei}^{2}\delta}{\sigma\delta}\Big],\\
\Rightarrow\theOpt(\sigma) & \leq\underbrace{2\Big[\max_{\sigma\in[\sigma_{\noisei},C_{1}]}\sigma\big(1+\tfrac{1}{\delta}\big)+\tfrac{\E B^{2}+\sigma_{\noisei}^{2}\delta}{\sigma\delta}\Big]}_{:=C_{2}},\,\forall\sigma\in[\sigma_{\noisei},C_{1}].
\end{align*}
As a result, $\max_{\theta\geq0}\Psi(\sigma,\theta)=\max_{\theta\in[0,C_{2}]}\Psi(\sigma,\theta)$
for $\sigma\in[\sigma_{\noisei},C_{1}]$. Therefore, by Berge Maximum
Theorem \cite[Theorem 17.31]{aliprantis06}, $\theOpt(\sigma)$ is
an upper hemicontinuous correspondence on $[\sigma_{\noisei},C_{1}]$.
By strong concavity of $\Psi(\sigma,\cdot)$, $\sigma\mapsto\theOpt(\sigma)$
is a function (\emph{i.e.}, single-valued correspondence). As a result,
one can easily check by definition that $\theOpt(\sigma)$ is continuous
on $[\sigma_{\noisei},C_{1}]$. Besides, we can get $\theOpt(\sigma)>0$
for any $\sigma\in[\sigma_{\noisei},C_{1}]$. Indeed from (\ref{eq:partial_Phi_partial_theta})
when $\P(\Lambda=0)<1$, $\frac{\partial\Psi(\sigma,0)}{\partial\theta}=\frac{1}{2}\Big(\frac{\sigma_{\noisei}^{2}}{\sigma}+\sigma\Big)+\frac{\E B^{2}}{2\sigma\delta}>0$
and when $\P(\Lambda=0)=1$, $\delta\geq1$, $\frac{\partial\Psi(\sigma,0)}{\partial\theta}\geq\tfrac{\sigma_{\noisei}^{2}}{2\sigma}+\tfrac{\sigma}{2}(1-\tfrac{1}{\delta})>0$.
Therefore, there exists $\theta_{\min}>0$ such that 
\begin{equation}
\theOpt(\sigma)\geq\theta_{\min},\label{eq:theta_star_lb}
\end{equation}
for any $\sigma\in[\sigma_{\noisei},C_{1}]$. Since $\theta^{*}(\sigma)\in[\theta_{\min},C_{2}]$
for any $\sigma\in[\sigma_{\noisei},C_{1}]$, we get $\max_{\theta\geq0}\Psi(\sigma,\theta)=\max_{\theta\in[\theta_{\min},C_{2}]}\Psi(\sigma,\theta)$
for any $\sigma\in[\sigma_{\noisei},C_{1}]$. On the other hand, $\Psi(\cdot,\theta)$
is $\tfrac{\sigma_{\noisei}^{2}\theta_{\min}}{C_{1}^{3}}$-strongly
convex on $[\sigma_{\noisei},C_{1}]$ for any fixed $\theta\in[\theta_{\min},C_{2}]$,
so we can check by definition that the function $\max_{\theta\in[\theta_{\min},C_{2}]}\Psi(\cdot,\theta)$
is also $\tfrac{\sigma_{\noisei}^{2}\theta_{\min}}{C_{1}^{3}}$-strongly
convex on $[\sigma_{\noisei},C_{1}]$. We conclude that $\sigma\mapsto\max_{\theta\geq0}\Psi(\sigma,\theta)$
is $\tfrac{\sigma_{\noisei}^{2}\theta_{\min}}{C_{1}^{3}}$-strongly
convex on $[\sigma_{\noisei},C_{1}]$, since $\max_{\theta\geq0}\Psi(\sigma,\theta)=\max_{\theta\in[\theta_{\min},C_{2}]}\Psi(\sigma,\theta)$.
Recall that any optimal solution $\sigOpt=\argmin{\sigma\geq\sigma_{\noisei}}\max_{\theta\geq0}\Psi(\sigma,\theta)$
should lie in $[\sigma_{\noisei},C_{1}]$, so the uniqueness holds.

Finally, we show $(\sigOpt,\theOpt)$ is a saddle point of $\Psi(\sigma,\theta)$
if and only if it is a solution of (\ref{eq:fixed_point_equation_sigmatheta}).
From (\ref{eq:partial_Phi_partial_sigma}) and (\ref{eq:partial_Phi_partial_theta}),
$\frac{\partial\Psi(\sigma_{\noisei},\theta)}{\partial\sigma}\leq0$
for any $\theta\geq0$ and $\frac{\partial\Psi(\sigma,0)}{\partial\theta}\geq0$
for any $\sigma\geq\sigma_{\noisei}$. Since $\Psi(\sigma,\theta)$
is convex-concave and continuously differentiable, by first order
optimality condition we know $(\sigOpt,\theOpt)$ is a saddle point
if and only if $\frac{\partial\Psi(\sigOpt,\theOpt)}{\partial\sigma}=\frac{\partial\Psi(\sigOpt,\theOpt)}{\partial\theta}=0$.
On the other hand, from (\ref{eq:Psi_sigma_theta_def}), (\ref{eq:F_deri1})
and (\ref{eq:F_deri2}) we can get
\begin{align}
\frac{\partial\Psi}{\partial\sigma} & =\frac{\theta}{2\sigma^{2}}\Big(\sigma^{2}-\sigma_{\noisei}^{2}-\frac{1}{\delta}\E\big(\eta(B+\sigma H;\mealim_{Y},\mealim_{\sigma\Lambda/\theta})-B\big)^{2}\Big),\label{eq:lim_partial_Phi_partial_sigma}\\
\frac{\partial\Psi}{\partial\theta} & =\frac{1}{2}\Big(\frac{\sigma_{\noisei}^{2}}{\sigma}+\sigma\Big)-\theta+\frac{1}{\delta}\Big[\frac{\E\big(\eta(B+\sigma H;\mealim_{Y},\mealim_{\sigma\Lambda/\theta})-B\big)^{2}}{2\sigma}-\sigma\E\eta'(B+\sigma H;\mealim_{Y},\mealim_{\sigma\Lambda/\theta})\Big].\label{eq:lim_partial_Phi_partial_theta}
\end{align}
Note that (\ref{eq:lim_partial_Phi_partial_sigma}) and (\ref{eq:lim_partial_Phi_partial_theta})
are actually the scalar representation of (\ref{eq:partial_Phi_partial_sigma})
and (\ref{eq:partial_Phi_partial_theta}). Setting the RHS of (\ref{eq:lim_partial_Phi_partial_sigma})
and (\ref{eq:lim_partial_Phi_partial_theta}) to be zero, we can get
(\ref{eq:fixed_point_equation_sigmatheta}).
\end{IEEEproof}

\subsection{Moreau Envelope of $\protect\regu$\label{subsec:Moreau-Envelope-lim}}

Recall that for $\tau>0$, the\emph{ }Moreau envelope of $\regu(\vx)$
is:
\begin{equation}
\mathcal{M}_{\vlambda}(\vy;\tau)=\min_{\vx}\frac{1}{2\tau}\|\vx-\vy\|^{2}+\regu(\vx)\label{eq:Moreau_f}
\end{equation}
and the corresponding optimal solution is the proximal operator $\tprox_{\tau\vlambda}(\vy)$.
In this section, we study the limiting behavior of the following function:
\begin{equation}
\begin{aligned}\Moreau_{p}(\sigma,\theta) & \bydef\frac{1}{p}\mathcal{M}_{\vlambda}(\sgl+\sigma\vh;\frac{\sigma}{\theta})\\
 & =\frac{1}{p}\min_{\vx}\frac{\theta}{2\sigma}\|\vx-(\sgl+\sigma\vh)\|^{2}+\regu(\vx),
\end{aligned}
\label{eq:Fp_ctau}
\end{equation}
where $(\sigma,\theta)\in\R_{>0}\times\R_{\geq0}$, $\vh\sim\mathcal{N}(\boldsymbol{0},\mI_{p})$
and $\sgl$ and $\vlambda$ are the same as in (\ref{eq:model}) and
(\ref{eq:slope_opt}).

\begin{lem}
\label{lem:conv_F} Consider a sequence of instances $\{\vh^{(p)},\sgl^{(p)},\vlambda^{(p)}\}_{p\in\mathbb{Z}^{+}}$,
where $\vh^{(p)}\sim\mathcal{N}(\boldsymbol{0},\mI_{p})$ are all
independent and $\{\sgl^{(p)}\}_{p\in\mathbb{Z}^{+}}$, $\{\vlambda^{(p)}\}_{p\in\mathbb{Z}^{+}}$
are both converging sequences with limiting measure $\mealim_{B}$
and $\mealim_{\Lambda}$. As $p\to\infty,$ for every $(\sigma,\theta)\in\R_{>0}\times\R_{\geq0}$,
\begin{equation}
\Moreau_{p}(\sigma,\theta)\asconv\Moreau(\sigma,\theta),\label{eq:F_asconv}
\end{equation}
and
\begin{equation}
\E_{\vh}\Moreau_{p}(\sigma,\theta)\to\Moreau(\sigma,\theta),\label{eq:F_lone_conv}
\end{equation}
where 
\begin{equation}
\Moreau(\sigma,\theta)\bydef\min_{g\in\Lipndset}\frac{\theta}{2\sigma}\E[Y-g(Y)]^{2}+\int_{0}^{1}F_{\Lambda}^{-1}(u)F_{|g(Y)|}^{-1}(u)du.\label{eq:F_lim_form}
\end{equation}
Here, $Y=B+\sigma H$, with $H\sim\mathcal{N}(0,1)$, $B\sim\mealim_{B}$
independent and $\eta(\cdot)$ is the limiting scalar function in
Proposition \ref{prop:prox}.
\end{lem}
\begin{IEEEproof}
For notational simplicity, we will omit the superscript ``$(p)$''
in $\vh^{(p)},\sgl^{(p)}$ and $\vlambda^{(p)}$. We first show the
empirical measure $\meajoint{\vh}{\sgl}$ converges almost surely
to $H\otimes B$ under Wasserstein-2 distance. Let $g(x,y)$ be a
bounded and continuous test function. By strong law of large number
for triangular array \cite[Theorem 2.1]{hu1997strong}, we can get
$\frac{1}{p}\sum_{i=1}^{p}[g(h_{i},\sgli_{i})-\E_{H}g(H,\sgli_{i})]\asconv0$.
For any $y$, $\E_{H}g(H,y)$ is bounded and it is not hard to show
$y\mapsto\E_{H}g(H,y)$ is continuous: we know $g(h,y)$ is uniformly
continuous over any compact set in $\R^{2}$, so for $C=\Phi^{-1}\big(1-\frac{\veps}{8\|g\|_{\infty}}\big)$
and any $y_{0}\in\R$, $\veps>0$, there exists $\delta>0$ such that
$|g(h,y)-g(h,y_{0})|\leq\frac{\veps}{2}$ whenever $|y-y_{0}|\leq\delta$
and $|h|\leq C$. Hence,
\[
\begin{aligned}|\E_{H}g(H,y)-\E_{H}g(H,y_{0})|\leq & \E_{H}\big[\indicatorfn_{|H|\leq C}|g(H,y)-g(H,y_{0})|\big]+2\|g\|_{\infty}\E_{H}(\indicatorfn_{|H|>C})\leq\veps\end{aligned}
.
\]
This shows the continuity of $y\mapsto\E_{H}[g(H,y)]$. Therefore,
$\frac{1}{p}\sum_{i=1}^{p}\E_{H}[g(H,\sgli_{i})]\to\E_{H,B}[g(H,B)]$
and we get for any bounded and continuous $g(x,y)$, $\frac{1}{p}\sum_{i=1}^{p}g(h_{i},\sgli_{i})\asconv\E_{H,B}[g(H,B)]$
indicating the almost sure weak convergence of $\meajoint{\vh}{\sgl}$
to $H\otimes B$. On the other hand, by strong law of large number
again, $\frac{1}{p}\sum_{i=1}^{p}h_{i}^{2}+\sgli_{i}^{2}\asconv1+\E B^{2}$.
Therefore, by Theorem 7.12 (iii) in \cite{villani2003topics}, 
\begin{equation}
W_{2}\big(\meajoint{\vh}{\sgl},H\otimes B\big)\asconv0.\label{eq:mu_B_H_W2_convergence}
\end{equation}
Then we can show $W_{2}\big(\meamarginal{\vy},\meamarginal Y\big)\asconv0$,
where $\vy:=\sgl+\sigma\vh$. Indeed,
\begin{align}
W_{2}\big(\meamarginal{\vy},\meamarginal Y\big)^{2}= & \inf_{\pi\in\Pi(\mu_{\vy},\mu_{Y})}\int(x-y)^{2}d\pi(x,y)\nonumber \\
= & \inf_{\pi\in\Pi(\meajoint{\vh}{\sgl},H\otimes B)}\int[(x_{2}+\sigma x_{1})-(y_{2}+\sigma y_{1})]^{2}d\pi(\vx,\vy)\nonumber \\
\leq & 2(\sigma^{2}+1)\inf_{\pi\in\Pi(\meajoint{\vh}{\sgl},H\otimes B)}\int\|\vx-\vy\|^{2}d\pi(\vx,\vy)\nonumber \\
= & 2(\sigma^{2}+1)W_{2}\big(\meajoint{\vh}{\sgl},H\otimes B\big)^{2}\label{eq:mu_y_W2_convergence}
\end{align}
and since $W_{2}\big(\meajoint{\vh}{\sgl},H\otimes B\big)\asconv0$,
we get $W_{2}\big(\meamarginal{\vy},\meamarginal Y\big)\asconv0$.
Similarly, for $\theta>0$ we can show $W_{2}\big(\meamarginal{\sigma\vlambda/\theta},\meamarginal{\sigma\Lambda/\theta}\big)\to0$
from our assumption that $W_{2}\big(\meamarginal{\vlambda},\meamarginal{\Lambda}\big)\to0$.

Now we can prove (\ref{eq:F_asconv}). For $\theta=0$, it directly
follows from (\ref{eq:Fp_ctau}) and (\ref{eq:F_lim_form}) that $\Moreau_{p}(\sigma,\theta)=\Moreau(\sigma,\theta)=0$.
For $\theta>0$, we have $W_{2}\big(\meamarginal{\vy},\meamarginal Y\big)\asconv0$
and $W_{2}\big(\meamarginal{\sigma\vlambda/\theta},\meamarginal{\sigma\Lambda/\theta}\big)\to0$.
Then from Proposition \ref{prop:prox}, we can get (\ref{eq:F_asconv})
by letting $\tau=\sigma/\theta$.

To prove (\ref{eq:F_lone_conv}), first observe that:
\begin{align*}
\Moreau_{p}(\sigma,\theta) & =\frac{\min_{\vx}\frac{\theta}{2\sigma}\|\sigma\vh+\sgl-\vx\|^{2}+\regu(\vx)}{p}\leq\theta\frac{\sigma^{2}\|\vh\|^{2}+\|\sgl\|^{2}}{\sigma p}.
\end{align*}
By strong law of large number (SLLN), we have 
\begin{align*}
\theta\frac{\sigma^{2}\|\vh\|^{2}+\|\sgl\|^{2}}{\sigma p} & \asconv\frac{\theta(\sigma^{2}+\E B^{2})}{\sigma}=\lim_{p\to\infty}\E_{\vh}\theta\frac{\sigma^{2}\|\vh\|^{2}+\|\sgl\|^{2}}{\sigma p}.
\end{align*}
In other words,
\[
\lim_{p\to\infty}\E_{\vh}\theta\frac{\sigma^{2}\|\vh\|^{2}+\|\sgl\|^{2}}{\sigma p}=\E\lim_{p\to\infty}\theta\frac{\sigma^{2}\|\vh\|^{2}+\|\sgl\|^{2}}{\sigma p}.
\]
Then by (\ref{eq:F_asconv}) and generalized dominated convergence
theorem (GDCT) (Theorem 19 in Sec. 4.4 of \cite{royden2010real}),
we have
\begin{equation}
\lim_{p\to\infty}\E\Moreau_{p}(\sigma,\theta)=\E\lim_{p\to\infty}\Moreau_{p}(\sigma,\theta),\label{eq:DCT3}
\end{equation}
which is exactly (\ref{eq:F_lone_conv}).
\end{IEEEproof}
\begin{lem}
\label{lem:F_sigma_theta_contdiff}$\Moreau(\sigma,\theta)$ is convex-concave
and continuously differentiable with respect to both $\sigma$ and
$\theta$ on $\R_{>0}\times\R_{\geq0}$, with partial derivatives:
\begin{align}
\frac{\partial\Moreau(\sigma,\theta)}{\partial\sigma} & =\frac{-\theta\E[B-\eta(Y)]^{2}}{2\sigma^{2}}+\frac{\theta}{2},\label{eq:F_deri1}\\
\frac{\partial\Moreau(\sigma,\theta)}{\partial\theta} & =\frac{\E\big[\eta(Y)-B\big]^{2}}{2\sigma}-\sigma\E\eta'(Y)+\frac{\sigma}{2},\label{eq:F_deri2}
\end{align}
where $B\sim\mealim_{B}$, $\Lambda\sim\mealim_{\Lambda}$, $Y=B+\sigma H$,
with $H\sim\calN(0,1)$ independent of $B$. Here, when $\theta>0$,
$\prox(\cdot)=\prox(\cdot;F_{Y},F_{\sigma\Lambda/\theta})$; when
$\theta=0$, we let $\prox(y)=0$, if $\P(\Lambda=0)<1$ and $\eta(y)=y$,
if $\P(\Lambda=0)=1$.
\end{lem}
\begin{IEEEproof}
When $\P(\Lambda=0)=1$, from (\ref{eq:F_lim_form}) we have $\Moreau(\sigma,\theta)=0$
for any $(\sigma,\theta)\in\R_{>0}\times\R_{\geq0}$, so $\frac{\partial\Moreau(\sigma,\theta)}{\partial\sigma}=\frac{\partial\Moreau(\sigma,\theta)}{\partial\theta}=0$.
In this case, all the results trivially hold. Therefore, it suffices
to consider $\P(\Lambda=0)<1$.

We first prove the convexity of $\Moreau(\sigma,\theta)$. Note $\Moreau_{p}(\sigma,\theta)$
{[}defined in (\ref{eq:Fp_ctau}){]} can be rewritten as:
\begin{equation}
\Moreau_{p}(\sigma,\theta)=\frac{1}{p}\min_{\vv}\frac{\sigma\theta}{2}\|\vv/\sigma-\vh\|^{2}+\regu(\vv+\sgl).\label{eq:Fp_ctau_1}
\end{equation}
Since $h(\vv)=\frac{\theta}{2}\|\vv-\vh\|^{2}$ is a convex function
(due to $\theta\geq0$), $(\sigma,\vv)\mapsto\frac{\sigma\theta}{2}\|\vv/\sigma-\vh\|^{2}$
is convex since it is the perspective function of $h(\vv)$. Therefore,
the objective function in (\ref{eq:Fp_ctau_1}) is jointly convex
w.r.t. $(\sigma,\vv)$ and after partial minimization over $\vv$,
$\Moreau_{p}(\cdot,\theta)$ is still convex. On the other hand, $\Moreau_{p}(\sigma,\theta)$
as a function of $\theta$ is the infimum of a family of linear functions
of $\theta$, so $\Moreau_{p}(\sigma,\cdot)$ is concave. Denote $\overline{\Moreau}_{p}(\sigma,\theta):=\E_{\vh}\Moreau_{p}(\sigma,\theta)$.
Clearly, $\overline{\Moreau}_{p}(\sigma,\theta)$ and $F(\sigma,\theta)$
are still convex-cancave, since taking expectation and limit preserves
convexity.

Then we show for fixed $\sigma\in\R_{>0}$, $\Moreau_{p}(\sigma,\cdot)$
is continuously differentiable on $\R_{\geq0}$. We follow the same
argument as Theorem 2.26 in \cite{rockafellar2009variational}. Denote
$\vy:=\sgl+\sigma\vh$ and 
\begin{equation}
g(\veps):=\Moreau_{p}(\sigma,\theta+\veps)-\Moreau_{p}(\sigma,\theta)-\tfrac{\|\vy-\tprox_{\sigma\vlambda/\theta}(\vy)\|^{2}}{2\sigma p}\veps,\label{eq:g_delta_def}
\end{equation}
where $\veps\geq-\theta$. On one hand, we have 
\begin{equation}
\begin{aligned}g(\veps) & \leq\tfrac{(\theta+\veps)\|\vy-\tprox_{\sigma\vlambda/\theta}(\vy)\|^{2}}{2\sigma p}+\tfrac{\regu\big(\tprox_{\sigma\vlambda/\theta}(\vy)\big)}{p}\\
 & \quad-\Big[\tfrac{\theta\|\vy-\tprox_{\sigma\vlambda/\theta}(\vy)\|^{2}}{2\sigma p}+\tfrac{\regu\big(\tprox_{\sigma\vlambda/\theta}(\vy)\big)}{p}\Big]-\tfrac{\|\vy-\tprox_{\sigma\vlambda/\theta}(\vy)\|^{2}}{2\sigma p}\veps=0
\end{aligned}
\label{eq:g_delta_ubd}
\end{equation}
On the other hand, 
\begin{align}
g(\veps) & \geq\tfrac{(\theta+\veps)\|\vy-\tprox_{\sigma\vlambda/(\theta+\veps)}(\vy)\|^{2}}{2\sigma p}+\tfrac{\regu\big(\tprox_{\sigma\vlambda/(\theta+\veps)}(\vy)\big)}{p}\nonumber \\
 & \quad-\Big[\tfrac{\theta\|\vy-\tprox_{\sigma\vlambda/(\theta+\veps)}(\vy)\|^{2}}{2\sigma p}+\tfrac{\regu\big(\tprox_{\sigma\vlambda/(\theta+\veps)}(\vy)\big)}{p}\Big]-\tfrac{\|\vy-\tprox_{\sigma\vlambda/\theta}(\vy)\|^{2}}{2\sigma p}\veps\nonumber \\
 & =\tfrac{\veps}{\sigma}\Big(\tfrac{\|\tprox_{\sigma\vlambda/(\theta+\veps)}(\vy)\|^{2}-\|\tprox_{\sigma\vlambda/\theta}(\vy)\|^{2}}{2p}\Big).\label{eq:g_delta_lbd}
\end{align}
From Lemma \ref{lem:partial_prox_finite} there exists $C>0$ such
that $\frac{1}{p}\left|\frac{\partial\|\tprox_{\sigma\vlambda/\theta}(\vy)\|^{2}}{\partial\theta}\right|\leq C$
for any $\theta\geq0$. Combined with (\ref{eq:g_delta_ubd}) and
(\ref{eq:g_delta_lbd}), this indicates $-\frac{C\veps^{2}}{2\sigma}\leq g(\veps)\leq0.$
Substituting the boundedness of $g(\veps)$ into (\ref{eq:g_delta_def})
yields for any $\theta\geq0$,
\begin{equation}
\tfrac{\partial\Moreau_{p}(\sigma,\theta)}{\partial\theta}=\tfrac{\|\vy-\tprox_{\sigma\vlambda/\theta}(\vy)\|^{2}}{2\sigma p}.\label{eq:par_Fp_theta}
\end{equation}
Also $\frac{\partial\Moreau_{p}(\sigma,\theta)}{\partial\theta}$
is continuous, which follows from (\ref{eq:par_Fp_theta}), (\ref{eq:par_proxnormsq_par_theta})
and (\ref{eq:par_proxinnerpd_par_theta}).

Following the same procedure as above, we can also show for fixed
$\theta\in\R_{\geq0}$, $\Moreau_{p}(\cdot,\theta)$ is continuously
differentiable on $\R_{>0}$ with
\begin{equation}
\tfrac{\partial\Moreau_{p}(\sigma,\theta)}{\partial\sigma}=-\tfrac{\theta\|\sgl-\tprox_{\sigma\vlambda/\theta}(\vy)\|^{2}}{2\sigma^{2}p}+\tfrac{\theta\|\vh\|^{2}}{2p}.\label{eq:par_Fp_sigma}
\end{equation}
From (\ref{eq:par_Fp_theta}) and (\ref{eq:par_Fp_sigma}), we can
get $\left|\frac{\partial\Moreau_{p}(\sigma,\theta)}{\partial\theta}\right|\leq\frac{4(\sigma^{2}\|\vh\|^{2}+\|\sgl\|^{2})}{\sigma p}$
and $\left|\frac{\partial\Moreau_{p}(\sigma,\theta)}{\partial\sigma}\right|\leq\frac{3\theta\|\sgl\|^{2}+\sigma^{2}(2+\theta)\|\vh\|^{2}}{\sigma^{2}p}.$
Since both bounds have finite expectation, by dominated convergence
theorem (DCT) we have: 
\begin{equation}
\tfrac{\partial\overline{\Moreau}_{p}(\sigma,\theta)}{\partial\theta}=\tfrac{\E\|\vy-\tprox_{\sigma\vlambda/\theta}(\vy)\|^{2}}{2\sigma p}\label{eq:par_EFp_theta}
\end{equation}
and 
\begin{equation}
\tfrac{\partial\overline{\Moreau}_{p}(\sigma,\theta)}{\partial\sigma}=\tfrac{-\theta\E\|\sgl-\tprox_{\sigma\vlambda/\theta}(\vy)\|^{2}}{2\sigma^{2}p}+\tfrac{\theta}{2}.\label{eq:par_EFp_sigma}
\end{equation}

We now show $\Moreau(\sigma,\cdot)$ and $\Moreau(\cdot,\theta)$
are continuously differentiable on $\R_{\geq0}$ and $\R_{>0}$, respectively.
In particular, we only present the proof for $\Moreau(\sigma,\cdot)$
and the case of $\Moreau(\cdot,\theta)$ can be derived following
same approach. The key is to establish the uniform convergence of
$\frac{\partial\overline{\Moreau}_{p}(\sigma,\theta)}{\partial\theta}$
to $\frac{\partial\overline{\Moreau}_{\infty}(\sigma,\theta)}{\partial\theta}$
on $\theta\geq0$, where $\frac{\partial\overline{\Moreau}_{\infty}(\sigma,\theta)}{\partial\theta}:=\lim_{p\to\infty}\tfrac{\partial\overline{\Moreau}_{p}(\sigma,\theta)}{\partial\theta}$.

First consider the case $\theta>0$. Let $[a,b]\subset\R_{>0}$ and
$\theta_{1},\theta_{2}\in[a,b]$. We have
\begin{align*}
\left|\tfrac{\partial\overline{\Moreau}_{p}(\sigma,\theta_{1})}{\partial\theta}-\tfrac{\partial\overline{\Moreau}_{p}(\sigma,\theta_{2})}{\partial\theta}\right| & \leq\E\left|\tfrac{\partial\Moreau_{p}(\sigma,\theta_{1})}{\partial\theta}-\tfrac{\partial\Moreau_{p}(\sigma,\theta_{2})}{\partial\theta}\right|\\
 & \tleq{\text{(a)}}\tfrac{1}{\sigma}\Big(\tfrac{1}{2p}\E\left|\tfrac{\partial\|\tprox_{\sigma\vlambda/\theta}(\vy)\|^{2}}{\partial\theta}\big|_{\theta=\theta'}\right|+\tfrac{1}{p}\E\left|\tfrac{\partial\vy^{\T}\tprox_{\sigma\vlambda/\theta}(\vy)}{\partial\theta}\big|_{\theta=\theta'}\right|\Big)\left|\theta_{1}-\theta_{2}\right|\\
 & \tleq{\text{(b)}}\tfrac{2}{\sigma\sqrt{p}}\tfrac{\E\|\vy\|}{a^{2}}\left|\theta_{1}-\theta_{2}\right|\\
 & \tleq{\text{(c)}}C\left|\theta_{1}-\theta_{2}\right|,
\end{align*}
where (a) follows from (\ref{eq:par_Fp_theta}) and intermediate value
theorem, with $\theta'\in[\theta_{1},\theta_{2}]$, (b) follows from
(\ref{eq:par_proxnormsq_par_theta}) and (\ref{eq:par_proxinnerpd_par_theta}),
and in (c), $C>0$ is some constant that only depends on $\sigma$,
$a$ and $b$. Therefore, $\tfrac{\partial\overline{\Moreau}_{p}(\sigma,\theta)}{\partial\theta}$
is $C$-Lipschitz continuous on $\theta\in[a,b]$ for any $p\in\mathbb{Z}^{+}$.
Meanwhile, we can show $\big\{\tfrac{\partial\overline{\Moreau}_{p}(\sigma,\theta)}{\partial\theta}\big\}_{p\in\mathbb{Z}^{+}}$
converges pointwise to $\frac{\partial\overline{\Moreau}_{\infty}(\sigma,\theta)}{\partial\theta}$
on $\theta\geq0$ following the proof of (\ref{eq:DCT3}). Then for
any $\veps>0$ we can construct a $\tfrac{\veps}{3C}$-epsilon net
$\mathcal{E}$ such that for any $\theta\in[a,b]$, there exists $z\in\mathcal{E}$
satisfying $|\theta-z|\leq\tfrac{\veps}{3C}$. Clearly, the cardinality
$|\mathcal{E}|<\infty$. Therefore, for any $\veps>0$ there exists
$p_{0}$ such that for any $m,p\geq p_{0}$, $\max_{z\in\mathcal{E}}\left|\tfrac{\partial\overline{\Moreau}_{m}(\sigma,z)}{\partial\theta}-\tfrac{\partial\overline{\Moreau}_{p}(\sigma,z)}{\partial\theta}\right|\leq\frac{\veps}{3}$.
Hence for any $\veps>0$, $\theta\in[a,b]$ and $m,p\geq p_{0}$,
\begin{align*}
\left|\tfrac{\partial\overline{\Moreau}_{m}(\sigma,\theta)}{\partial\theta}-\tfrac{\partial\overline{\Moreau}_{p}(\sigma,\theta)}{\partial\theta}\right| & \leq\left|\tfrac{\partial\overline{\Moreau}_{m}(\sigma,\theta)}{\partial\theta}-\tfrac{\partial\overline{\Moreau}_{m}(\sigma,z)}{\partial\theta}\right|+\left|\tfrac{\partial\overline{\Moreau}_{m}(\sigma,z)}{\partial\theta}-\tfrac{\partial\overline{\Moreau}_{p}(\sigma,z)}{\partial\theta}\right|+\left|\tfrac{\partial\overline{\Moreau}_{p}(\sigma,z)}{\partial\theta}-\tfrac{\partial\overline{\Moreau}_{p}(\sigma,\theta)}{\partial\theta}\right|\\
 & \leq C\times\tfrac{\veps}{3C}+\tfrac{\veps}{3}+C\times\tfrac{\veps}{3C}=\veps,
\end{align*}
which implies the uniform convergence of $\frac{\partial\overline{\Moreau}_{p}(\sigma,\theta)}{\partial\theta}$
when $\theta\in[a,b]$. Meanwhile, by DCT and continuity of $\frac{\partial\Moreau_{p}(\sigma,\theta)}{\partial\theta}$
for any fixed $\vh$, $\frac{\partial\overline{\Moreau}_{p}(\sigma,\theta)}{\partial\theta}$
is also continuous w.r.t. $\theta$. Then we can apply Theorem 7.12
in \cite{rudin1976principles} to obtain $\frac{\partial\overline{\Moreau}_{\infty}(\sigma,\theta)}{\partial\theta}$
is continuous on $[a,b]$. Since $[a,b]$ above can be arbitrary subset
of $\R_{>0},$ we actually obtain that $\frac{\partial\overline{\Moreau}_{\infty}(\sigma,\theta)}{\partial\theta}$
is continuous when $\theta>0$.

Next we handle the case when $\theta=0$. From (\ref{eq:par_EFp_theta}),
(\ref{eq:par_proxnormsq_par_theta}) and (\ref{eq:par_proxinnerpd_par_theta}),
we can get for any $\theta\geq0$
\[
\tfrac{\partial^{2}\overline{\Moreau}_{p}(\sigma,\theta)}{\partial\theta^{2}}=\tfrac{2}{\theta^{2}}\E\Big\{\sum_{i=1}^{p}\indicatorfn_{[\tprox_{\sigma\vlambda/\theta}(\vy)]_{i}\neq0}\big[[\tprox_{\sigma\vlambda/\theta}(|\vy|)]_{i}-|y_{i}|\big]\Big\}\leq0,
\]
where in the first equality we use DCT and the last inequality follows
from Fact \ref{fact:SLOPE_prox_properties}. Hence, $\frac{\partial\overline{\Moreau}_{p}(\sigma,\theta)}{\partial\theta}$
is non-increasing on $\theta\geq0$, with $\frac{\partial\overline{\Moreau}_{p}(\sigma,0)}{\partial\theta}=\E\frac{\|\vy\|^{2}}{2\sigma p}$
and $\lim_{\theta\to\infty}\frac{\partial\overline{\Moreau}_{p}(\sigma,\theta)}{\partial\theta}=0$.
Since $\frac{\partial\overline{\Moreau}_{p}(\sigma,\theta)}{\partial\theta}$
converges to $\frac{\partial\overline{\Moreau}_{\infty}(\sigma,\theta)}{\partial\theta}$
pointwise, $\frac{\partial\overline{\Moreau}_{\infty}(\sigma,\theta)}{\partial\theta}$
is also non-increasing with $\frac{\partial\overline{\Moreau}_{\infty}(\sigma,0)}{\partial\theta}=\frac{\E B^{2}+\sigma^{2}}{2\sigma}$
and $\lim_{\theta\to\infty}\frac{\partial\overline{\Moreau}_{\infty}(\sigma,\theta)}{\partial\theta}=0$.
On the other hand, from (\ref{eq:par_EFp_theta})
\begin{align*}
\tfrac{\partial\overline{\Moreau}_{p}(\sigma,\theta)}{\partial\theta} & \geq\E\Big(\tfrac{\|\vy\|^{2}-2\|\vy\|\|\tprox_{\sigma\vlambda/\theta}(\vy)\|}{2\sigma p}\Big)\\
 & \tgeq{\text{(a)}}\E\Big(\tfrac{\|\vy\|^{2}-2\|\vy\|\|\tprox_{\sigma\overline{\vlambda}/\theta}(\vy)\|}{2\sigma p}\Big)\\
 & \geq\tfrac{\E\|\vy\|^{2}}{2\sigma p}-\tfrac{1}{\sigma}\Big(\tfrac{\E\|\vy\|^{2}}{p}\Big)^{1/2}\Big[\tfrac{1}{p}\sum_{i=1}^{p}\E(|y_{i}|-\tfrac{\sigma}{\theta}\bar{\lambda})_{+}^{2}\Big]^{1/2},
\end{align*}
where step (a) follows from (\ref{eq:prox_norm_bd_by_LASSO}), with
$\overline{\vlambda}=\frac{\sum_{i=1}^{p}\lambda_{i}}{p}\boldsymbol{1}_{p}$.
Hence taking $p\to\infty$ on both sides above, we have
\begin{align*}
\tfrac{\partial\overline{\Moreau}_{\infty}(\sigma,\theta)}{\partial\theta} & \tgeq{\text{}}\tfrac{1}{2\sigma}\Big[\E B^{2}+\sigma^{2}-2\sqrt{(\E B^{2}+\sigma^{2})\E(|Y|-\tfrac{\sigma}{\theta}\E\Lambda)_{+}^{2}}\Big].
\end{align*}
Then taking $\theta\to0^{+}$, we get 
\[
\begin{aligned}\lim_{\theta\to0^{+}}\tfrac{\partial\overline{\Moreau}_{\infty}(\sigma,\theta)}{\partial\theta} & \geq\lim_{\theta\to0^{+}}\tfrac{1}{2\sigma}\Big[\E B^{2}+\sigma^{2}-\tfrac{1}{\sigma}\sqrt{(\E B^{2}+\sigma^{2})\E(|Y|-\tfrac{\sigma}{\theta}\E\Lambda)_{+}^{2}}\Big]\\
 & \teq{\text{(a)}}\tfrac{\E B^{2}+\sigma^{2}}{2\sigma}=\tfrac{\partial\overline{\Moreau}_{\infty}(\sigma,0)}{\partial\theta},
\end{aligned}
\]
where in step (a) we use DCT and $\E\Lambda>0$. Since $\frac{\partial\overline{\Moreau}_{\infty}(\sigma,\theta)}{\partial\theta}$
is non-increasing on $\theta\geq0$, we can get $\lim_{\theta\to0^{+}}\frac{\partial\overline{\Moreau}_{\infty}(\sigma,\theta)}{\partial\theta}=\frac{\partial\overline{\Moreau}_{\infty}(\sigma,0)}{\partial\theta}.$
Recall that we have already proved $\frac{\partial\overline{\Moreau}_{\infty}(\sigma,\theta)}{\partial\theta}$
is continuous when $\theta>0$, so $\frac{\partial\overline{\Moreau}_{\infty}(\sigma,\theta)}{\partial\theta}$
is continuous on $\R_{\geq0}$.

Up to now, we have shown $\Big\{\frac{\partial\overline{\Moreau}_{p}(\sigma,\theta)}{\partial\theta}\Big\}_{p\in\mathbb{Z}^{+}}$,
$\frac{\partial\overline{\Moreau}_{\infty}(\sigma,\theta)}{\partial\theta}$
are all bounded, non-increasing and continuous functions on $\R_{\geq0}$
and also $\Big\{\frac{\partial\overline{\Moreau}_{p}(\sigma,\theta)}{\partial\theta}\Big\}_{p\in\mathbb{Z}^{+}}$
converges to $\frac{\partial\overline{\Moreau}_{\infty}(\sigma,\theta)}{\partial\theta}$
pointwise on $\theta\geq0$. Therefore, following the proof of Glivenko-Cantelli
theorem (for a reference, see Theorem 19.1 in \cite{van2000asymptotic})
we can show $\frac{\partial\overline{\Moreau}_{p}(\sigma,\theta)}{\partial\theta}\to\frac{\partial\overline{\Moreau}_{\infty}(\sigma,\theta)}{\partial\theta}$
converges uniformly on $\theta\geq0$ . In (\ref{eq:F_lone_conv})
we prove the pointwise convergence $\overline{\Moreau}_{p}(\sigma,\theta)\to\Moreau(\sigma,\theta)$.
Then using Theorem 7.17 and 7.12 in \cite{rudin1976principles} together
with (\ref{eq:par_EFp_theta}), we get $\Moreau(\sigma,\cdot)$ is
continuously differentiable on $\R_{\geq0}$, with 
\begin{align}
\tfrac{\partial\Moreau(\sigma,\theta)}{\partial\theta} & =\lim_{p\to\infty}\tfrac{\E\|\vy-\tprox_{\sigma\vlambda/\theta}(\vy)\|^{2}}{2\sigma p},\label{eq:lim_par_EFp_theta}
\end{align}
where $\vy=\sgl+\sigma\vh$. Repeating the same procedure, we can
also prove that $\Moreau(\cdot,\theta)$ is continuously differentiable
on $\R_{>0}$, with 
\begin{equation}
\tfrac{\partial\Moreau(\sigma,\theta)}{\partial\sigma}=\lim_{p\to\infty}\tfrac{-\theta\E\|\sgl-\tprox_{\sigma\vlambda/\theta}(\vy)\|^{2}}{2\sigma^{2}p}+\tfrac{\theta}{2}.\label{eq:lim_par_EFp_sigma}
\end{equation}

Finally, we compute the limit in (\ref{eq:lim_par_EFp_theta}) and
(\ref{eq:lim_partial_Phi_partial_sigma}). From Proposition \ref{prop:prox}
and (\ref{eq:mu_y_W2_convergence}), we have $\tfrac{\|\tprox_{\sigma\vlambda/\theta}(\vy)-\eta(\vy)\|_{2}^{2}}{p}\asconv0$
as $p\to\infty$, where $\prox:=\prox(\cdot;\mealim_{B+\sigma H},\mealim_{\sigma\Lambda/\tau})$.
In addition,
\[
\begin{aligned}\tfrac{1}{p}\left|\|\tprox_{\sigma\vlambda/\theta}(\vy)-\sgl\|_{2}^{2}-\|\eta(\vy)-\sgl\|^{2}\right| & \leq\tfrac{2}{p}\|\tprox_{\sigma\vlambda/\theta}(\vy)-\eta(\vy)\|(\|\vy\|+\|\sgl\|),\end{aligned}
\]
so we can get
\begin{equation}
\tfrac{1}{p}\|\tprox_{\sigma\vlambda/\theta}(\vy)-\sgl\|^{2}\asconv\tfrac{1}{p}\|\eta(\vy)-\sgl\|^{2}\asconv\E[\eta(Y)-B]^{2},\label{eq:fixedequa_lim1_as}
\end{equation}
where the last step follows from Theorem 7.12 (iv) in \cite{villani2003topics}
after combining (\ref{eq:mu_B_H_W2_convergence}) with the fact that
$[\prox(b+\sigma h)-b]^{2}\leq C(1+h^{2}+b^{2})$ for any $h,b\in\R$
and some $C>0$. Then similar as (\ref{eq:DCT3}), we can obtain
\begin{equation}
\tfrac{1}{p}\E\|\tprox_{\sigma\vlambda/\theta}(\vy)-\sgl\|^{2}\to\E[\eta(Y)-B]^{2}.\label{eq:fixedequa_lim1}
\end{equation}
On the other hand, we can also get
\begin{equation}
\tfrac{1}{p}\vh^{\T}\tprox_{\sigma\vlambda/\theta}(\vy)\asconv\tfrac{1}{p}\vh^{\T}\eta(\vy)\asconv\E[\eta(Y)H]\label{eq:fixedequa_lim2_as}
\end{equation}
and
\begin{align}
\tfrac{1}{p}\E[\vh^{\T}\tprox_{\sigma\vlambda/\theta}(\vy)] & \to\E[\eta(Y)H]=\sigma\E\eta'(Y).\label{eq:fixedequa_lim2}
\end{align}
where in the last step we use Stein's lemma. Substituting (\ref{eq:fixedequa_lim1})
and (\ref{eq:fixedequa_lim2}) into (\ref{eq:lim_par_EFp_theta})
and (\ref{eq:lim_par_EFp_sigma}), we reach at (\ref{eq:F_deri1})
and (\ref{eq:F_deri2}).
\end{IEEEproof}
\begin{lem}
\label{lem:partial_prox_finite}For any $\theta\geq0$, $\vy\in\R^{p}$
and $\vlambda\in\R_{\geq0}^{p}$, we have
\begin{equation}
\tfrac{\partial\|\tprox_{\sigma\vlambda/\theta}(\vy)\|^{2}}{\partial\theta}=\begin{cases}
0 & \vlambda=\boldsymbol{0},\\
\tfrac{2\cdot\boldsymbol{1}^{\T}\tprox_{\sigma\vlambda/\theta}(|\vy|)}{\theta^{2}} & \vlambda\neq\boldsymbol{0},
\end{cases}\label{eq:par_proxnormsq_par_theta}
\end{equation}
and
\begin{equation}
\tfrac{\partial\vy^{\T}\tprox_{\sigma\vlambda/\theta}(\vy)}{\partial\theta}=\begin{cases}
0 & \vlambda=\boldsymbol{0},\\
\tfrac{1}{\theta^{2}}\sum_{i=1}^{p}|y_{i}|\indicatorfn_{[\tprox_{\sigma\vlambda/\theta}(\vy)]_{i}\neq0} & \vlambda\neq\boldsymbol{0}.
\end{cases}.\label{eq:par_proxinnerpd_par_theta}
\end{equation}
Here when $\vlambda\neq\boldsymbol{0}$ and $\theta=0$, we let $\tprox_{\sigma\vlambda/\theta}(\vy):=\boldsymbol{0}$.
\end{lem}
\begin{IEEEproof}
When $\vlambda=\boldsymbol{0}$, we have $\tprox_{\sigma\vlambda/\theta}(\vy)=\vy$,
so $\|\tprox_{\sigma\vlambda/\theta}(\vy)\|^{2}=\vy^{\T}\tprox_{\sigma\vlambda/\theta}(\vy)=\|\vy\|^{2}$
and $\tfrac{\partial\|\tprox_{\sigma\vlambda/\theta}(\vy)\|^{2}}{\partial\theta}=\tfrac{\partial\vy^{\T}\tprox_{\sigma\vlambda/\theta}(\vy)}{\partial\theta}=0$.

Next we consider $\vlambda\neq\boldsymbol{0}$. From Lemma 2.3 and
2.4 of \cite{bogdan2013statistical}, for any $\va\in\R^{p}$ satisfying
$0\leq a_{1}\leq a_{2}\leq\cdots\leq a_{p}$,
\[
\tfrac{\partial[\tprox_{\vlambda}(\va)]_{i}}{\partial\lambda_{j}}=-\tfrac{\indicatorfn_{i\in I_{j}}}{\max\{|I_{j}|,1\}},
\]
where $I_{j}\bydef\big\{ k\in[p]\mid|[\tprox_{\vlambda}(\va)]_{k}|=|[\tprox_{\vlambda}(\va)]_{j}|\text{ and }[\tprox_{\vlambda}(\va)]_{k}\neq0\big\}$.
Therefore,
\begin{equation}
\begin{aligned}\tfrac{\partial\|\tprox_{\vlambda}(\va)\|^{2}}{\partial\lambda_{j}} & =\sum_{i=1}^{p}2[\tprox_{\vlambda}(\va)]_{i}\tfrac{\partial[\tprox_{\vlambda}(\va)]_{i}}{\partial\lambda_{j}}=-2[\tprox_{\vlambda}(\va)]_{j}\end{aligned}
\label{eq:par_proxnorm2_par_lam}
\end{equation}
and 
\begin{equation}
\begin{aligned}\tfrac{\partial\va^{\T}\tprox_{\vlambda}(\va)}{\partial\lambda_{j}} & =\sum_{i=1}^{p}a_{i}\tfrac{\partial[\tprox_{\vlambda}(\va)]_{i}}{\partial\lambda_{j}}=-\tfrac{1}{\max\{|I_{j}|,1\}}\sum_{i\in I_{j}}a_{i}.\end{aligned}
\label{eq:par_yprox_par_lam}
\end{equation}
On the other hand, by Fact \ref{fact:SLOPE_prox_properties}, $\|\tprox_{\vlambda}(\va)\|^{2}$
and $\va^{\T}\tprox_{\vlambda}(\va)$ only depend on $\mealim_{\vlambda}$
and $\mealim_{|\va|}$. Therefore, for any $\vy\in\R^{p}$ and $\theta>0,$
it holds that $\frac{\partial\|\tprox_{\sigma\vlambda/\theta}(\vy)\|^{2}}{\partial\theta}=\frac{2}{\theta^{2}}\boldsymbol{1}^{\T}\tprox_{\sigma\vlambda/\theta}(|\vy|)$
and $\frac{\partial\vy^{\T}\tprox_{\sigma\vlambda/\theta}(\vy)}{\partial\theta}=\frac{1}{\theta^{2}}\sum_{i=1}^{p}|y_{i}|\indicatorfn_{[\tprox_{\sigma\vlambda/\theta}(\vy)]_{i}\neq0}$
by chain rule. For $\theta=0$, we need to study the behavior of $\|\tprox_{\sigma\vlambda/\theta}(\vy)\|$
when $\theta$ is closed to 0. It can be shown (for a proof, see (A.11)
in \cite{su2016slope})
\begin{equation}
\tprox_{\vlambda}(\vy)=\vy-\text{Proj}_{\dualball}(\vy),\label{eq:prox_as_projection}
\end{equation}
where $\dualball=\{\boldsymbol{\nu}\in\R^{p}\mid\boldsymbol{\nu}\majorized\vlambda\}$
(``$\majorized$'' denotes majorization, see Definition \ref{def:majorization})
is the unit ball of the dual norm of $\regu$ \cite[Proposition 1.1]{bogdan2013statistical}
and $\text{Proj}_{C_{\vlambda}}$ is the orthogonal projection onto
$\dualball$. Take $\overline{\lambda}=\frac{1}{p}\sum_{i=1}^{p}\lambda_{i}$
and it is not hard to show $\overline{\vlambda}:=\overline{\lambda}\boldsymbol{1}_{p}$
satisfies $\overline{\vlambda}\majorized\vlambda$. Clearly, $\frac{\sigma}{\theta}\overline{\vlambda}\majorized\frac{\sigma}{\theta}\vlambda$
for $\sigma,\theta>0$ and $C_{\sigma\overline{\vlambda}/\theta}\subset C_{\sigma\vlambda/\theta}$,
so from (\ref{eq:prox_as_projection}) we have
\begin{align}
\|\tprox_{\sigma\vlambda/\theta}(\vy)\|^{2} & \leq\|\tprox_{\sigma\overline{\vlambda}/\theta}(\vy)\|^{2}=\sum_{i=1}^{p}\max\big\{|y_{i}|-\tfrac{\sigma\bar{\lambda}}{\theta},0\big\}^{2}.\label{eq:prox_norm_bd_by_LASSO}
\end{align}
Since $\overline{\lambda}>0$ (due to $\vlambda\neq\boldsymbol{0}$),
(\ref{eq:prox_norm_bd_by_LASSO}) indicates that when $0<\theta\leq\frac{\sigma\overline{\lambda}}{\max(\{|y_{i}|\},1)}$,
$\tprox_{\sigma\vlambda/\theta}(\vy)=\boldsymbol{0}$. On the other
hand, we let $\tprox_{\sigma\vlambda/\theta}(\vy):=\boldsymbol{0}$,
when $\vlambda\neq\boldsymbol{0}$ and $\theta=0$. As a result, $\tprox_{\sigma\vlambda/\theta}(\vy)=\boldsymbol{0}$
for $\theta\in\big[0,\frac{\sigma\overline{\lambda}}{\max(\{|y_{i}|\},1)}\big]$
and combining the partial derivatives of $\|\tprox_{\sigma\vlambda/\theta}(\vy)\|^{2}$
and $\vy^{\T}\tprox_{\sigma\vlambda/\theta}(\vy)$ on $\theta>0$
obtained above, we can get (\ref{eq:par_proxnormsq_par_theta}) and
(\ref{eq:par_proxinnerpd_par_theta}).
\end{IEEEproof}

\subsection{Proof of Proposition \ref{prop:function_space}\label{subsec:Proof-of-Proposition-functionspace}}

It directly follows from Proposition \ref{prop:prox} that $\Realizableset_{\mealim_{Y}}\subseteq\Lipndset$,
so we just need to show $\Lipndset\subseteq\Realizableset_{\mealim_{Y}}$.

For any $\proxfree\in\Lipndset$, consider the function $r(y)=y-\proxfree(y)$.
It can be easily verified that $r(y)\in\Lipndset$. We claim that
if we choose $\Lambda\sim r(|Y|)$ with $Y\sim\mealim_{Y}$, then
$\proxfree$ is the optimal solution of (\ref{eq:infinite_SLOPE})
(when $\tau=1$). Indeed, when $\tau=1$ and $\Lambda\sim r(|Y|)$,
(\ref{eq:infinite_SLOPE}) can be equivalently written as
\begin{align}
\text{Problem }\eqref{eq:infinite_SLOPE}\teq{\text{(a)}} & \min_{g\in\Lipndset}\frac{1}{2}\E_{\mu_{Y}}[|Y|-g(|Y|)]^{2}+\int_{0}^{1}F_{r(|Y|)}^{-1}(u)F_{g(|Y|)}^{-1}(u)du\nonumber \\
\teq{\text{(b)}} & \min_{g\in\Lipndset}\frac{1}{2}\E_{\mu_{Y}}[|Y|-g(|Y|)]^{2}+\E_{\mu_{Y}}[|Y|-\proxfree(|Y|)]g(|Y|)\nonumber \\
= & \min_{g\in\Lipndset}\frac{1}{2}\E_{\mu_{Y}}[\proxfree(|Y|)-g(|Y|)]^{2}+\frac{1}{2}\E_{\mu_{Y}}[Y^{2}-\proxfree^{2}(|Y|)],\label{eq:opti_lambda_problem_1}
\end{align}
where (a) follows from $g\in\Lipndset$ and in (b) we substitute $r(y)=y-\proxfree(y)$
and use the fact that $r\in\Lipndset$. From (\ref{eq:opti_lambda_problem_1}),
we can see $\proxfree$ is the optimal solution of (\ref{eq:infinite_SLOPE}).
On the other hand, since $f\in\Lipndset$ and $\E Y^{2}<\infty$,
we can verify that $\E\Lambda^{2}<\infty$ and hence $\Lambda\in\Wspace(\R)$.
In conclusion, for any $\proxfree\in\Lipndset$, we can always choose
$\Lambda\sim|Y|-\proxfree(|Y|)$ satisfying $\Lambda\in\Wspace(\R)$,
such that $\proxfree$ is the optimal solution of (\ref{eq:infinite_SLOPE})
(when $\tau=1$). By Proposition \ref{prop:prox}, this means $f(y)=\prox(y;\mealim_{Y},\mealim_{\Lambda})$
and hence $\proxfree\in\Realizableset_{\mealim_{Y}}$. Therefore,
$\Lipndset\subseteq\Realizableset_{\mealim_{Y}}$.

\subsection{Auxiliary Results for Proving Proposition \ref{prop:min_MSE}\label{subsec:Auxiliary-Results-for-estimation}}
\begin{lem}
\label{lem:f_sigma_unique_Lsigma_conti}For any $\sigma>0$, we have:
(I) optimization problem (\ref{eq:opt_esti_prox}) is convex and always
has a unique optimal solution $\proxfree_{\sigma}\in\Lipndset$, (II)
${\cal L}(\sigma)$ defined in (\ref{eq:opt_esti_prox}) is continuous
at $\sigma$.
\end{lem}
\begin{IEEEproof}
(I) Optimization problem (\ref{eq:opt_esti_prox}) can be equivalently
written as:
\[
\begin{aligned}\min_{\proxfree\in\Lipndset} & \E_{\mealim_{Y}}[\proxfree(Y)-\E(B\mid Y)]^{2}+\E_{\mealim_{Y}}[\text{Var}(B\mid Y)]\\
\st & \E_{\mealim_{Y}}\proxfree'(Y)\leq\delta.
\end{aligned}
\]
Then by the same arguments in the proof of Lemma \ref{lem:gstar_uniqueness},
it is not hard to check for any $\sigma>0$, it is (strongly) convex
and has a unique solution $\proxfree_{\sigma}\in\Lipndset$.

(II) Next, we prove the continuity of ${\cal L}(\sigma)$ at any $\sigma>0$.
Define the following set
\begin{equation}
\Lipndset_{\sigma}\bydef\{\proxfree\mid f\in\Lipndset\text{ and }\E\proxfree^{\prime}(B+\sigma H)\leq\delta\},\label{eq:set_Lipbd_sigma}
\end{equation}
where $H\sim\mathcal{N}(0,1)$ and $B\sim\mealim_{B}$ are independent.
Note that for any $\sigma>0$, we have $\Lipndset_{\sigma}\neq\emptyset$,
since $\{f=0\}\in\Lipndset_{\sigma}$.

The first step is to show for any $\sigma,r>0$, there exists $\veps\in(0,\sigma/2)$
such that whenever $\h{\sigma}\in\bdball_{\veps}(\sigma)$ and $\h{\proxfree}\in\Lipndset_{\hat{\sigma}}$,
we can always find a $\proxfree\in\Lipndset_{\sigma}$ satisfying
$|\proxfree(x)-\h{\proxfree}(x)|<r$ almost everywhere on $\R$. This
can be proved as follows. If $\h{\proxfree}\in\Lipndset_{\sigma}$,
we can choose $\proxfree=\h{\proxfree}$, which trivially satisfies
$|\proxfree(x)-\h{\proxfree}(x)|<r$; if $\h{\proxfree}\notin\Lipndset_{\sigma}$,
then $\E\h{\proxfree}'(B+\sigma H)>\delta$. Since $\hat{\proxfree}\in\Lipndset_{\hat{\sigma}}\subseteq\Lipndset$,
it follows that $|\h{\proxfree}'|\leq1$ almost everywhere on $\R$.
Meanwhile, since $\sigma,\h{\sigma}>0$, by the properties of Gaussian
convolution we know both $B+\hat{\sigma}H$ and $B+\sigma H$ have
smooth density functions supported on $\R$. Denote their density
functions as $q_{1}$ and $q_{2}$. Then we have 
\begin{align}
|\E\h{\proxfree}'(B+\h{\sigma}H)-\E\h{\proxfree}'(B+\sigma H)|= & \big|\int\h{\proxfree}'(y)[q_{1}(y)-q_{2}(y)]dy\big|\nonumber \\
\leq & \int|q_{1}(y)-q_{2}(y)|dy\nonumber \\
= & 2\text{TV}\big(\mealim_{B+\h{\sigma}H},\mealim_{B+\sigma H}\big)\nonumber \\
\tleq{\text{(a)}} & 2\text{TV}\big(\mealim_{\h{\sigma}H},\mealim_{\sigma H}\big)\nonumber \\
\tleq{\text{(b)}} & \sqrt{2\text{KL}(\mealim_{\h{\sigma}H},\mealim_{\sigma H})}\nonumber \\
\tleq{\text{(c)}} & C|\h{\sigma}-\sigma|,\label{eq:prox_hat_deri_diff_0}
\end{align}
where $\text{TV}(\cdot,\cdot)$ and $\text{KL}(\cdot,\cdot)$ denote
the total variation distance and KL-divergence between two probability
measures and $C>0$ is a fixed constant only depending on $\sigma$.
In reaching (\ref{eq:prox_hat_deri_diff_0}), (b) follows from Pinsker's
inequality and (c) follows from standard results of KL-divergence
between Gaussian random variables and the fact that $\veps\in(0,\sigma/2)$
and $\h{\sigma}\in\bdball_{\veps}(\sigma)$. Step (a) in (\ref{eq:prox_hat_deri_diff_0})
can be obtained as follows. Recall that $\Pi(\mealim_{1},\mealim_{2})$
denotes the set of all couplings between measures $\mealim_{1}$ and
$\mealim_{2}$. Then for any $(\h{\sigma}H_{1},\sigma H_{2})\in\Pi(\mealim_{\h{\sigma}H},\mealim_{\sigma H})$
and $B_{0}\sim\mealim_{B}$ independent of $(\h{\sigma}H_{1},\sigma H_{2})$,
we have $(B_{0}+\h{\sigma}H_{1},B_{0}+\sigma H_{2})\in\Pi(\mealim_{B+\h{\sigma}H},\mealim_{B+\sigma H})$.
Therefore, 
\begin{align}
\text{TV}\big(\mealim_{B+\h{\sigma}H},\mealim_{B+\sigma H}\big)\teq{\text{(a)}} & \inf_{\substack{(Y_{1},Y_{2})\in\Pi(\mealim_{B+\h{\sigma}H},\mealim_{B+\sigma H})}
}\P(Y_{1}\neq Y_{2})\nonumber \\
\leq & \inf_{\substack{(\h{\sigma}H_{1},\sigma H_{2})\in\Pi(\mealim_{\h{\sigma}H},\mealim_{\sigma H})\\
B_{0}\sim\mealim_{B}\text{ indep. of }(\h{\sigma}H_{1},\sigma H_{2})
}
}\P(B_{0}+\h{\sigma}H_{1}\neq B_{0}+\sigma H_{2})\nonumber \\
= & \inf_{\substack{(\h{\sigma}H_{1},\sigma H_{2})\in\Pi(\mealim_{\h{\sigma}H},\mealim_{\sigma H})}
}\P(\h{\sigma}H_{1}\neq\sigma H_{2})\nonumber \\
\teq{\text{(b)}} & \text{ TV}\big(\mealim_{\h{\sigma}H},\mealim_{\sigma H}\big),\label{eq:TV_bound_1}
\end{align}
where (a) and (b) follow from Strassen's Theorem \cite[p.7]{villani2003topics}.
From (\ref{eq:prox_hat_deri_diff_0}), when $\veps\in(0,\sigma/2)$
and $\h{\sigma}\in\bdball_{\veps}(\sigma)$,
\begin{equation}
\begin{aligned}|\E\h{\proxfree}'(B+\h{\sigma}H)-\E\h{\proxfree}'(B+\sigma H)| & \tleq{}C\veps.\end{aligned}
\label{eq:prox_hat_deri_diff}
\end{equation}
Since $\h{\proxfree}\in\Lipndset_{\hat{\sigma}}$, from (\ref{eq:set_Lipbd_sigma})
and (\ref{eq:prox_hat_deri_diff}) we get
\begin{equation}
\E\h{\proxfree}'(B+\sigma H)\leq\delta+C\veps.\label{eq:f_deri_aver_bd_1}
\end{equation}
We now slightly shrink $\h{\proxfree}$ to obtain the $\proxfree\in\Lipndset_{\sigma}$
satisfying $|\proxfree(x)-\h{\proxfree}(x)|<r$ almost everywhere.
On one hand, we know $B+\sigma H$ has a density function supported
on $\R$ for $\sigma>0$. On the other hand, since $\E\h{\proxfree}'(B+\sigma H)>\delta$
(due to $\h{\proxfree}\notin\Lipndset_{\sigma}$) and $|\h{\proxfree}'|\leq1$
almost everywhere, it is not hard to show if $\veps\leq\frac{\delta}{2C}$,
there exists some $\mathcal{A}\subset\R_{>0}$ such that $\E[\h{\proxfree}'(Y)\indicatorfn_{Y\in\mathcal{A}}]=C\veps$.
Accordingly, we set 
\[
\proxfree'(y)=\begin{cases}
0 & \pm y\in{\cal A},\\
\h{\proxfree}'(y) & \otherwise
\end{cases}
\]
and choose $\proxfree(y)$ to be:
\begin{equation}
\proxfree(y)=\int_{0}^{y}\proxfree'(t)dt.\label{eq:shrink_y_2}
\end{equation}
With this choice, from (\ref{eq:f_deri_aver_bd_1}) we have $\E f'(Y)\leq\delta-C\veps$
and thus $\proxfree\in\Lipndset_{\sigma}$. In addition, $0\leq\h{\proxfree}(x)-f(x)\leq2C\veps$
almost everywhere. Hence we can choose $\veps\leq\frac{r}{3C}$ so
that $0\leq\h{\proxfree}(x)-f(x)<r$ almost everywhere. Summing up,
for any $\sigma,r>0$, there exists $C$ depending only on $\sigma$
such that whenever $\veps\leq\min\{\frac{r}{3C},\tfrac{\sigma}{2}\}$,
$\h{\sigma}\in\bdball_{\veps}(\sigma)$ and $\h{\proxfree}\in\Lipndset_{\hat{\sigma}}$,
we can always find a $\proxfree\in\Lipndset_{\sigma}$ satisfying
$0\leq\h{\proxfree}(x)-f(x)<r$ almost everywhere on $\R$.

Define ${\cal L}(\proxfree,\sigma)\bydef\E[\proxfree(B+\varsigma H)-B]^{2}$,
which is the objective function in (\ref{eq:opt_esti_prox}). For
any compact interval $\minsigmaI\subseteq\R_{>0}$ and any $\sigma_{1},\sigma_{2}\in\minsigmaI$,
$\proxfree\in\Lipndset$, we have
\begin{align}
|\calL(\proxfree,\sigma_{1})-\calL(\proxfree,\sigma_{2})| & \leq\E\left|\proxfree^{2}(B+\sigma_{1}H)-\proxfree^{2}(B+\sigma_{2}H)\right|\nonumber \\
 & \quad+2\E|B||\proxfree(B+\sigma_{1}H)-\proxfree(B+\sigma_{2}H)|\nonumber \\
 & \leq\big(\sqrt{\E(|B+\sigma_{1}H|+|B+\sigma_{2}H|)^{2}}+2\sqrt{\E B^{2}}\big)|\sigma_{1}-\sigma_{2}|\nonumber \\
 & \leq C_{1}|\sigma_{1}-\sigma_{2}|,\label{eq:uniform_continuous_L}
\end{align}
where $C_{1}$ is a constant that only depends on $\minsigmaI$. Therefore,
for any $\proxfree\in\Lipndset$, $\calL(\proxfree,\sigma)$ is uniformly
Lipschitz continuous w.r.t. $\sigma$ on $\minsigmaI$. Consider $\proxfree_{\hat{\sigma}}$
which is the optimal solution of (\ref{eq:opt_esti_prox}) under $\hat{\sigma}$.
From the discussion in the last paragraph, for any $\sigma\in\minsigmaI$
and $r>0$, if $\h{\sigma}\in\bdball_{\veps}(\sigma)\bigcap\minsigmaI$
under small enough $\veps$, then there exists some $\proxfree\in\Lipndset_{\sigma}$
satisfying $|\proxfree_{\hat{\sigma}}(x)-f(x)|<r$ almost everywhere
on $\R$. Therefore, for this $\proxfree$ we have
\begin{equation}
\begin{aligned}|\calL(\proxfree_{\hat{\sigma}},\h{\sigma})-\calL(\proxfree,\sigma)| & \leq|\calL(\proxfree_{\hat{\sigma}},\h{\sigma})-\calL(\proxfree_{\hat{\sigma}},\sigma)|+|\calL(\proxfree_{\hat{\sigma}},\sigma)-\calL(\proxfree,\sigma)|\\
 & \leq C_{2}r,
\end{aligned}
\label{eq:L_f_sigma_perturbation}
\end{equation}
where $C_{2}$ is some constant that does not depend on $\veps,r$
and in the last step we use (\ref{eq:uniform_continuous_L}) and the
fact that $|\proxfree_{\hat{\sigma}}(x)-f(x)|<r$ almost everywhere.
Since $\calL(\h{\sigma})=\calL(\proxfree_{\hat{\sigma}},\h{\sigma})$,
we have from (\ref{eq:L_f_sigma_perturbation})
\begin{align*}
\calL(\h{\sigma}) & \tgeq{\text{(a)}}\calL(\proxfree,\sigma)-C_{2}r\tgeq{\text{(b)}}\calL(\sigma)-C_{2}r,
\end{align*}
where (a) follows from (\ref{eq:L_f_sigma_perturbation}) and (b)
follows from definition of $\calL(\sigma)$ in (\ref{eq:opt_esti_prox}).
By exchanging $\sigma$ and $\hat{\sigma}$, we can also get $\calL(\sigma)\geq\calL(\h{\sigma})-C_{2}r$.
In conclusion, for any compact interval $\minsigmaI\subseteq\R_{>0}$
and $r>0$, there exists $\veps>0$ such that for any $\sigma,\hat{\sigma}\in\minsigmaI$
satisfying $|\hat{\sigma}-\sigma|\leq\veps$, we have $|\calL(\h{\sigma})-\calL(\sigma)|\leq r$.
This proves the continuity of $\calL(\sigma)$ over $\R_{>0}$.
\end{IEEEproof}
\begin{lem}
\label{lem:min_sol_infA_lbd}Equation (\ref{eq:sigma_min_equation})
satisfies the following: (I) it always has a solution and the minimum
solution $\sigma_{0}\in[\sigma_{\noisei},(\sigma_{\noisei}^{2}+\delta^{-1}\E B^{2})^{1/2}]$,
(II) $\sigma_{0}=\inf\mathcal{A}$, where set $\mathcal{A}$ is defined
in (\ref{eq:sigmaset_def_1}).
\end{lem}
\begin{IEEEproof}
(I) We first prove $\sigma_{0}$ always exists and $\sigma_{0}\in\minsigmaI$,
where $\minsigmaI:=[\sigma_{\noisei},(\sigma_{\noisei}^{2}+\delta^{-1}\E B^{2})^{1/2}]$.
It is not hard to verify that at the boundary of $\minsigmaI$, ${\cal L}(\sigma)$
satisfies 
\begin{equation}
\begin{cases}
{\cal L}(\sigma)\leq\delta(\sigma^{2}-\sigma_{\noisei}^{2}) & \sigma=\big(\sigma_{\noisei}^{2}+\tfrac{\E B^{2}}{\delta}\big)^{1/2},\\
{\cal L}(\sigma)\geq\delta(\sigma^{2}-\sigma_{\noisei}^{2}) & \sigma=\sigma_{\noisei}.
\end{cases}\label{eq:L_sigma_bd_value}
\end{equation}
Indeed, when $\sigma=\big(\sigma_{\noisei}^{2}+\tfrac{\E B^{2}}{\delta}\big)^{1/2}$,
set $\proxfree=0$ in (\ref{eq:opt_esti_prox}) and $\E[\proxfree(B+\sigma H)-B]^{2}=\E B^{2}=\delta(\sigma^{2}-\sigma_{\noisei}^{2})$,
so ${\cal L}(\sigma)\leq\delta(\sigma^{2}-\sigma_{\noisei}^{2})$.
On the other hand, since ${\cal L}(\sigma)\geq0$, the second inequality
of (\ref{eq:L_sigma_bd_value}) immediately follows. Then by (\ref{eq:L_sigma_bd_value}),
the continuity of ${\cal L}(\sigma)$ shown in Lemma \ref{lem:f_sigma_unique_Lsigma_conti}
and the fact that $\sigma_{0}\geq\sigma_{\noisei}$, we know $\sigma_{0}$
always exists and $\sigma_{0}\in\minsigmaI$.

(II) To prove $\inf\mathcal{A}=\sigma_{0}$, we proceed as follows:
\begin{enumerate}
\item[(i)] Show
\begin{equation}
\inf\mathcal{A}\in\minsigmaI.\label{eq:inf_A_bounded}
\end{equation}
\item[(ii)] Prove the following membership certificate of set $\mathcal{A}$
for those $\sigma\in\minsigmaI$:
\begin{equation}
\sigma\in\mathcal{A}\Longleftrightarrow{\cal L}(\sigma)\leq\delta(\sigma^{2}-\sigma_{\noisei}^{2}).\label{eq:membership_cert}
\end{equation}
\end{enumerate}
It is not hard to see the above steps will imply $\inf\mathcal{A}=\sigma_{0}$.
Indeed, combining (\ref{eq:membership_cert}) with (\ref{eq:inf_A_bounded})
yields the following characterization of $\inf\mathcal{A}$: 
\begin{equation}
\inf\mathcal{A}=\inf\big\{\sigma\mid\sigma\in\minsigmaI\text{ and }{\cal L}(\sigma)\leq\delta(\sigma^{2}-\sigma_{\noisei}^{2})\big\}.\label{eq:char_inf_A}
\end{equation}
By (\ref{eq:L_sigma_bd_value}) and the continuity of ${\cal L}(\sigma)$,
the infimum on the RHS of (\ref{eq:char_inf_A}) is reached by $\sigma_{0}$,
which always exists. Therefore, $\inf\mathcal{A}=\sigma_{0}$. Step
(i)-(ii) can be proved as follows:

{[}Proof of (i){]} From the first equation of (\ref{eq:sigma2_cond1}),
we have $\inf\mathcal{A}\geq\sigma_{\noisei}$. On the other hand,
since $\proxfree=0$, $\sigma=\big(\sigma_{\noisei}^{2}+\tfrac{\E B^{2}}{\delta}\big)^{1/2}$,
$\tau=1$ is a solution of (\ref{eq:fixed_point_equation_f_sigmatau_1})-(\ref{eq:fixed_point_equation_f_sigmatau_2}),
from (\ref{eq:sigopt_2}) we have $\optsig\leq\big(\sigma_{\noisei}^{2}+\tfrac{\E B^{2}}{\delta}\big)^{1/2}$.
This together with the lower bound of $\optsig$ in (\ref{eq:sig_opt_lbd_1})
indicates: $\inf\mathcal{A}\leq\optsig\leq\big(\sigma_{\noisei}^{2}+\tfrac{\E B^{2}}{\delta}\big)^{1/2}$.
Therefore, $\inf\mathcal{A}\in\minsigmaI$.

{[}Proof of (ii){]} The ``$\Rightarrow$'' direction of (\ref{eq:membership_cert})
immediately follows from the definition of $\mathcal{A}$ in (\ref{eq:sigmaset_def_1}).
For the other direction, suppose we have a $\sigma\in\minsigmaI$
satisfying ${\cal L}(\sigma)\leq\delta(\sigma^{2}-\sigma_{\noisei}^{2})$.
Then 
\begin{equation}
\E[\proxfree_{\sigma}(B+\sigma H)-B]^{2}={\cal L}(\sigma)\leq\delta(\sigma^{2}-\sigma_{\noisei}^{2}),\label{eq:onepoint_small}
\end{equation}
where the first equality is due to that ${\cal L}(\sigma)$ can be
achieved by $\proxfree_{\sigma}$. Now consider the shrinkage of $\proxfree_{\sigma}$
as: $\alpha\proxfree_{\sigma}$, where $\alpha\in[0,1]$. Clearly,
for any $\alpha\in[0,1]$, $\alpha\proxfree_{\sigma}$ still satisfies
$\E[\alpha\proxfree_{\sigma}'(B+\sigma H)]\leq\delta$. Also we have:
\begin{equation}
\E[0\cdot\proxfree_{\sigma}(B+\sigma H)-B]^{2}=\E B^{2}\geq\delta(\sigma^{2}-\sigma_{\noisei}^{2}),\label{eq:zeropoint_large}
\end{equation}
where the last inequality follows from the condition that $\sigma\in\minsigmaI$.
On the other hand, it can be easily checked that $\alpha\mapsto\E[\alpha\proxfree_{\sigma}(B+\sigma H)-B]^{2}$
is continuous, so from (\ref{eq:zeropoint_large}), (\ref{eq:onepoint_small}),
there exists $\alpha_{0}\in[0,1]$ such that $(\alpha_{0}\proxfree_{\sigma},\sigma)$
is a solution of (\ref{eq:sigma2_cond1}), indicating $\sigma\in\mathcal{A}$.
\end{IEEEproof}

\subsection{Auxiliary Results for Proving Proposition \ref{prop:max_power}\label{subsec:Auxiliary-Results-for-testing}}
\begin{lem}
\label{lem:function_space1}For a probability measure $\mealim_{Y}\in\Wspace(\R)$,
define
\begin{equation}
\begin{aligned}\widetilde{\mathcal{M}}_{\mealim_{Y}}\bydef\big\{\prox(\cdot\,;\mealim_{Y},\mealim_{\Lambda}) & \mid\mealim_{\Lambda}\in\Wspace(\R)\text{ and }\text{if }\probthreshzero>0,\\
 & {\textstyle \int_{t}^{\probthreshzero}F_{\Lambda}^{-1}(u)du>\int_{t}^{\probthreshzero}F_{|Y|}^{-1}(u)du,\forall t\in[0,\probthreshzero)\big\}}
\end{aligned}
\label{eq:feasible_denoiser_2}
\end{equation}
where $\probthreshzero\bydef\P\big(\prox(Y;\mealim_{Y},\mealim_{\Lambda})=0\big)$.
Then for any $\mealim_{Y}\in\Wspace(\R)$, we have $\widetilde{\mathcal{M}}_{\mealim_{Y}}=\Lipndset$.
Correspondingly, for any $\proxfree(y)\in\Lipndset$, we can take
$\mealim_{\Lambda}$ as the law of $\max\{\ythreshzero,|Y|-\proxfree(|Y|)\}$
($Y\sim\mealim_{Y}$), so that $\prox(y\,;\mealim_{Y},\mealim_{\Lambda})=\proxfree(y)$.
Here $\ythreshzero\bydef\sup_{y\geq0}\left\{ y\mid\proxfree(y)=0\right\} $,
\end{lem}
\begin{IEEEproof}
The proof is similar as Proposition \ref{prop:function_space}. When
$\mealim_{\Lambda}$ is the law of $\max\{\ythreshzero,|Y|-\proxfree(|Y|)\}$,
optimization (\ref{eq:infinite_SLOPE}) ($\tau=1$) becomes:
\begin{equation}
\begin{aligned} & \min_{g\in\Lipndset}\frac{1}{2}\E\big\{[\proxfree(|Y|)-g(|Y|)]^{2}\indicatorfn_{|Y|\geq\ythreshzero}\big\}+\frac{1}{2}\E\big\{[(|Y|-\ythreshzero)-g(|Y|)]^{2}\indicatorfn_{|Y|<\ythreshzero}\big\}\\
 & \qquad+\frac{1}{2}\E Y^{2}-\E[|Y|-f(|Y|)]^{2}.
\end{aligned}
\label{eq:opti_lambda_problem_2}
\end{equation}
Since the feasible set in (\ref{eq:opti_lambda_problem_2}) is $\Lipndset$,
we know the optimal solution of (\ref{eq:opti_lambda_problem_2})
is exactly $\proxfree$. Recall that $\prox(y;\mealim_{Y},\mealim_{\Lambda})$
is the optimal solution of (\ref{eq:infinite_SLOPE}) ($\tau=1$),
so we have $\prox(y;\mealim_{Y},\mealim_{\Lambda})=\proxfree(y)$.

Finally, we show the law of $\max\{\ythreshzero,|Y|-\proxfree(|Y|)\}$
satisfies the constraint in (\ref{eq:feasible_denoiser_2}). Since
$\mealim_{Y}\in\Wspace(\R)$ and $\proxfree\in\Lipndset$, we can
easily get $\mealim_{\Lambda}\in\Wspace(\R)$. On the other hand,
suppose $\probthreshzero>0$. For any $t\in[0,\probthreshzero)$,
we have
\[
\begin{aligned}\int_{t}^{\probthreshzero}F_{|Y|}^{-1}(u)du & =\E\big(\indicatorfn_{F_{|Y|}^{-1}(t)\leq|Y|\leq F_{|Y|}^{-1}(\probthreshzero)}\cdot|Y|\big)\\
 & \tleq{\text{(a)}}\E\big(\indicatorfn_{F_{|Y|}^{-1}(t)\leq|Y|\leq\ythreshzero}\cdot|Y|\big)\\
 & \tle{\text{(b)}}\ythreshzero\P\big(F_{|Y|}^{-1}(t)\leq|Y|\leq\ythreshzero\big)\\
 & =\ythreshzero(\probthreshzero-t)\\
 & \teq{\text{(c)}}\int_{t}^{\probthreshzero}F_{\Lambda}^{-1}(u)du,
\end{aligned}
\]
where in (a) we use the fact that $F_{|Y|}^{-1}(\probthreshzero)\leq\ythreshzero$,
since $\prox(y;\mealim_{Y},\mealim_{\Lambda})=\proxfree(y)$ and $\probthreshzero=\P\big(\prox(Y;\mealim_{Y},\mealim_{\Lambda})=0\big)$,
(b) follows from $F_{|Y|}^{-1}(t)<F_{|Y|}^{-1}(\probthreshzero)=\ythreshzero$,
since $t<q_{0}$ and (c) is due to our choice of $\Lambda$, which
yields $F_{\Lambda}^{-1}(u)=\ythreshzero$ for any $0\leq u\leq\probthreshzero$.
\end{IEEEproof}
\begin{lem}
\label{lem:f_sigma_unique_L_alpha_sigma_conti}For $\alpha\in[0,1]$
and $\sigma>0$, we have: (I) optimization problem (\ref{eq:opt_testing_prox})
is convex and always has a unique optimal solution $\proxfree_{\alpha,\sigma}\in\Lipndset$,
(II) $\mathcal{L}_{\alpha}(\sigma)$ defined in (\ref{eq:opt_testing_prox})
is a continuous function over $\R_{>0}$, (III) equation $\mathcal{L}_{\alpha}(\sigma)=\delta(\sigma^{2}-\sigma_{\noisei}^{2})$
always has a solution and the minimum solution $\tminsigsol\in\big[\sigma_{\noisei},\,\sqrt{\sigma_{\noisei}^{2}+\delta^{-1}\E B^{2}}\big]$.
\end{lem}
\begin{IEEEproof}
(I) Comparing with (\ref{eq:opt_esti_prox}) and (\ref{eq:opt_testing_prox}),
we can find the only difference is that in (\ref{eq:opt_testing_prox}),
we add a constraint $\proxfree\in\softfunset$. It is not hard to
check for any $\alpha\in[0,1]$ and $\sigma>0$, the set $\softfunset$
is convex and closed in the $L^{2}$ space $\Lyspace_{\mu_{Y}}$ {[}definition
can be found in (\ref{eq:L2space_mealim_Y}){]}, so the uniqueness
of optimal solution of (\ref{eq:opt_testing_prox}) still holds using
the same arguments.

(II) The case of $\alpha=0\text{ or }1$ is easy. When $\alpha=1$,
we have $\Lipndset\subseteq\softfunset$, so $\mathcal{L}_{\alpha}(\sigma)=\mathcal{L}(\sigma)$
and its continuity is proved in the last part of Lemma \ref{lem:f_sigma_unique_Lsigma_conti};
when $\alpha=0$, $\softfunset$ contains only one element: $\proxfree(x)=0$
and $\mathcal{L}_{\alpha}(\sigma)=\E B^{2}$, which is trivially continuous.
Therefore, it only remains to verify for the case when $\alpha\in(0,1)$.

The proof for case $\alpha\in(0,1)$ is similar to the proof of continuity
of $\mathcal{L}(\sigma)$ in Lemma \ref{lem:f_sigma_unique_Lsigma_conti}.
For any $\alpha\in(0,1)$ and $\sigma>0$, define the following set
\begin{equation}
\Lipndset_{\alpha,\sigma}\bydef\{\proxfree\mid f\in\Lipndset\cap\softfunset\text{ and }\E\proxfree^{\prime}(B+\sigma H)\leq\delta\},\label{eq:set_Lipbd_sigma-1}
\end{equation}
where $H\sim\mathcal{N}(0,1)$ and $B\sim\mealim_{B}$ are independent.
We always have $\Lipndset_{\alpha,\sigma}\neq\emptyset$, since $\{f=0\}\in\Lipndset_{\alpha,\sigma}$.
Next we show that for any $\alpha\in(0,1)$ and $\sigma,r>0$, there
exists $\veps\in(0,\sigma/2)$ such that whenever $\h{\sigma}\in\bdball_{\veps}(\sigma)$
and $\h{\proxfree}\in\Lipndset_{\alpha,\hat{\sigma}}$, we can always
find a $\proxfree\in\Lipndset_{\alpha,\sigma}$ satisfying $|\proxfree(x)-\h{\proxfree}(x)|<r$
almost everywhere on $\R$. First, for any $\hat{\sigma}>0$ and $\h{\proxfree}\in\Lipndset_{\alpha,\hat{\sigma}}$,
we have $|\h{\proxfree}(y)|=0$, when $|y|\leq\Phi^{-1}(1-\tfrac{\alpha}{2})\hat{\sigma}$.
We can then apply the following shrinkage to $\h{\proxfree}$:
\begin{equation}
\h{\proxfree}_{T}(y)=\sgn(y)\max\{0,\h{\proxfree}(|y|)-\Phi^{-1}(1-\tfrac{\alpha}{2})|\sigma-\hat{\sigma}|\}.\label{eq:f_shrunk_1}
\end{equation}
One can check $\h{\proxfree}_{T}$ in (\ref{eq:f_shrunk_1}) satisfies:
$\h{\proxfree}_{T}\in\Lipndset_{\alpha,\hat{\sigma}}\cap\softfunset$
and 
\begin{equation}
0\leq\hat{\proxfree}(y)-\h{\proxfree}_{T}(y)\leq\Phi^{-1}(1-\tfrac{\alpha}{2})|\sigma-\hat{\sigma}|\label{eq:fhat_fT_diff}
\end{equation}
almost everywhere on $\R$. For small enough $\veps>0$ and $\h{\sigma}\in\bdball_{\veps}(\sigma)$,
we can then follow the same steps leading to (\ref{eq:shrink_y_2})
in the proof of continuity of $\mathcal{L}(\sigma)$ in Lemma \ref{lem:f_sigma_unique_Lsigma_conti}
to obtain a $\proxfree\in\Lipndset$ satisfying 
\begin{equation}
0\leq\h{\proxfree}_{T}(y)-\proxfree(y)\leq\frac{r}{2}\label{eq:fT_f_diff}
\end{equation}
almost everywhere on $\R$ and $\E\proxfree^{\prime}(B+\sigma H)\leq\delta$.
Meanwhile, since $\h{\proxfree}_{T}\in\softfunset$, we also have
$\proxfree\in\softfunset$. As a result, $\proxfree\in\Lipndset_{\alpha,\sigma}$.
Besides, combining (\ref{eq:fhat_fT_diff}) and (\ref{eq:fT_f_diff})
we have for small enough $\veps>0$ and $\h{\sigma}\in\bdball_{\veps}(\sigma)$,
$0\leq\hat{\proxfree}(y)-\proxfree(y)\leq r$ almost everywhere on
$\R$. The remaining proof is completely same as the last part of
the proof of $\mathcal{L}(\sigma)$'s continuity and we omit the details
here.

(III) The proof is the same as Lemma \ref{lem:min_sol_infA_lbd}.
Using the same argument, it can be verified that $\mathcal{L}_{\alpha}(\sigma)$
also satisfies 
\[
\begin{cases}
\mathcal{L}_{\alpha}(\sigma)\leq\delta(\sigma^{2}-\sigma_{\noisei}^{2}) & \sigma=\big(\sigma_{\noisei}^{2}+\tfrac{\E B^{2}}{\delta}\big)^{1/2},\\
\mathcal{L}_{\alpha}(\sigma)\geq\delta(\sigma^{2}-\sigma_{\noisei}^{2}) & \sigma=\sigma_{\noisei}.
\end{cases}
\]
Then by the continuity of $\mathcal{L}_{\alpha}(\sigma)$ and the
fact that $\tminsigsol\geq\sigma_{\noisei}$, we get the desired result.
\end{IEEEproof}
\begin{lem}
\label{lem:testing_tightness_cond_1}For any given $\alpha\in[0,1]$,
we have the following.
\begin{enumerate}
\item[(a)] It is always true that $\toptsig\geq\tminsigsol$.
\item[(b)]  If $\delta^{-1}\E\big[\proxfree_{\alpha}'(\toptY)\big]<1$, then
$\toptsig=\tminsigsol$ and the infimum of (\ref{eq:slope_test_opt_ubd1})
can be achieved by choosing $\mealim_{\Lambda}=\mealim_{\text{opt},\alpha}$.
\end{enumerate}
\end{lem}
\begin{IEEEproof}
The proof is similar to that of Proposition \ref{prop:min_MSE} (c).
Recall that a key step in the proof is the conversion of the optimization
over $\mealim_{\Lambda}$ into an equivalent optimization over realizable
limiting scalar functions $\proxfree$. We want to adopt the same
strategy here, but since an additional constraint on $\mealim_{\Lambda}$
is added, we need to determine the new realizable set of $\proxfree$,
as we did in Proposition \ref{prop:function_space}. It turns out
that the new realizable set is still $\Lipndset$. This result is
proved in In Lemma \ref{lem:function_space1} and it enables us to
follow the same steps leading to (\ref{eq:sigopt_2}) to show that
(\ref{eq:slope_test_opt_ubd1}) is equivalent to
\begin{equation}
\begin{aligned}\toptsig= & \inf\{\sigma\mid(\sigma,\tau)\in\mathcal{D}_{F}(\alpha),\text{ for some }\text{\ensuremath{\tau>0}}\},\end{aligned}
\label{eq:slope_test_opt_ubd2}
\end{equation}
where $\mathcal{D}_{F}(\alpha)$ is defined as:
\[
\mathcal{D}_{F}(\alpha)\bydef\big\{(\sigma,\tau)\in\R_{>0}^{2}:\,\exists\proxfree\in\Lipndset\cap\softfunset\text{ s.t. }(\proxfree,\sigma,\tau)\text{ satisfies \eqref{eq:fixed_point_equation_f_sigmatau_1}-\eqref{eq:fixed_point_equation_f_sigmatau_2}}\big\}.
\]
Comparing (\ref{eq:sigopt_2}) and (\ref{eq:slope_test_opt_ubd2}),
it can be seen that the only difference is that in (\ref{eq:slope_test_opt_ubd2})
we require $f\in\softfunset$, which is needed to control the type-I
error level. Then similar as (\ref{eq:sigmaset_def_1}) we define
\[
\mathcal{A}(\alpha)\bydef\big\{\sigma\in\R_{>0}:\,\exists\proxfree\in\Lipndset\cap\softfunset,\text{s.t. \ensuremath{(\proxfree,\sigma)}\text{ satisfies \eqref{eq:sigma2_cond1}}}\big\}
\]
and it holds that $\toptsig\geq\inf\mathcal{A}(\alpha)$. Meanwhile,
by the same reasoning in Lemma \ref{lem:min_sol_infA_lbd}, we can
also show $\inf\mathcal{A}(\alpha)=\tminsigsol$. This gives us $\toptsig\geq\tminsigsol$.

Now consider the scenario where $\delta^{-1}\E\big[\proxfree_{\alpha}'(\toptY)\big]<1$.
In this case, we have $\tmintausol\in(0,\infty)$. Then it is not
hard to check $(\proxfree_{\alpha},\tminsigsol,\tmintausol)$ satisfies
equation (\ref{eq:fixed_point_equation_f_sigmatau_1})-(\ref{eq:fixed_point_equation_f_sigmatau_2}).
Therefore, $(\tminsigsol,\tmintausol)\in\mathcal{D}_{F}(\alpha)$.
By (\ref{eq:slope_test_opt_ubd2}), we have $\tminsigsol\geq\toptsig$.
Since we already get $\toptsig\geq\tminsigsol$, we can conclude that
$\toptsig=\tminsigsol$. Let us denote $(\sigOpt,\tauOpt)$ as the
solution of fixed-point equation (\ref{eq:sigmatau_1})-(\ref{eq:sigmatau_2}),
when $\mealim_{\Lambda}=\mealim_{\text{opt},\alpha}$. Using Lemma
\ref{lem:function_space1}, it is not hard to check 
\begin{equation}
(\sigOpt,\tauOpt)=(\tminsigsol,\tmintausol)\label{eq:sigmastar_reached}
\end{equation}
and 
\begin{equation}
\prox(y;\mealim_{\YOpt},\mealim_{\tauOpt\Lambda})=\proxfree_{\alpha}(y).\label{eq:falpha_y_reached}
\end{equation}
Meanwhile, we have $\mealim_{\text{opt},\alpha}\in\tLamspace$. Recall
that the infimum of $\sigOpt$ in (\ref{eq:slope_test_opt_ubd1})
is $\tminsigsol$ {[}c.f. (\ref{eq:slope_test_opt_ubd2}){]}. As a
result, the infimum of $\sigOpt$ in (\ref{eq:slope_test_opt_ubd1})
is reached when $\mealim_{\Lambda}=\mealim_{\text{opt},\alpha}$.
\end{IEEEproof}
\begin{lem}
\label{lem:testing_tightness_cond_2}For $\alpha\in(0,1]$, if $\delta^{-1}\E\big[\proxfree_{\alpha}'(\toptY)\big]<1$
and $\tythreshzero=\Phi^{-1}(1-\frac{\alpha}{2})\tminsigsol$, then
$\overline{\mathcal{\limpower}}(\alpha)=\limpower(\alpha)$ and $\lim_{p\to\infty}\text{Power}=\limpower(\alpha)$,
when $\mealim_{\Lambda}=\mealim_{\text{opt},\alpha}$.
\end{lem}
\begin{IEEEproof}
In the proof, let $(\sigOpt,\tauOpt)$ be the solution of fixed-point
equation (\ref{eq:sigmatau_1})-(\ref{eq:sigmatau_2}), when $\mealim_{\Lambda}=\mealim_{\text{opt},\alpha}$.
Also denote $\ythresh:=\sup_{y\geq0}\{y\mid\prox(y;\mealim_{\YOpt},\mealim_{\tauOpt\Lambda})=0\}$.

Assume $\delta^{-1}\E\big[\proxfree_{\alpha}'(\toptY)\big]<1$ and
$\tythreshzero=\Phi^{-1}(1-\frac{\alpha}{2})\tminsigsol$. Recall
from the proof of Lemma \ref{lem:testing_tightness_cond_1}, when
$\mealim_{\Lambda}=\mealim_{\text{opt},\alpha}$, we have
\[
\ythresh\teq{\text{(a)}}\tythreshzero\teq{\text{(b)}}\Phi^{-1}(1-\frac{\alpha}{2})\tminsigsol\teq{\text{(c)}}\Phi^{-1}(1-\frac{\alpha}{2})\sigOpt,
\]
where (a) follows from (\ref{eq:falpha_y_reached}), (b) follows from
assumption $\tythreshzero=\Phi^{-1}(1-\frac{\alpha}{2})\tminsigsol$
and (c) follows from (\ref{eq:sigmastar_reached}). Therefore, 
\begin{align}
\P(|B+\sigOpt H|\geq\ythresh\mid B\neq0) & =\P\big(|B+\sigOpt H|\geq\Phi^{-1}(1-\tfrac{\alpha}{2})\sigOpt\mid B\neq0\big)\nonumber \\
 & =\overline{\mathcal{\limpower}}(\alpha),\label{eq:P_ubd_tight_2}
\end{align}
where the last equality is due to that $\mealim_{\text{opt},\alpha}$
is the optimal solution of (\ref{eq:slope_test_opt_ubd1}), as is
proved by Lemma \ref{lem:testing_tightness_cond_1}. Therefore, from
(\ref{eq:P_ubd_tight_2}) we know when $\mealim_{\Lambda}=\mealim_{\text{opt},\alpha}$,
the objective value of (\ref{eq:slope_test_opt}) equals to $\overline{\mathcal{\limpower}}(\alpha)$.
This implies $\limpower(\alpha)\geq\overline{\mathcal{\limpower}}(\alpha)$,
since $\limpower(\alpha)$ is the optimal value of (\ref{eq:slope_test_opt}).
Combined with the fact that $\overline{\mathcal{\limpower}}(\alpha)$
is the upper bound of $\limpower(\alpha)$ {[}c.f. (\ref{eq:power_ubd2}){]},
it then follows that $\overline{\mathcal{\limpower}}(\alpha)=\limpower(\alpha)$.
Also when $\mealim_{\Lambda}=\mealim_{\text{opt},\alpha}$, $\lim_{p\to\infty}\text{Power}=\overline{\mathcal{\limpower}}(\alpha)=\limpower(\alpha)$.

\end{IEEEproof}



\bibliographystyle{IEEEtran}
\bibliography{refs}

\begin{thebibliography}{10}
\providecommand{\url}[1]{#1}
\csname url@samestyle\endcsname
\providecommand{\newblock}{\relax}
\providecommand{\bibinfo}[2]{#2}
\providecommand{\BIBentrySTDinterwordspacing}{\spaceskip=0pt\relax}
\providecommand{\BIBentryALTinterwordstretchfactor}{4}
\providecommand{\BIBentryALTinterwordspacing}{\spaceskip=\fontdimen2\font plus
\BIBentryALTinterwordstretchfactor\fontdimen3\font minus
  \fontdimen4\font\relax}
\providecommand{\BIBforeignlanguage}[2]{{%
\expandafter\ifx\csname l@#1\endcsname\relax
\typeout{** WARNING: IEEEtran.bst: No hyphenation pattern has been}%
\typeout{** loaded for the language `#1'. Using the pattern for}%
\typeout{** the default language instead.}%
\else
\language=\csname l@#1\endcsname
\fi
#2}}
\providecommand{\BIBdecl}{\relax}
\BIBdecl

\bibitem{Hu2019AsymptoticsAO}
H.~Hu and Y.~M. Lu, ``Asymptotics and optimal designs of {SLOPE} for sparse
  linear regression,'' \emph{2019 IEEE International Symposium on Information
  Theory (ISIT)}, pp. 375--379, 2019.

\bibitem{bogdan2015slope}
M.~Bogdan, E.~Van Den~Berg, C.~Sabatti, W.~Su, and E.~J. Cand{\`e}s,
  ``{SLOPE}-adaptive variable selection via convex optimization,'' \emph{The
  annals of applied statistics}, vol.~9, no.~3, p. 1103, 2015.

\bibitem{zhong2012efficient}
L.~W. Zhong and J.~T. Kwok, ``Efficient sparse modeling with automatic feature
  grouping,'' \emph{IEEE transactions on neural networks and learning systems},
  vol.~23, no.~9, pp. 1436--1447, 2012.

\bibitem{zeng2014decreasing}
X.~Zeng and M.~A. Figueiredo, ``Decreasing weighted sorted $\ell_1$
  regularization,'' \emph{IEEE Signal Processing Letters}, vol.~21, no.~10, pp.
  1240--1244, 2014.

\bibitem{bondell2008simultaneous}
H.~D. Bondell and B.~J. Reich, ``Simultaneous regression shrinkage, variable
  selection, and supervised clustering of predictors with {OSCAR},''
  \emph{Biometrics}, vol.~64, no.~1, pp. 115--123, 2008.

\bibitem{figueiredo2016ordered}
M.~Figueiredo and R.~Nowak, ``Ordered weighted $\ell_1$ regularized regression
  with strongly correlated covariates: Theoretical aspects,'' in
  \emph{Artificial Intelligence and Statistics}, 2016, pp. 930--938.

\bibitem{su2016slope}
W.~Su, E.~Candes \emph{et~al.}, ``{SLOPE} is adaptive to unknown sparsity and
  asymptotically minimax,'' \emph{The Annals of Statistics}, vol.~44, no.~3,
  pp. 1038--1068, 2016.

\bibitem{bellec2018slope}
P.~C. Bellec, G.~Lecu{\'e}, A.~B. Tsybakov \emph{et~al.}, ``{SLOPE} meets
  {LASSO}: improved oracle bounds and optimality,'' \emph{The Annals of
  Statistics}, vol.~46, no.~6B, pp. 3603--3642, 2018.

\bibitem{bogdan2013statistical}
M.~Bogdan, E.~v.~d. Berg, W.~Su, and E.~Candes, ``Statistical estimation and
  testing via the sorted $\ell_{1}$ norm,'' \emph{arXiv preprint
  arXiv:1310.1969}, 2013.

\bibitem{bayati2012lasso}
M.~Bayati and A.~Montanari, ``The {LASSO} risk for gaussian matrices,''
  \emph{IEEE Transactions on Information Theory}, vol.~58, no.~4, pp.
  1997--2017, 2012.

\bibitem{su2017false}
W.~Su, M.~Bogdan, E.~Candes \emph{et~al.}, ``False discoveries occur early on
  the {LASSO} path,'' \emph{The Annals of Statistics}, vol.~45, no.~5, pp.
  2133--2150, 2017.

\bibitem{oymak2013squared}
S.~Oymak, C.~Thrampoulidis, and B.~Hassibi, ``The squared-error of generalized
  lasso: A precise analysis,'' in \emph{2013 51st Annual Allerton Conference on
  Communication, Control, and Computing (Allerton)}.\hskip 1em plus 0.5em minus
  0.4em\relax IEEE, 2013, pp. 1002--1009.

\bibitem{el2018impact}
N.~El~Karoui, ``On the impact of predictor geometry on the performance on
  high-dimensional ridge-regularized generalized robust regression
  estimators,'' \emph{Probability Theory and Related Fields}, vol. 170, no.
  1-2, pp. 95--175, 2018.

\bibitem{thrampoulidis2016ber}
C.~Thrampoulidis, E.~Abbasi, W.~Xu, and B.~Hassibi, ``{BER} analysis of the box
  relaxation for {BPSK} signal recovery,'' in \emph{2016 IEEE International
  Conference on Acoustics, Speech and Signal Processing (ICASSP)}.\hskip 1em
  plus 0.5em minus 0.4em\relax IEEE, 2016, pp. 3776--3780.

\bibitem{zheng2017does}
L.~Zheng, A.~Maleki, H.~Weng, X.~Wang, and T.~Long, ``Does
  $\ell_{p}$-minimization outperform $\ell_{1}$-minimization?'' \emph{IEEE
  Transactions on Information Theory}, vol.~63, no.~11, pp. 6896--6935, 2017.

\bibitem{barbier2016mutual}
J.~Barbier, M.~Dia, N.~Macris, and F.~Krzakala, ``The mutual information in
  random linear estimation,'' in \emph{2016 54th Annual Allerton Conference on
  Communication, Control, and Computing (Allerton)}.\hskip 1em plus 0.5em minus
  0.4em\relax IEEE, 2016, pp. 625--632.

\bibitem{reeves2016replica}
G.~Reeves and H.~D. Pfister, ``The replica-symmetric prediction for compressed
  sensing with gaussian matrices is exact,'' in \emph{2016 IEEE International
  Symposium on Information Theory (ISIT)}.\hskip 1em plus 0.5em minus
  0.4em\relax IEEE, 2016, pp. 665--669.

\bibitem{donoho2009observed}
D.~Donoho and J.~Tanner, ``Observed universality of phase transitions in
  high-dimensional geometry, with implications for modern data analysis and
  signal processing,'' \emph{Philosophical Transactions of the Royal Society of
  London A: Mathematical, Physical and Engineering Sciences}, vol. 367, no.
  1906, pp. 4273--4293, 2009.

\bibitem{donoho2009message}
D.~L. Donoho, A.~Maleki, and A.~Montanari, ``Message-passing algorithms for
  compressed sensing,'' \emph{Proceedings of the National Academy of Sciences},
  vol. 106, no.~45, pp. 18\,914--18\,919, 2009.

\bibitem{bayati2011dynamics}
M.~Bayati and A.~Montanari, ``The dynamics of message passing on dense graphs,
  with applications to compressed sensing,'' \emph{IEEE Transactions on
  Information Theory}, vol.~57, no.~2, pp. 764--785, 2011.

\bibitem{chandrasekaran2012convex}
V.~Chandrasekaran, B.~Recht, P.~A. Parrilo, and A.~S. Willsky, ``The convex
  geometry of linear inverse problems,'' \emph{Foundations of Computational
  mathematics}, vol.~12, no.~6, pp. 805--849, 2012.

\bibitem{krzakala2012statistical}
F.~Krzakala, M.~M{\'e}zard, F.~Sausset, Y.~Sun, and L.~Zdeborov{\'a},
  ``Statistical-physics-based reconstruction in compressed sensing,''
  \emph{Physical Review X}, vol.~2, no.~2, p. 021005, 2012.

\bibitem{javanmard2013state}
A.~Javanmard and A.~Montanari, ``State evolution for general approximate
  message passing algorithms, with applications to spatial coupling,''
  \emph{Information and Inference: A Journal of the IMA}, vol.~2, no.~2, pp.
  115--144, 2013.

\bibitem{el2013robust}
N.~El~Karoui, D.~Bean, P.~J. Bickel, C.~Lim, and B.~Yu, ``On robust regression
  with high-dimensional predictors,'' \emph{Proceedings of the National Academy
  of Sciences}, vol. 110, no.~36, pp. 14\,557--14\,562, 2013.

\bibitem{amelunxen2014living}
D.~Amelunxen, M.~Lotz, M.~B. McCoy, and J.~A. Tropp, ``Living on the edge:
  Phase transitions in convex programs with random data,'' \emph{Information
  and Inference: A Journal of the IMA}, vol.~3, no.~3, pp. 224--294, 2014.

\bibitem{thrampoulidis2015regularized}
C.~Thrampoulidis, S.~Oymak, and B.~Hassibi, ``Regularized linear regression: A
  precise analysis of the estimation error,'' in \emph{Conference on Learning
  Theory}, 2015, pp. 1683--1709.

\bibitem{donoho2016high}
D.~Donoho and A.~Montanari, ``High dimensional robust m-estimation: Asymptotic
  variance via approximate message passing,'' \emph{Probability Theory and
  Related Fields}, vol. 166, no.~3, pp. 935--969, 2016.

\bibitem{dhifallah2017phase}
O.~Dhifallah, C.~Thrampoulidis, and Y.~M. Lu, ``Phase retrieval via linear
  programming: Fundamental limits and algorithmic improvements,'' in \emph{2017
  55th Annual Allerton Conference on Communication, Control, and Computing
  (Allerton)}.\hskip 1em plus 0.5em minus 0.4em\relax IEEE, 2017, pp.
  1071--1077.

\bibitem{thrampoulidis2018precise}
C.~Thrampoulidis, E.~Abbasi, and B.~Hassibi, ``Precise error analysis of
  regularized {$M$}-estimators in high-dimensions,'' \emph{IEEE Transactions on
  Information Theory}, 2018.

\bibitem{sur2019modern}
P.~Sur and E.~J. Cand{\`e}s, ``A modern maximum-likelihood theory for
  high-dimensional logistic regression,'' \emph{Proceedings of the National
  Academy of Sciences}, vol. 116, no.~29, pp. 14\,516--14\,525, 2019.

\bibitem{tanaka2002statistical}
T.~Tanaka, ``A statistical-mechanics approach to large-system analysis of cdma
  multiuser detectors,'' \emph{IEEE Transactions on Information theory},
  vol.~48, no.~11, pp. 2888--2910, 2002.

\bibitem{kabashima2009typical}
Y.~Kabashima, T.~Wadayama, and T.~Tanaka, ``A typical reconstruction limit for
  compressed sensing based on $\ell_p$-norm minimization,'' \emph{Journal of
  Statistical Mechanics: Theory and Experiment}, vol. 2009, no.~09, p. L09003,
  2009.

\bibitem{gordon1985some}
Y.~Gordon, ``Some inequalities for gaussian processes and applications,''
  \emph{Israel Journal of Mathematics}, vol.~50, no.~4, pp. 265--289, 1985.

\bibitem{stojnic2013framework}
M.~Stojnic, ``A framework to characterize performance of lasso algorithms,''
  \emph{arXiv preprint arXiv:1303.7291}, 2013.

\bibitem{salehi2018precise}
F.~Salehi, E.~Abbasi, and B.~Hassibi, ``A precise analysis of phasemax in phase
  retrieval,'' in \emph{2018 IEEE International Symposium on Information Theory
  (ISIT)}.\hskip 1em plus 0.5em minus 0.4em\relax IEEE, 2018, pp. 976--980.

\bibitem{deng2019model}
Z.~Deng, A.~Kammoun, and C.~Thrampoulidis, ``A model of double descent for
  high-dimensional binary linear classification,'' \emph{arXiv preprint
  arXiv:1911.05822}, 2019.

\bibitem{montanari2019generalization}
A.~Montanari, F.~Ruan, Y.~Sohn, and J.~Yan, ``The generalization error of
  max-margin linear classifiers: High-dimensional asymptotics in the
  overparametrized regime,'' \emph{arXiv preprint arXiv:1911.01544}, 2019.

\bibitem{kini2020analytic}
G.~R. Kini and C.~Thrampoulidis, ``Analytic study of double descent in binary
  classification: The impact of loss,'' in \emph{2020 IEEE International
  Symposium on Information Theory (ISIT)}.\hskip 1em plus 0.5em minus
  0.4em\relax IEEE, 2020, pp. 2527--2532.

\bibitem{bean2013optimal}
D.~Bean, P.~J. Bickel, N.~El~Karoui, and B.~Yu, ``Optimal m-estimation in
  high-dimensional regression,'' \emph{Proceedings of the National Academy of
  Sciences}, vol. 110, no.~36, pp. 14\,563--14\,568, 2013.

\bibitem{taheri2021sharp}
H.~Taheri, R.~Pedarsani, and C.~Thrampoulidis, ``Sharp guarantees and optimal
  performance for inference in binary and gaussian-mixture models,''
  \emph{Entropy}, vol.~23, no.~2, p. 178, 2021.

\bibitem{advani2016statistical}
M.~Advani and S.~Ganguli, ``Statistical mechanics of optimal convex inference
  in high dimensions,'' \emph{Physical Review X}, vol.~6, no.~3, p. 031034,
  2016.

\bibitem{aubin2020generalization}
B.~Aubin, F.~Krzakala, Y.~M. Lu, and L.~Zdeborov{\'a}, ``Generalization error
  in high-dimensional perceptrons: Approaching bayes error with convex
  optimization,'' \emph{arXiv preprint arXiv:2006.06560}, 2020.

\bibitem{taheri2021fundamental}
H.~Taheri, R.~Pedarsani, and C.~Thrampoulidis, ``Fundamental limits of
  ridge-regularized empirical risk minimization in high dimensions,'' in
  \emph{International Conference on Artificial Intelligence and
  Statistics}.\hskip 1em plus 0.5em minus 0.4em\relax PMLR, 2021, pp.
  2773--2781.

\bibitem{weng2018overcoming}
H.~Weng, A.~Maleki, L.~Zheng \emph{et~al.}, ``Overcoming the limitations of
  phase transition by higher order analysis of regularization techniques,''
  \emph{Annals of Statistics}, vol.~46, no.~6A, pp. 3099--3129, 2018.

\bibitem{wang2020bridge}
S.~Wang, H.~Weng, A.~Maleki \emph{et~al.}, ``Which bridge estimator is the best
  for variable selection?'' \emph{Annals of Statistics}, vol.~48, no.~5, pp.
  2791--2823, 2020.

\bibitem{celentano2019fundamental}
M.~Celentano and A.~Montanari, ``Fundamental barriers to high-dimensional
  regression with convex penalties,'' \emph{arXiv preprint arXiv:1903.10603},
  2019.

\bibitem{bu2020algorithmic}
Z.~Bu, J.~M. Klusowski, C.~Rush, and W.~J. Su, ``Algorithmic analysis and
  statistical estimation of slope via approximate message passing,'' \emph{IEEE
  Transactions on Information Theory}, vol.~67, no.~1, pp. 506--537, 2020.

\bibitem{wang2019does}
S.~Wang, H.~Weng, and A.~Maleki, ``Does slope outperform bridge regression?''
  \emph{arXiv preprint arXiv:1909.09345}, 2019.

\bibitem{celentano2019approximate}
M.~Celentano, ``Approximate separability of symmetrically penalized least
  squares in high dimensions: characterization and consequences,'' \emph{arXiv
  preprint arXiv:1906.10319}, 2019.

\bibitem{miolane2018distribution}
L.~Miolane and A.~Montanari, ``The distribution of the {Lasso}: Uniform control
  over sparse balls and adaptive parameter tuning,'' \emph{arXiv preprint
  arXiv:1811.01212}, 2018.

\bibitem{mousavi2018consistent}
A.~Mousavi, A.~Maleki, and R.~G. Baraniuk, ``Consistent parameter estimation
  for {LASSO} and approximate message passing,'' \emph{The Annals of
  Statistics}, vol.~46, no.~1, pp. 119--148, 2018.

\bibitem{donoho2011noise}
D.~L. Donoho, A.~Maleki, and A.~Montanari, ``The noise-sensitivity phase
  transition in compressed sensing,'' \emph{IEEE Transactions on Information
  Theory}, vol.~57, no.~10, pp. 6920--6941, 2011.

\bibitem{van2000asymptotic}
A.~W. Van~der Vaart, \emph{Asymptotic statistics}.\hskip 1em plus 0.5em minus
  0.4em\relax Cambridge university press, 2000, vol.~3.

\bibitem{montanari2017universality}
A.~Montanari and P.-M. Nguyen, ``Universality of the elastic net error,'' in
  \emph{2017 IEEE International Symposium on Information Theory (ISIT)}.\hskip
  1em plus 0.5em minus 0.4em\relax IEEE, 2017, pp. 2338--2342.

\bibitem{panahi2017universal}
A.~Panahi and B.~Hassibi, ``A universal analysis of large-scale regularized
  least squares solutions,'' in \emph{Advances in Neural Information Processing
  Systems}, 2017, pp. 3381--3390.

\bibitem{villani2003topics}
C.~Villani, \emph{Topics in optimal transportation}.\hskip 1em plus 0.5em minus
  0.4em\relax American Mathematical Soc., 2003, no.~58.

\bibitem{royden2010real}
H.~L. Royden, \emph{Real analysis}.\hskip 1em plus 0.5em minus 0.4em\relax
  Prentice Hall, 2010.

\bibitem{hu1997strong}
T.-C. Hu and R.~Taylor, ``On the strong law for arrays and for the bootstrap
  mean and variance,'' \emph{International Journal of Mathematics and
  Mathematical Sciences}, vol.~20, no.~2, pp. 375--382, 1997.

\bibitem{sion1958general}
M.~Sion, ``On general minimax theorems.'' \emph{Pacific Journal of
  mathematics}, vol.~8, no.~1, pp. 171--176, 1958.

\bibitem{bertsekas2009convex}
D.~P. Bertsekas, \emph{Convex optimization theory}.\hskip 1em plus 0.5em minus
  0.4em\relax Athena Scientific, 2009.

\bibitem{aliprantis06}
C.~D. Aliprantis and K.~C. Border, \emph{Infinite Dimensional Analysis: a
  Hitchhiker's Guide}.\hskip 1em plus 0.5em minus 0.4em\relax Springer, 2006.

\bibitem{rockafellar2009variational}
R.~T. Rockafellar and R.~J.-B. Wets, \emph{Variational analysis}.\hskip 1em
  plus 0.5em minus 0.4em\relax Springer Science \& Business Media, 2009, vol.
  317.

\bibitem{rudin1976principles}
W.~Rudin \emph{et~al.}, \emph{Principles of mathematical analysis}.\hskip 1em
  plus 0.5em minus 0.4em\relax McGraw-hill New York, 1976, vol.~3, no. 4.2.

\end{thebibliography}

\end{document}